\documentclass[pra,aps,amsmath,amssymb,twocolumn,eqsecnum]{revtex4-2}

\usepackage{epsfig}
\usepackage{graphicx}
\usepackage{bm}
\usepackage{relsize}
\DeclareMathOperator\erf{erf}
\DeclareMathOperator\e{e}
\usepackage{color}

\begin{document}

\title{ Orbital liquid in the $e_g$ orbital Hubbard model in $d=\infty$ dimensions }

\author {     Louis Felix Feiner }
\affiliation{\mbox{Institute for Theoretical Physics, Utrecht University,
                 PO Box 80.000, 3508 TA Utrecht, Netherlands}}

\author {     Andrzej M. Ole\'{s} }
\affiliation{ Max Planck Institute for Solid State Research,
              Heisenbergstrasse 1, D-70569 Stuttgart, Germany }
\affiliation{\mbox{Institute of Theoretical Physics, Jagiellonian University,
              Prof. Stanis\l{}awa \L{}ojasiewicza 11, PL-30348 Krak\'ow, Poland}}

\date{31 October, 2022}

\begin{abstract}
We demonstrate that the three-dimensional $e_g$ orbital Hubbard model
can be generalized to arbitrary dimension $d$, and that the form of
the result is determined uniquely by the requirements that
 (i) the two-fold degeneracy of the $e_g$ orbital be retained, and
(ii) the cubic lattice be turned into a hypercubic lattice.
While the local Coulomb interaction $U$ is invariant for each basis of
orthogonal orbitals, the form of the kinetic energy depends on the
orbital basis and takes the most symmetric form for the so-called
complex-orbital basis. Characteristically, with respect to this basis,
the model has two hopping channels, one that is orbital-flavor
conserving, and a second one that is orbital-flavor non-conserving.
We show that the noninteracting electronic structure consists of two
nondegenerate bands of plane-wave real-orbital single-particle states
for which the orbital depends on the wave vector. Due to the latter
feature each band is unpolarized at any filling, and has a non-Gaussian
density of states at $d=\infty$. The \textit{orbital liquid} state is
obtained by filling these two bands up to the same Fermi energy. We
investigate the $e_g$ orbital Hubbard model in the limit $d\to\infty$,
treating the on-site Coulomb interaction $U$ within the Gutzwiller
approximation, thus determining the correlation energy of the orbital
liquid and the (disordered) para-orbital states. In perfect analogy
with the case of the spin Hubbard model, the Gutzwiller approximation
is demonstrated to be exact at $d=\infty$ for the orbital Hubbard
model, because of the collapse of electron correlations to a single
site. At half-filling (one electron per site on average, $n=1$) one
finds a Brinkman-Rice type ``metal-insulator'' transition in the
orbital liquid, which is analogous to the transition for a paramagnetic
state in the spin model, but occurs at stronger Hubbard interaction $U$
due to the enhanced kinetic energy provided by the non-conserving
hopping channel. We show that the orbital liquid is the ground state
everywhere in the $(n,U)$ phase diagram except close to half-filling at
sufficiently large $U$, where ferro-orbital order with real orbitals
occupied is favored. The latter feature is shown to be specific for
$d=\infty$, being of mathematical nature due to the exponential tails
in the density of states.
\end{abstract}


\maketitle

\section{Introduction}
\label{sec:intro}

Our understanding of electron correlations in solids stems to a
considerable extent from the Hubbard model \cite{Hub63}. This describes
electrons which propagate on a lattice and interact by an on-site
Coulomb interaction $U$ when two electrons with opposite spins occupy
the same site. At half-filling (one electron per site) it provides a
picture of a metal-insulator transition due to electron localization
and of magnetism caused by superexchange according to ideas going back
to Anderson \cite{And59}. Away from half-filling it represents the
competition between the kinetic energy of the electrons and the
localizing effect of the Coulomb interaction. An important development
regarding the Hubbard model was the discovery by Metzner and Vollhardt
that in the limit of infinite dimension ($d\!=\!\infty$) only the
on-site correlations survive \cite{Vol89,Met88}. They thus proved that
in this limit the Gutzwiller approximation to the variational ground
state wave function \cite{Gut63,Gut65,Vol84} becomes exact \cite{Met89},
and that a similar property applies \cite{Met91} to the resonating
valence-bond (RVB) state \cite{And73}. It was then shown that the ground
state phase diagram of the doped Hubbard model at $d=\infty$ includes
ferromagnetic (FM) and antiferromagnetic (AF) order stabilized at large
Coulomb interaction \cite{Faz90}.

The $e_g$ orbital Hubbard model \cite{Ish97,Fei05,Tan05,Tan07,Cza17},
discussed in the present paper, describes spinless fermions which
propagate on a lattice of atoms with twofold degenerate orbitals and
which interact by an on-site Coulomb interaction $U$ when two electrons
occupy both orbitals on a site. Importantly, the hopping strength $t$
is dependent on the orientation of the orbitals with respect to the
hopping direction. Obviously, with twofold orbital degeneracy replacing
twofold spin degeneracy, the orbital model closely resembles the
spin model, but there is a crucial difference:
in the spin model the local symmetry is SU(2) but in the orbital model
it is the threefold rotation group $C_3$ and moreover this is linked to
the hopping. As an immediate consequence the model contains, in addition
to the familiar orbital-conserving hopping channel (analogous to the
spin-conserving hopping channel in the (spin) Hubbard model), also an
orbital-flipping hopping channel. This leads in the 3D case to the
existence of an orbital liquid (OL) state \cite{Ish97,Fei05}, in analogy
to a spin liquid \cite{Bal10,Sav17}, which is
disordered in a way different from paramagnetism. Our aim is to analyze
the limit $d\to\infty$ along the lines set out by Metzner and Vollhardt
and explore whether this gives further insight into the properties of
the $e_g$ orbital Hubbard model and the OL state in particular.

The $e_g$ orbital Hubbard model comes from the field of spin-orbital
physics, which has developed over the last two decades into a very
active and challenging part of solid state physics which unifies
frustrated magnetism and the phenomena in strongly correlated electron
systems. It arose from the pioneering ideas of Kugel and Khomskii
\cite{Kug82} who recognized that in transition metal oxides with partly
filled degenerate $3d$ orbitals, both the spin and the orbital degrees
of freedom are quantum variables and have to be treated on equal
footing. At large local Coulomb interaction $U$ most of these compounds
are Mott (or charge-transfer) insulators due to electron localization,
and superexchange arises from coupled spins and orbitals together
\cite{Kho14,Ima98,Tok00}. More recently similar phenomena have been
found in ultracold fermion systems in optical lattices \cite{Gor10}.

In the degenerate-orbital case \cite{Kog04}, quantum fluctuations in
the insulating state are enhanced with respect to the nondegenerate
case \cite{Fei97,Fei98,Kha97}.
In some cases these joint fluctuations could even trigger a new state
of quantum matter---a spin-orbital liquid in model \cite{Karlo} or real
systems \cite{Kha00,Kha05,Kit18}. In most cases, however, one finds
spin-orbital order due to spin interactions of a Heisenberg form with
SU(2) symmetry, coupled to anisotropic orbital superexchange
\mbox{\cite{Fei99,vdB99,Ole05,Rei05,Cha08,Sol08,Nor08,Zaa09,Lv10,Nor11,
Cha11,WohEPL,Woh11,Sch12,Woh13,Bis15,CCC15,Brz15,Brz16,Brz17},}
in agreement with the Goodenough-Kanamori rules \cite{Goode}.
These interactions give in general entangled ground states
\cite{Ole06,Ole12,Brz12,Brz13,Sna19,Brz20}. We remark that the physical
properties of such systems are very rich and depend on whether the
orbital degrees of freedom are of $e_g$, $t_{2g}$, or $p$ type.
The main difference is the spatial character of the orbitals which
makes their interactions directional.

An issue of high interest is how such systems, characterized by the
presence of orbital and spin degrees of freedom, behave under doping,
{\it i.e.\/}, how their properties compare with the more familiar
behavior of doped spin systems, such as the high-$T_c$ cuprates
\cite{Lee06,Oga08,Kei15}. In the latter systems remarkable progress has
been achieved by employing dynamical mean-field theory (DMFT) \cite{Geo96},
which gave very valuable insights into the transport properties
\cite{Mer00} and spectral functions \cite{Kyu06}. This theory, which is
currently used widely for electronic structure calculations of strongly
correlated materials where the one-electron description breaks down
\cite{Kot06}, is in fact based on the above-mentioned discovery by
Metzner and Vollhardt regarding the Hubbard model \cite{Vol89,Met88}.

A realistic description of transition metal oxides with degenerate $3d$
orbitals requires the treatment of spin-orbital physics in at least a
degenerate Hubbard model, involving both orbital and spin degrees of
freedom \cite{Tok00}. However, in particular situations the problem
becomes somewhat simplified. In spin-orbital systems with FM order and
weak spin-orbit coupling, the spins may be ignored because the spin
flavor is conserved in the hopping processes, and the spins disentangle
from the orbitals \cite{Got20,Got20a} and do not contribute to the
dynamics. The degenerate Hubbard model appropriate for the spin-orbital
system \cite{Ole83} then reduces to an orbital Hubbard model with
direction-dependent hopping.

When the active orbitals are of $t_{2g}$ type this orbital Hubbard model
is rather similar to the standard spin Hubbard model because the
orbital flavor is then conserved in the hopping processes
(like for spin) but the hopping is two-dimensional (2D)
\cite{Kha00,Har03}. This gives almost localized orbital polarons for
single holes \cite{Dag08,Woh08} or orbital stripes due to
self-organization of the doped Mott insulator \cite{Wro10,Wro12}.

However, in doped FM systems with active $e_g$ orbitals
\cite{Hor99,Hor99a} the orbital flavor is not conserved, in contrast
to the $t_{2g}$ case above, and the kinetic energy consists of all
possible hopping processes including those which change the orbital
flavor. Obviously, this represents a qualitative difference with the
spin Hubbard model. Moreover, it suggests that disordered phases are
favored more in doped $e_g$ than in doped $t_{2g}$ systems.

In doped manganites the FM metallic phase occurs due to the kinetic
energy gain by the double exchange mechanism \cite{Zen51}, because
antiferromagnetic (AF) bonds hinder electron hopping while hopping is
unrenormalized along FM bonds, and the $e_g$ electrons reorient the
$t_{2g}$ spins coupled to them by Hund's rule exchange. Double exchange
for strongly correlated $e_g$ electrons is indeed responsible for the
spectacular properties of doped perovskite manganites
\cite{Dag01,Dag05,Ram07}, including colossal magnetoresistance in the
FM metallic phase \cite{Tok06}. At low doping the $A$-type AF order in
La$_{1-x}$Sr$_x$MnO$_3$ favors $e_g$ electron transport within 2D
$(a,b)$ planes, and gradually changes the spin order to FM along the
$c$ axis. Increasing doping generates orbital polarons \cite{Kil99}.
Already a single polaron triggers \cite{Dag04} an insulator-to-metal
transition that occurs here to a FM metallic phase.
Eventually, the orbitals decouple from the spins and a disordered
orbital liquid (OL) phase arises in a 3D system
\cite{Ish97,Fei05,Tan05,Tan07}. This OL phase may explain the cubic
dispersion relation of magnons in the FM phase of some manganites which
indicates that magnetic interactions are isotropic \cite{Khaki,Ole02}
unless they couple strongly to orbital excitations \cite{Sna19}.
Specifically this compound, La$_{1-x}$Sr$_x$MnO$_3$, provides a prime
example of the physical situation represented by the orbital model studied
in this paper, with the spins integrated out and the dynamics determined
solely by the orbitals.

So the $e_g$ orbital Hubbard model, while still directly relevant for
a class of doped FM insulators, is the {\it simplest nontrivial
multi-orbital model representing interacting lattice fermions\/}, and as
such presents an opportunity for studying fundamental issues in orbital
dynamics. Moreover, because of their mathematical similarity one can
readily compare its behavior with that of the (spin) Hubbard model, at
the same time exploiting the wealth of results available for the latter
one.

As stated, our overall aim is to investigate whether the limit of
infinite dimension elucidates the characteristic features of the $e_g$
orbital Hubbard model, like it does for the spin Hubbard model
\cite{Faz90}. In particular we want to establish which features found in
the 3D case \cite{Ish97,Fei05,Tan05,Tan07} are characteristic for the
model as such, {\it i.e.\/}, continue to hold for arbitrary dimension
up to $d=\infty$, and which are specific for $d=3$ {\it per se\/} or
are caused by the fact that at finite dimension the Gutzwiller approach
is an approximation.

The purpose of this paper is therefore\hfill\break
(i)~to derive the orbital $e_g$ model at arbitrary dimension $d$ from
the requirements that the twofold degeneracy of the $e_g$ orbitals be
retained and that the cubic lattice symmetry be turned into a hypercubic
symmetry;
\hfill\break
(ii)~to highlight the fundamental difference with the spin Hubbard model
which is that the orbital flavor is not conserved and the kinetic energy
is therefore potentially larger, which should hinder orbital order and
favor states with disordered orbitals;
\hfill\break
(iii)~to demonstrate that the Gutzwiller approximation is again exact in
the limit \mbox{$d\to\infty$} like it is for the (spin) Hubbard model,
\hfill\break
and next to investigate
\hfill\break
(iv)~in what way the OL state avoids orbital polarization, and the role
played therein by symmetry;
\hfill\break
(v)~whether at half-filling electron localization induces a
metal-insulator transition in the OL state and, if so, at which
critical value of the on-site interaction $U$
\hfill\break
(vi)~what the nature is of the ground state (long-range order of ferro
type or alternating type, or rather disordered of para-orbital type or
orbital-liquid type), dependent on the strength of the interaction $U$
and the particle density~$n$.

The paper is organized as follows. In Sec. \ref{sec:3d} we introduce
the orbital Hubbard model for spin-polarized $e_g$ electrons for a
cubic lattice ($d=3$), and discuss its symmetry properties. We show
that the cubic symmetry of the hopping may be better appreciated when a
particular basis consisting of two orbitals with complex coefficients is
used. This serves to introduce the orbital model at general dimension
$d$ and at $d=\infty$ in Sec.~\ref{sec:doo}. Next we use the
translational invariance to present the noninteracting Hamiltonian in
momentum space in Sec.~\ref{sec:wf}. Band dispersion and densities of
states for complex and real single-particle states are presented in
Secs.~\ref{comwf} and \ref{reawf}, respectively. In Sec.~\ref{sec:sen}
we address the single-particle eigenstates and show that they exhibit a
generic splitting into a lower and an upper band. The orbital model is
compared at dimension $d=\infty$ with the spin model in
Sec.~\ref{sec:compar} and we argue from the densities of states that
its general feature is an enhanced kinetic energy.
In Sec.~\ref{sec:ekin} we analyze the kinetic energy in dependence of
filling and compare it for the simplest symmetry-broken states,
disordered (para-orbital) states, and the OL state built from the
single-particle eigenstates.

The electron correlations are described
using the Gutzwiller approximation within a generalization of the
Metzner approach \cite{Met89,Met91}, analyzed in Sec.~\ref{sec:gen},
and we evaluate the renormalized propagator at $d=\infty$
using the collapse of diagrams to a single site in Sec.~\ref{sec:ssa}.
The general formalism for uniform and two-sublattice states is
introduced in Sec.~\ref{sec:inv}.
In Sec.~\ref{sec:gure} we treat specific trial variational states:
(i) ordered states in Sec.~\ref{sec:oo},
(ii) the OL phase in Sec.~\ref{sec:ol}, concluding the demonstration
that the Metzner approach remains valid in the orbital case. In the
next Section \ref{sec:pol} we compare the OL in the orbital Hubbard
model with the paramagnet in the spin Hubbard model, and show that a
Brinkman-Rice transition takes place in the OL just like in the
paramagnet but for a considerably larger value of $U$. The results of
the numerical analysis and the phase diagram of the orbital Hubbard
model at $d=\infty$ are presented in Sec.~\ref{sec:phd}. At the end
we focus on some general aspects of the orbital physics and suggest
possible extensions of the present study (Sec.~\ref{sec:summa}).
The paper is concluded in Sec.~\ref{sec:con} by pointing out the
differences between the spin and orbital Hubbard model in infinite
dimension. In the Appendix we present a proof that the orbital liquid
phase is unpolarized.

\section{The $e_g$ orbital Hubbard model}
\label{sec:ohm}

\subsection{The model at dimension $d=3$}
\label{sec:3d}

The usual choice of basis for $e_g$ orbitals in a 3D
cubic lattice is to take
\begin{equation}
\label{real}\textstyle{
|z\rangle\equiv \frac{1}{\sqrt{6}}(2z^2-x^2-y^2),
\hspace{0.4cm}
|\bar{z}\rangle\equiv \frac{1}{\sqrt{2}}(x^2-y^2),}
\end{equation}
called \textit{real orbitals}.
However, because this basis is the natural one only for the bonds
parallel to the $c$ axis but not for those in the $(a,b)$ plane, the
kinetic energy takes then  a rather nonsymmetric form, having a very
different appearance depending on the bond direction. It is thus
preferred to use instead the basis of \textit{complex orbitals} at
each site \cite{notesign},
\begin{equation}\textstyle{
|+\rangle=\frac{1}{\sqrt{2}}\big(|z\rangle + i|\bar{z}\rangle\big),
\hspace{0.5cm}
|-\rangle=\frac{1}{\sqrt{2}}\big(|z\rangle - i|\bar{z}\rangle\big),}
\label{complex}
\end{equation}
corresponding to ``up'' and ``down'' pseudospin flavors, with the local
pseudospin operators defined as
\begin{eqnarray}
{\hat T}_i^+&=&{\hat c}_{i +}^{\dagger}{\hat c}_{i -}^{},    \hspace{1.2cm}
{\hat T}_i^- = {\hat c}_{i -}^{\dagger}{\hat c}_{i +}^{},    \nonumber \\
{\hat T}_i^z&=&\textstyle{\frac{1}{2}}
({\hat c}_{i +}^{\dagger}{\hat c}_{i+}^{}-{\hat c}_{i-}^{\dagger}{\hat c}_{i-}^{} )
= \textstyle{\frac{1}{2}}({\hat n}_{i+}  -{\hat n}_{i-}).
\label{pseudospin}
\end{eqnarray}
For later reference it is convenient to introduce also electron
creation operators ${\hat c}_i^{\dagger}(\psi_i,\theta_i)$ which
create $e_g$ electrons in orbital coherent states $|\Omega_i\rangle
\equiv |\Omega_i(\psi_i,\theta_i)\rangle$ at site $i$, defined as
\begin{eqnarray}
|\Omega_i\rangle\equiv
 \e^{-i\theta_i/2}\cos\left(\frac{\psi_i}{2}\right)|i+\rangle
+\e^{+i\theta_i/2}\sin\left(\frac{\psi_i}{2}\right)|i-\rangle, \nonumber\\
\label{cohorb}
\end{eqnarray}
in analogy with the well-known spin coherent states \cite{Kla79}.
The local pseudospin operator in this state is
\begin{equation}
\langle \Omega_i| {\bm {\hat T}}_i |\Omega_i\rangle =
\textstyle{\frac{1}{2}} (\sin\psi_i\cos\theta_i,
                         \sin\psi_i\sin\theta_i, \cos\psi_i ),
\label{vector}
\end{equation}
{\it i.e.\/}, it behaves like a vector \cite{notecoh}.

\begin{figure}[t!]
\begin{center}
\includegraphics[width=\columnwidth]{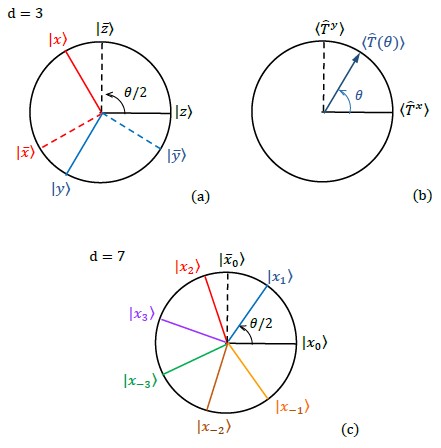}
\end{center}
\caption{
Schematic representation of the real orbital wave functions in the
'equatorial' plane at dimensions
(a) $d=3$,
(c) $d=7$,
and (b) of the pseudospin vector at any dimension. The solid lines in
(a) correspond to directional $(3z^2-r^2)$-like orbitals (the azimuthal
angle $\theta_i/2$ in Eq.~(\ref{realcohorb}) equals one of the
$\chi_{\alpha}/2$), while the dashed lines correspond
to the $(x^2-y^2)$-like orbitals orthogonal to them ($\theta_i/2$
equalling $\chi_{\alpha}/2+\pi/2$). In (c) the solid lines correspond to
the $|x_n\rangle_d$ orbitals (the azimuthal angle $\theta_i/2$ equals
one of the $\chi_n/2$), while only the dashed line corresponding to the
central orthogonal orbital $|\bar{x}_0\rangle_d$ is shown.
}
\label{fig:orbi}
\end{figure}

The parameter space of the coherent orbital (\ref{cohorb}) is a sphere,
with  the ``poles'' ($\psi_i=0$ and $\psi_i=\pi$) corresponding to
the complex orbitals $|i+\rangle$ and $|i-\rangle$,
and the ``equator'' ($\psi_i=\pi/2$) corresponding to
(linear combinations of) the real orbitals,
\begin{equation}
\left|\Omega_i\left(\frac{\pi}{2},\theta_i\right)\right\rangle =
\cos\left(\frac{\theta_i}{2}\right)\left|iz\right\rangle
   + \sin\left(\frac{\theta_i}{2}\right)\left|i\bar{z}\right\rangle.
\label{realcohorb}
\end{equation}
The three real bases,
$\{|ix\rangle,|i\bar{x}\rangle\}$,
$\{|iy\rangle,|i\bar{y}\rangle\}$, and
$\{|iz\rangle,|i\bar{z}\rangle\}$,
associated with the cubic axes $a$, $b$, and $c$, respectively,
are obtained by setting the in-plane angle $\theta_i/2$ equal to
$\chi_a/2=+2\pi/3$, $\chi_b/2=-2\pi/3$, and $\chi_c/2=0$,
for $|ix\rangle, |iy\rangle$, and $|iz\rangle, $ and to $\chi_\alpha + \pi/2$
($\alpha=a,b,c$)
for $|i\bar{x}\rangle, |i\bar{y}\rangle, |i\bar{z}\rangle$ ,
as illustrated in Fig.~\ref{fig:orbi}(a).

In the complex-orbital representation the {\it orbital Hubbard model\/}
for $e_g$ electrons in $d=3$ takes the form \cite{Fei05}
\begin{eqnarray}
{\cal H}_3&=& -\frac{1}{2}\; t
\sum_{\langle ij\rangle\parallel{\hat\alpha}}
  \Big\{\Big({\hat c}_{i+}^{\dagger}{\hat c}_{j+}^{}
 +{\hat c}_{i-}^{\dagger}{\hat c}_{j-}^{}\Big) \nonumber \\
  & & \hskip 0.7cm
  + \gamma \Big(\e^{-i\chi_{\alpha}}{\hat c}_{i+}^{\dagger}{\hat c}_{j-}^{}
    +\e^{+i\chi_{\alpha}}{\hat c}_{i-}^{\dagger}{\hat c}_{j+}^{}\Big)\Big\}
                                \nonumber \\
  &+ & U \sum_i{\hat n}_{i+}^{}{\hat n}_{i-}^{},
\label{Hoo}
\end{eqnarray}
where the
parameter $\gamma$ actually takes the value $\gamma=1$. This parameter
is introduced here as a device by which one may interpolate between the
standard 2-flavor Hubbard model with hopping $t/2$ at $\gamma=0$ and
the orbital $e_g$ model at \mbox{$\gamma=1$.} The appearance of the
phase factors $\e^{\pm i\chi_{\alpha}}$ is characteristic for the
orbital problem --- they occur because the orbitals
have an actual shape in real space so that each hopping process depends
on the bond direction and on the orbitals between which hopping occurs.
Moreover, Eq.~(\ref{Hoo}) exhibits the crucial feature that
(except at $\gamma=0$) {\it orbital flavor is not conserved\/}, or,
equivalently, that {\it pseudospin is not conserved\/}, compare
Eq.~\eqref{pseudospin}.

The model (\ref{Hoo}) consists of the kinetic energy
$H_{\rm kin}\propto t$ and the interorbital Coulomb interaction
$H_{\rm int}\propto U$. The interaction is invariant under any local
basis transformation to a pair of orthogonal orbitals, {\it i.e.\/}, it
gives an energy $U$ when a double occupancy occurs in any representation,
{\it i.e.\/}, either when both real orbitals are simultaneously occupied
or when both complex orbitals are occupied,
\begin{equation}
H_{\rm int}=U\sum_i{\hat n}_{iz}{\hat n}_{i\bar{z}}
           =U\sum_i{\hat n}_{i+}{\hat n}_{i-}.
\end{equation}
We emphasize that this simple Hubbard-like form of the Coulomb
interaction operator obtained here corresponds to the high-spin ($S=1$)
charge excitations, \mbox{$d^1_id^1_j\rightarrow d^2_id^0_j$}, --- these
are the only excited states in a FM system. The parameter $U$ stands
then for the onsite interorbital Coulomb repulsion, and is in fact lower
than the intraorbital Coulomb element (Kanamori parameter) $U_0$ by
$3J_H$ due to Hund's exchange in the triplet state, {\it i.e.\/},
$U=U_0-3J_H$ \cite{vdB99,Cza17}.

Importantly, as noted in \cite{Fei05}, the equivalence of the cubic
axes shows up in the kinetic energy term in Eq.~(\ref{Hoo}) in a
transparent way, viz. as a formal invariance under threefold rotations
of the phase angles $\chi_\alpha \mapsto \chi_\alpha\! -\! 4\pi/3$
in conjunction with a phase shift of the electron operators
${\hat c}_{i\pm}^{\dagger} \mapsto
\e^{\mp 2\pi/3}{\hat c}_{i\pm}^{\dagger}$.
So the relevant symmetry group is $C_3$, with the $e_g$ doublets
$\{|iz\rangle,|i\bar{z}\rangle\}$ transforming as real $E$
representations, or equivalently the pairs $\{|i+\rangle,|i-\rangle\}$
transforming as conjugate complex $A_1$ representations.

\subsection{Generalization to large dimension $d$}
\label{sec:doo}

This form, Eq.~(\ref{Hoo}), therefore lends itself to a natural
generalization of the {\it $e_g$ orbital Hubbard model\/} with two
orbital flavors $\{|+\rangle,|-\rangle\}$ to large dimension $d$ as
follows,
\begin{eqnarray}
{\cal H}_d&=& -\; \frac{1}{\sqrt{2d}}\,\frac{t}{2}
\sum_{\langle ij\rangle\parallel\hat{x}_n}
  \Big\{\Big({\hat c}_{i+}^{\dagger}{\hat c}_{j+}^{}
  +{\hat c}_{i-}^{\dagger}{\hat c}_{j-}^{}\Big)
                                \nonumber \\
  & & \hskip 1.5cm
  + \gamma \Big(\e^{-i\chi_n}{\hat c}_{i+}^{\dagger}{\hat c}_{j-}^{}
    +\e^{+i\chi_n}{\hat c}_{i-}^{\dagger}{\hat c}_{j+}^{}\Big)\Big\}
                                \nonumber \\
  & & + U \sum_i{\hat n}_{i+}^{}{\hat n}_{i-}^{},
\label{Hd}
\end{eqnarray}
with $\chi_n=4\pi n/d$, where $n=-m,-m+1,\cdots,m-1,m$, and
$d\equiv 2m+1$ ({\it i.e.\/}, taken odd for convenience but $d$ could
also be even with only slightly different derivations below),
and we note here that \mbox{$\sum_n\e^{i\chi_n}=0$}.
So instead of three cubic axes we now have $d$ hypercubic axes
$\{\hat{x}_n\}$, labelled by the index $n$. Importantly, the number of
orbital flavors remains two (as for spin $S=1/2$).

The real basis $\{|x_n\rangle_d,|\bar{x}_n\rangle_d\}$ associated with
axis $\hat{x}_n$ is given by Eq.~(\ref{realcohorb}) with $\theta_i/2$
set equal to $\chi_n/2$ and to $\chi_n/2+\pi/2$, respectively. This is
illustrated in Fig.~\ref{fig:orbi}(c) for $d=7$ in comparison with the
case for $d=3$ shown in Fig.~\ref{fig:orbi}(a). Figure \ref{fig:orbi}(b)
shows the associated behavior of the pseudospin vector, valid both for
$d=3$ and for general $d$.

The hopping parameter $t$ is scaled in the standard fashion
\cite{Vol89,Met89,Faz90} by $\sqrt{2d}$ in order that the average
kinetic energy remains finite as $d$ becomes large. Below we will use
the abbreviation $\tilde{t}=t/\sqrt{2d}$ whenever convenient. Again,
the parameter $\gamma$ takes the value $\gamma=1$ for the orbital
problem, in which {\it the orbital flavor (pseudospin) is not
conserved\/}, while $\gamma=0$ for the corresponding spin problem with
hopping $t/2$, in which the (spin) flavor is conserved.
The interorbital Coulomb interaction $\propto U$ is the same for any
dimension $d$ as it stands for the on-site Coulomb repulsion
when both orbital (or spin) flavors are occupied.

Obviously, the equivalence of the hypercubic axes manifests itself in
the Hamiltonian ${\cal H}_d$ as a $d$-fold rotational symmetry,
{\it i.e.\/}, by being invariant under phase shifts
${\hat c}_{i\pm}^{\dagger} \mapsto
\e^{\pm i 2\pi/d}{\hat c}_{i\pm}^{\dagger}$
of the electron operators in conjunction with shifts of the
angles $\chi_n \mapsto \chi_n\! -\! 4\pi/d$, in perfect analogy with
the 3-fold rotational symmetry of ${\cal H}_3$. So ${\cal H}_d$ is
invariant under $C_d$, with the $e_g$ doublets
$\{|i x_n\rangle_d,|i \bar{x}_n\rangle_d\}$ still transforming as real
$E$ representations. In the limit $d\to\infty$ the symmetry group $C_d$
turns into $C_{\infty}\equiv \mathbb{Z}$, which for practical purposes
approaches continuous rotational symmetry U(1).

The generalization to dimension $d=2m+1$ for the explicit form of the
normalized real basis orbitals associated with the central axis
$\hat{x}_0$ in terms of the coordinates $x_n$ is obtained
straightforwardly,
\begin{eqnarray}
\label{zd}
|x_0\rangle_d&\equiv& \frac{2}{\sqrt{2d}}
\left(\sum_{n=-m}^m \cos(\chi_n/2)\, x_n^2\right),\\
\label{xd}
|\bar{x}_0\rangle_d&\equiv& \frac{2}{\sqrt{2d}}
\left(\sum_{n=-m}^m \sin(\chi_n/2)\, x_n^2\right),
\label{realorbslarged}
\end{eqnarray}
of which Eq.~(\ref{real}) is seen to be the special case for $d=3$.
We observe that for general dimension $d$, the ``directional" orbital
$|x_0\rangle_d$ has a less pronounced directional shape than
$|z\rangle=|x_0\rangle_3$ in the 3D case, because of the many finite
contributions from all the coordinates $x_n$ with $n\neq0$, although
the square of the central coordinate $x_0$ still has the largest
coefficient. The orthogonal orbital $|\bar{x}_0\rangle_d$
is perpendicular to the $\hat{x}_0$ axis as in the 3D  case
(the coefficient of ${x_0}^2$ in $|\bar{x}_0\rangle_d$ is zero as
$\chi_0=0$). When $d$ gets large the shape of $|\bar{x}_0\rangle_d$
resembles that of $|x_0\rangle_d$ but with permuted coordinates
(because Eq.~(\ref{xd}) is obtained by replacing $\chi_n/2$ in
Eq.~(\ref{zd}) by $\chi_n/2\!-\!\pi/2$). Clearly, at each site all
real-orbital basis sets $\{|x_n\rangle_d, |\bar{x}_n\rangle_d\}$ are
equivalent. We have made the arbitrary choice to use
the set $\{|x_0\rangle_d, |\bar{x}_0\rangle_d\}$ as the reference
basis for real orbitals, and for convenience to drop the subscript $d$.

\subsection{Momentum space representation}
\label{sec:wf}

Using translational invariance, we introduce creation and
annihilation operators in momentum space,
\begin{equation}
\hat{c}_{{\bf k}\lambda}^{\dagger}=\frac{1}{\sqrt{N}}\sum_j
\mathrm{e}^{i{\bf k}{\rm R}_j}\,\hat{c}_{j\lambda}^{\dagger},
\label{ck}
\end{equation}
where $\lambda$ labels the orbital flavors, e.g. $\{x_0,\bar{x}_0\}$ or
$\{+,-\}$, and $N$ is the number of sites.
The free-electron part of ${\cal H}_{d}$, Eq.~(\ref{Hd}), describing
the kinetic energy ${\cal H}_{\rm kin}$, may be represented
upon Fourier transformation in the form
\begin{equation}
\label{Hk}
{\cal H}_{\rm kin} =
-  \tilde{t} \, \sum_{\bf k}
\left({\hat c}_{{\bf k}+}^{\dagger}\; {\hat c}_{{\bf k}-}^{\dagger}
\right)\!
\left( \begin{array}{cc}
A_{\bf k} &  \gamma G_{\bf k}^{\ast}
   \\[0.2cm]
\gamma G_{\bf k}^{} &   A_{\bf k}
\end{array}
\right)\!
  \left(\begin{array}{c} {\hat c}_{{\bf k}+}^{} \\[0.2cm]
                      {\hat c}_{{\bf k}-}^{} \end{array}\right),
\end{equation}
where the dispersion is given by orbital-conserving and
orbital-non-conserving terms,
\begin{eqnarray}
\label{Ak}
 A_{\bf k} &=&\sum_n \cos k_n \; ,   \\
\label{Gk}
 G_{\bf k} &=&\sum_n e^{+ i \chi_n} \cos k_n
        \equiv |G_{\bf k}|\, \e^{+ i \phi_{\bf k}} \; .
\label{AkGk}
\end{eqnarray}
It will turn out convenient to introduce the following definitions,
\begin{eqnarray}
\label{Bk}
\hspace{-0.5cm}
B_{\bf k}\!&\equiv& | G_{\bf k} |\!
  =\!\!\left(\sum_{n,n'}\cos(\chi_{n}\!-\chi_{n'})
    \cos k_{n}\!\cos k_{n'}\!\!\right)^{\!\!\!{1/2}}\!\! , \\
\label{Ck}
\hspace{-0.5cm}
C_{\bf k}\!&\equiv& \Re(G_{\bf k})= B_{\bf k} \cos \phi_{\bf k}
   = \sum_n  {\cos \chi_{n} \cos k_{n}} \, , \\
\label{Dk}
\hspace{-0.5cm}
D_{\bf k}\!&\equiv& \Im(G_{\bf k})= B_{\bf k} \sin \phi_{\bf k}
   = \sum_n  {\sin \chi_{n} \cos k_{n}} \, .
\label{BkCkDkphik}
\end{eqnarray}

One may now absorb the phase factors from the offdiagonal elements
of the matrix in Eq.~(\ref{Hk}) into the operators, by defining what we
will call ``phased'' complex-orbital single-particle operators,
\begin{equation}
\hat{\tilde{c}}_{{\bf k}\pm}^{\dagger}=
\mathrm{e}^{\mp i{\phi_{\bf k}}/2} \,\hat{c}_{{\bf k}\pm}^{\dagger},
\label{tck}
\end{equation}
and rewrite ${\cal H}_{\rm kin}$ as
\begin{equation}
\label{Hkt}
{\cal H}_{\rm kin} =
-  \tilde{t} \, \sum_{\bf k}
\left({\hat{\tilde c}}_{{\bf k}+}^{\dagger}\; {\hat{\tilde c}}_{{\bf k}-}^{\dagger}
\right)\!
\left( \begin{array}{cc}
A_{\bf k} &  \gamma B_{\bf k}
   \\[0.2cm]
\gamma B_{\bf k} &   A_{\bf k}
\end{array}
\right)\!
  \left(\begin{array}{c} {\hat{\tilde c}}_{{\bf k}+}^{} \\[0.2cm]
                      {\hat{\tilde c}}_{{\bf k}-}^{} \end{array}\right).
\end{equation}

If desired, the kinetic energy may also be rewritten from
Eq.~(\ref{Hk}) in terms of the real-orbital operators,
\begin{eqnarray}
\hspace{+0.1cm}
\label{Hkreal}
{\cal H}_{\rm kin} &=&
-  \tilde{t} \, \sum_{\bf k}
\left[ A_{\bf k} \,  {\hat{\cal A}}_{\bf k}
-  \gamma \left( C_{\bf k}  \, {\hat{\cal C}}_{\bf k}
                + D_{\bf k} \, {\hat{\cal D}}_{\bf k} \, \right)
\right]    \nonumber
\\
                            &=&
-  \tilde{t} \, \sum_{\bf k}
\left[ A_{\bf k} \,  {\hat{\cal A}}_{\bf k}
 -  \gamma B_{\bf k}
           \left( \cos \phi _{\bf k}  \, {\hat{\cal C}}_{\bf k}
                + \sin \phi_{\bf k} \, {\hat{\cal D}}_{\bf k} \, \right)
\right]  ,
\nonumber \\
\end{eqnarray}
where
\begin{eqnarray}
{\hat{\cal A}}_{\bf k} &=&
   \left( {\hat c}_{{\bf k}x_0}^{\dagger}{\hat c}_{{\bf k}x_0}^{}
   + {\hat c}_{{\bf k}{\bar x}_0}^{\dagger}{\hat c}_{{\bf k}{\bar x}_0}^{} \right) ,
      \\
{\hat{\cal C}}_{\bf k} &=&
   \left( {\hat c}_{{\bf k}{\bar x}_0}^{\dagger}{\hat c}_{{\bf k}{\bar x}_0}^{}
   - {\hat c}_{{\bf k}x_0}^{\dagger}{\hat c}_{{\bf k}x_0}^{}\right) ,
       \\
{\hat{\cal D}}_{\bf k} &=&
   \left( {\hat c}_{{\bf k}x_0}^{\dagger}{\hat c}_{{\bf k}{\bar x}_0}^{}
   + {\hat c}_{{\bf k}{\bar x}_0}^{\dagger}{\hat c}_{{\bf k}x_0}^{} \right) .
\end{eqnarray}
The invariance with respect to the symmetry group
$C_d$ of this real-orbital representation
of ${\cal H}_{\rm kin}$ is warranted by $A_{\bf k}$ and
${\hat{\cal A}}_{\bf k}$ both transforming as $A_1$, and by
$\{C_{\bf k}, D_{\bf k}\}$ and
$\{ {\hat{\cal C}}_{\bf k}, {\hat{\cal D}}_{\bf k} \}$
both transforming as real $E$ representations.

We note that whereas the kinetic-energy Hamiltonian (\ref{Hk}),
(\ref{Hkt}), or (\ref{Hkreal}), is block-diagonal in ${\bf k}$-space,
the orbital flavors $\{+,-\}$ or $\{x_0,{\bar x}_0\}$  are still
coupled. The further analysis of the kinetic energy depends on whether
some form of symmetry breaking takes place, or whether we deal with an
unpolarized orbital state.

\section{Noninteracting electrons}
\label{sec:free}

Below, in Sec. \ref{sec:gure}, we will calculate the electron
correlations due to the on-site Hubbard repulsion $U$ using the type
of wave function introduced by Gutzwiller \cite{Gut65,Vol84},
\begin{equation}
|\Psi\rangle=g^{\hat D}\big|\Phi_0\big\rangle
=\prod_{i} \left(1-\left(1-g\right)\hat D_i\right)\big|\Phi_0\big\rangle\,,
\label{psi}
\end{equation}
where $|\Phi_0\rangle$ is an arbitrary uncorrelated trial state,
\begin{equation}
\hat D = \sum_{i} \hat D_{i}, \quad \quad \quad
\hat D_{i} = \hat n_{i+} \hat n_{i-} ,
\label{hatD}
\end{equation}
and $g$ is a variational parameter, $0\leq g\leq 1$.
Frequently $|\Phi_0\rangle$ is chosen to be a Gutzwiller wave function
proper, {\it i.e.\/}, a Fermi-sea state of noninteracting particles,
\begin{equation}
|\Phi_0\rangle=
\prod_{{\bf k}\in{\cal K}_{\lambda}}\hat{c}_{{\bf k}\lambda}^\dagger
\prod_{{\bf q}\in{\cal K}_{\bar{\lambda}}}
\hat{c}_{{\bf q}\bar{\lambda}}^\dagger\,|0\rangle,
\label{ckk}
\end{equation}
where ${\cal K}_{\lambda}$ and ${\cal K}_{\bar{\lambda}}$ are the parts
of the $d$-dimensional Brillouin zone occupied by particles of opposite
orbital flavors $\lambda$ and $\bar{\lambda}$,
determined from the condition that the respective single-particle
kinetic energies be smaller than Fermi energies
$E_{\mathrm{F},\lambda}$ and $E_{\mathrm{F},\bar{\lambda}}$.

For the orbital model, because it has only $C_{\infty}$
symmetry, even for a Fermi-sea-like $|\Phi_0\rangle$
the free-particle kinetic energy depends on the particular set of
orbitals that are occupied, as in the 3D case \cite{Fei05}. This is
fundamentally different from the situation for the spin Hubbard model,
where because of the SU(2) symmetry of the spins one has just two
(spin) flavors quantized along an arbitrary direction in spin space,
and both kinetic energy and renormalization depend only on the filling
of the spin subbands. Therefore, we analyze in the present Section the
dispersions and the densities of states (DOSs) for the single-particle
states used later on in building variational states $|\Phi_0\rangle$.

\subsection{Complex-orbital single-particle states}
\label{comwf}

\begin{figure}[b!]
\vskip -.5cm
\includegraphics[width=1.17\columnwidth]{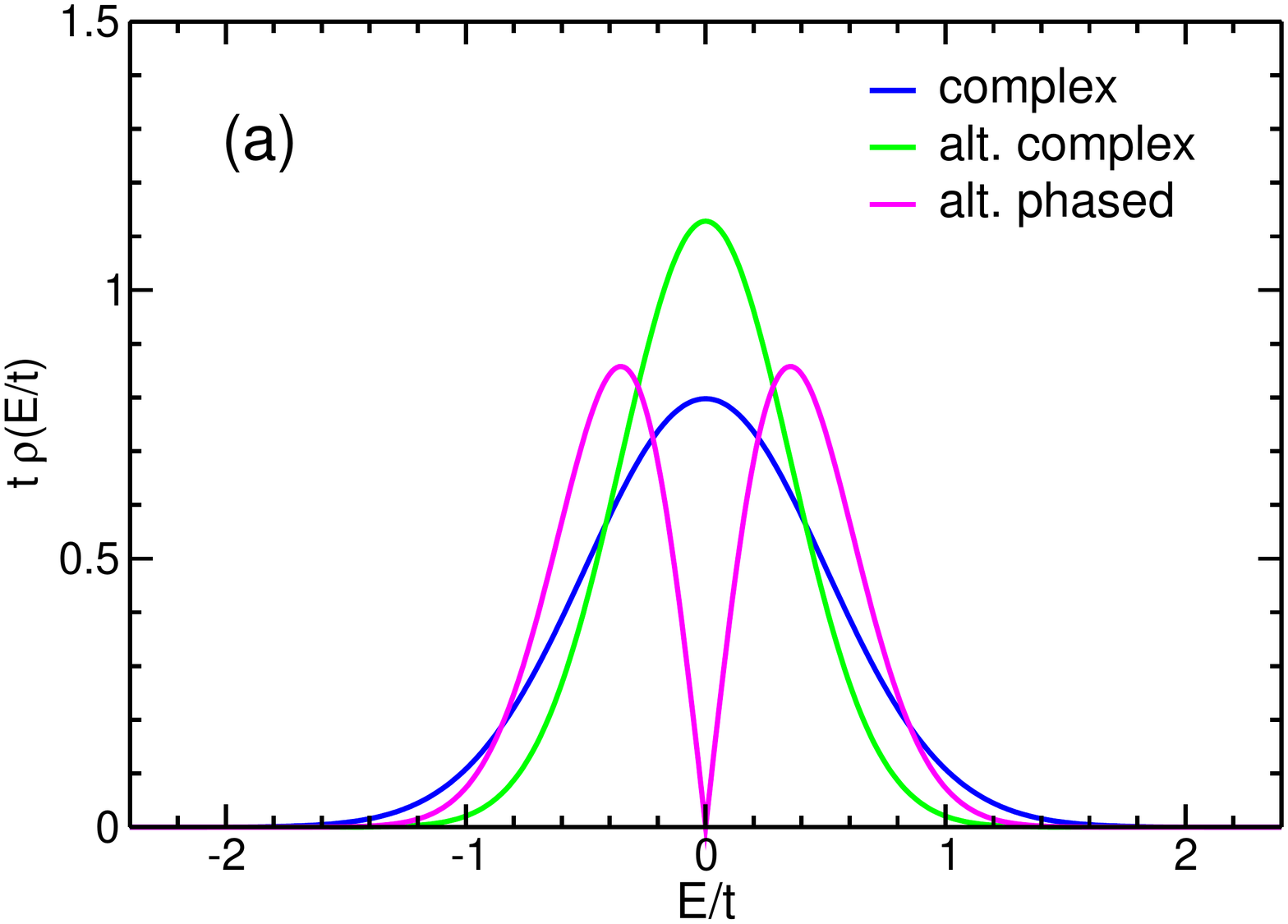}
\vskip -.3cm
\includegraphics[width=1.17\columnwidth]{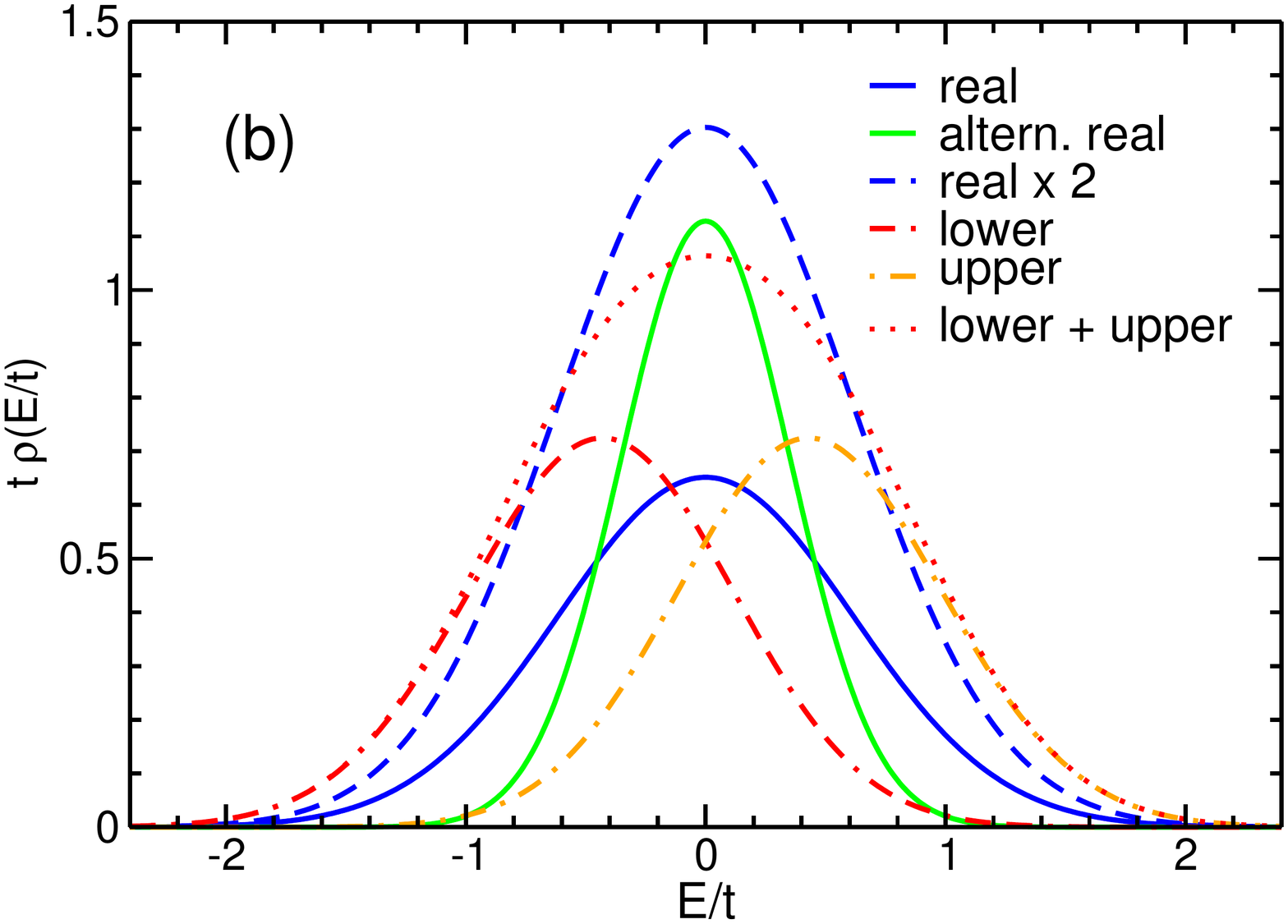}
\caption{
DOSs of the various single-particle states:
(a) for the complex-orbital single-particle states,
$\rho_{\rm c}(E)$,
$\rho_{\rm c,ao}(E)$, and
$\rho_{\tilde{\rm c},{\rm ao}}(E)$;
(b) for the real-orbital single-particle states,
$\rho_{\rm r}(E)$,
$\rho_{\rm r,ao}(E)$, and
$2 \rho_{\rm r}(E)$,
and for the single-particle eigenstates,
$\rho_{\ell}(E)$,
$\rho_{u}(E)$, and
$\rho_{\ell+u}(E)$.
}
\label{fig:dos}
\end{figure}

Particularly simple is the dispersion for the single-particle states with
complex orbitals as defined by Eq.~\eqref{ck}) for $\lambda \! = \! +$
or $\lambda \! = \! -$,
\begin{equation}
\label{epsc}
\varepsilon_{+}({\bf k}) =
\varepsilon_{-}({\bf k}) \equiv
\varepsilon_{\rm c} ({\bf k}) =
   - \tilde{t} \, A_{\bf k}  \, .
\end{equation}
We also consider single-particle states adapted to G-type partitioning of
the lattice into A and B sublattices. With the sublattice operators given by
\begin{eqnarray}
\hat{c}_{\mathrm{A}{\bf k}\lambda}^{\dagger}&=&\frac{1}{\sqrt{2}}
\left(\hat{c}_{{\bf k}\lambda}^{\dagger} +
\hat{c}_{{\bf k}+{\bf Q}\lambda}^{\dagger}   \right) ,
\label{ckA}
   \\
\hat{c}_{\mathrm{B}{\bf k}\lambda}^{\dagger}&=&\frac{1}{\sqrt{2}}
\left(\hat{c}_{{\bf k}\lambda}^{\dagger} -
\hat{c}_{{\bf k}+{\bf Q}\lambda}^{\dagger}   \right) ,
\label{ckB}
\end{eqnarray}
where ${\bf Q}=(\pi,\pi,\cdots,\pi)$ and $\lambda$ is equal to either
$+$ or $-$, we introduce single-particle operators with alternating
orbitals of opposite flavor $\lambda$ (on A sites) and $\bar{\lambda}$
(on B sites) according to
\begin{eqnarray}
\hat{c}_{{\bf k}\lambda\bar{\lambda}, \ell}^{\dagger} &=& \frac{1}{\sqrt{2}}
\left( \hat{c}_{\mathrm{A}{\bf k}\lambda}^{\dagger} -
\hat{c}_{\mathrm{B}{\bf k}\bar{\lambda}}^{\dagger} \right) ,
\label{ckABl}
\\
\hat{c}_{{\bf k}\lambda\bar{\lambda},u}^{\dagger} &=& \frac{1}{\sqrt{2}}
\left( \hat{c}_{\mathrm{A}{\bf k}\lambda}^{\dagger} +
\hat{c}_{\mathrm{B}{\bf k}\bar{\lambda}}^{\dagger} \right) .
\label{ckABu}
\end{eqnarray}
From Eq. (\ref{Hk}) and making use of the relations
\mbox{$A_{\bf k+Q} = - A_{\bf k}$,}
$B_{\bf k+Q} =   B_{\bf k}$, and
$\phi_{\bf k+Q} = \phi_{\bf k} + \pi$,
one readily finds that they correspond to a lower ($\ell$)
and an upper ($u$) band in the (reduced) Brillouin zone, both doubly
degenerate, with dispersions given by
\begin{eqnarray}
\label{epscalt}
\hspace{-0.5cm}
\varepsilon_{+-,\ell}({\bf k}) &=&
-\varepsilon_{+-,u}({\bf k})  =
\varepsilon_{-+,\ell}({\bf k}) =
-\varepsilon_{-+,u}({\bf k})  \nonumber\\
&\equiv& \varepsilon_{\rm c,ao}({\bf k}) =
-\gamma \tilde{t} \, C_{\bf k} =
-\gamma \tilde{t} \, B_{\bf k} \cos \phi_{\bf k} ,
\end{eqnarray}
where the first (second) subscript in the first line gives the
occupation of the A (B) sublattice. Following the same procedure we
also obtain operators for single-particle states with phased complex
orbitals and alternating flavor, and find their dispersions,
\begin{eqnarray}
\label{epscaltphase}
{\tilde \varepsilon}_{+-,\ell}({\bf k}) &=&
-{\tilde \varepsilon}_{+-,u}({\bf k})  =
{\tilde \varepsilon}_{-+,\ell}({\bf k}) =
-{\tilde \varepsilon}_{-+,u}({\bf k})  \nonumber\\
&\equiv& {\tilde \varepsilon}_{\rm c,ao}({\bf k}) =
-\gamma \tilde{t} \, B_{\bf k} \, .
\end{eqnarray}
It is noteworthy that the dispersions \eqref{epscalt} and
\eqref{epscaltphase} of the two alternating single-particle states are
proportional to $\gamma$, {\it i.e.\/} they originate entirely from the
orbital-flavor non-conserving part of the kinetic energy.

We now proceed to the DOSs associated with the above single-particle
states. In the limit $d\to\infty$
an exact analytical expression can be derived for all single-particle
states considered, also for those discussed below. This was well known
to be the case for the rather simple dispersion, identical to
Eq.~(\ref{epsc}), of the spin Hubbard model \cite{Vol89}, but it holds
generally because, from a mathematical point of view \cite{Pet21}, DOSs
are probability distributions and in the limit $d \to \infty$ are
governed by the law of large numbers. It follows that every DOS is
either of Gaussian type,
\begin{equation}
\label{rhog}
\rho_{\rm}(E) = \frac{1}{\sqrt{2 \pi}} \; \frac{1}{wt} \;
  \mathrm{e}^{-\frac{1}{2} \left(\frac{E}{wt}\right)^2} \, ,
\end{equation}
fully characterized by its (dimensionless) width $w$, or is derivable
fairly simply from Gaussians.
For the above dispersions Eqs.~(\ref{epsc}) and (\ref{epscalt}) the
DOSs are Gaussians with widths $w_{\rm c} = 1/2$ and
$w_{\rm c,ao} = \gamma / 2 \sqrt{2}$,
respectively, while for the dispersion (\ref{epscaltphase}) one finds
that the DOS is a symmetrized Rayleigh distribution \cite{Pet21},
\begin{equation}
\label{rhocaltphase}
\rho_{\tilde{\rm c},{\rm ao}}(E) = \frac{2}{\gamma t} \;
\left|\frac{2E}{\gamma t}\right| \;
\mathrm{e}^{-\left(\frac{2E}{\gamma t}\right)^2} \, .
\end{equation}
These three complex-orbital DOSs are shown for the full orbital case,
{\it i.e.\/} $\gamma=1$, in Fig.~\ref{fig:dos}(a)
by the blue, green, and purple line, respectively.

\subsection{Real-orbital single-particle states}
\label{reawf}

For the single-particle states with real orbitals as defined by
Eq.~(\ref{ck}) with $\lambda\!=\!x_0$ or $\lambda\!=\!\bar{x}_0$, and
for the single-particle states describing alternating real flavors $x_0$
and $\bar{x}_0$ on the sublattices defined in the same way as above by
Eqs.~(\ref{ckA})--(\ref{ckABu}), one finds the dispersions
\begin{eqnarray}
\label{epsrz}
\varepsilon_{x_0}({\bf k}) &=&
   - \tilde{t} \, \left( A_{\bf k} +\gamma C_{\bf k} \right)
   \equiv \varepsilon_r({\bf k}) \, ,   \\
\label{epsrx}
\varepsilon_{\bar{x}_0}({\bf k}) &=&
   - \tilde{t} \, \left( A_{\bf k} -\gamma C_{\bf k} \right)
     \equiv \varepsilon_r({\bf - k}) \, ,   \\
\label{epsralt}
\varepsilon_{{x_0}{\bar{x}_0},\ell}({\bf k}) &=&
- \varepsilon_{{x_0}{\bar{x}_0},u}({\bf k}) =
\varepsilon_{{\bar{x}_0}{x_0},\ell}({\bf k}) =
- \varepsilon_{{\bar{x}_0}{x_0},u}({\bf k})         \nonumber \\
   &\equiv& \varepsilon_{\rm r,ao}({\bf k})   =
   \gamma \tilde{t} D_{\bf k}   =
   \gamma \tilde{t} \, B_{\bf k} \sin \phi_{\bf k}  .
\end{eqnarray}
One observes that, as already indicated at the very end of Sec.
\ref{sec:doo}, the sets of real-orbital single-particle states
$\{|{\bf k}x_0\rangle\}$ and $\{|{\bf k}\bar{x}_0\rangle\}$ are
equivalent in the limit $d\to\infty$, in the sense that they have the
same energies be it at opposite ${\bf k}$. As above for the
complex-orbital single-particle states the
alternating real-orbital single-particle states correspond to a lower
($\ell$) and an upper ($u$) band in the (reduced) Brillouin zone, both
doubly degenerate, and their dispersions are proportional to $\gamma$,
stemming entirely from the orbital-flavor non-conserving part of the
kinetic energy.

The corresponding DOSs are again Gaussians, with widths
$w_{\rm r} = \sigma/2$ and $w_{\rm r,ao} = \gamma / 2 \sqrt{2}$,
respectively, where $\sigma=\sqrt{1+{\gamma^2}/2}$). They are shown,
for $\gamma=1$, in Fig.~\ref{fig:dos}(b) by the blue and green line
respectively. One may note that $w_{\rm r,ao}=w_{\rm c,ao}$, implying
that the DOS of the alternating real-orbital single-particle states
is identical to that of its complex-orbital counterpart
shown in the upper panel.

\subsection{Single-particle eigenstates}
\label{sec:sen}

%
\begin{figure}[t!]
\vskip-.5cm
\includegraphics[width=1.14\columnwidth]{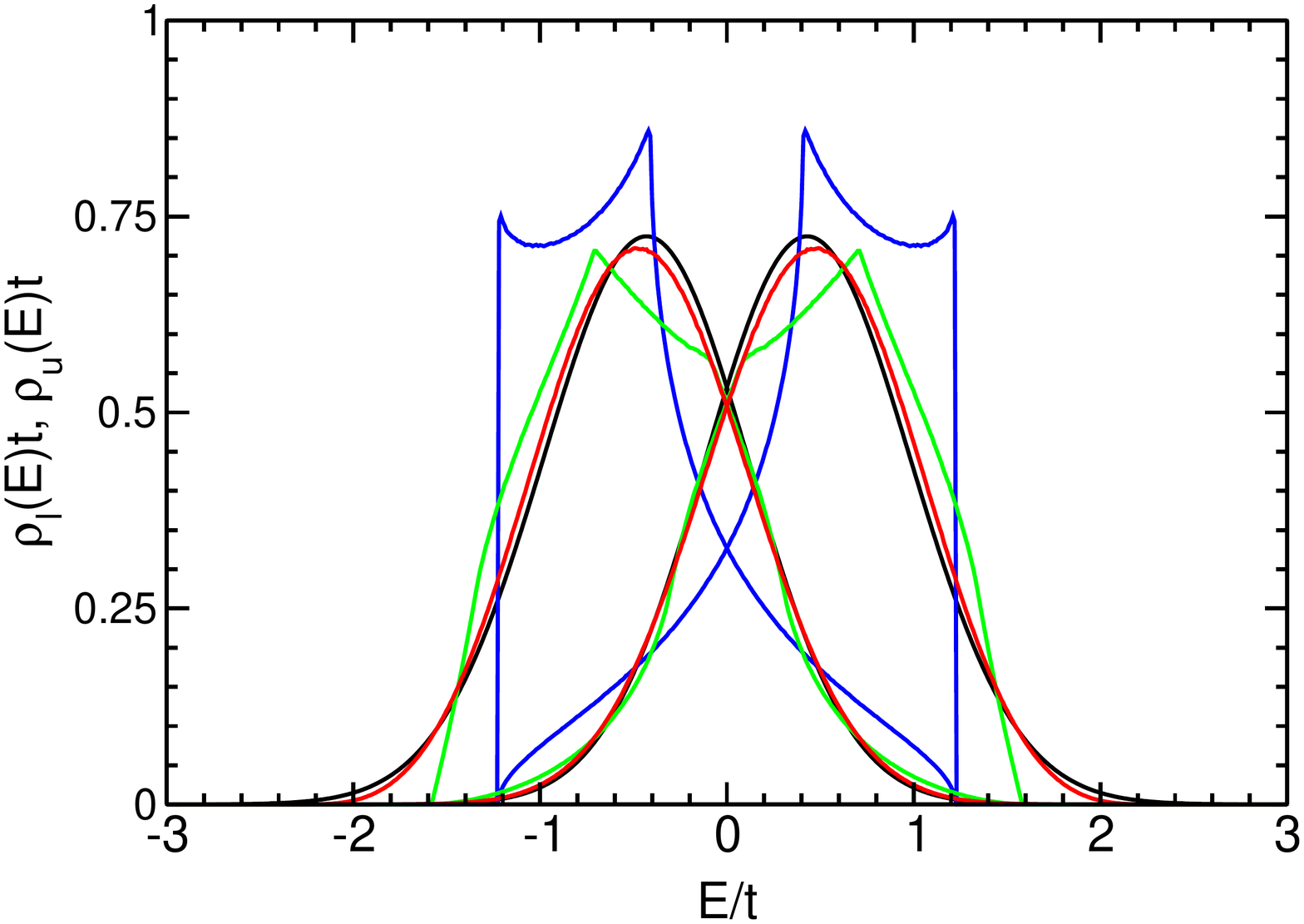}
\caption{
Evolution of the DOS for the single-particle eigenstates $\ell$ and
$u$ of the $e_g$ orbital model at $U=0$ with increasing dimension:
$d=3$ (blue), $d=5$ (green), $d=11$ (red), and $d=\infty$ (black).
The lines at finite $d$ have been calculated by ${\bf k}$-space
sampling, those at $d=\infty$ from Eq.~(\ref{rholu}).
}
\label{fig:dosdim}
\end{figure}

The creation operators for the single-particle {\it eigen\/}-states of
the kinetic energy are seen from Eq.~(\ref{Hkt}) to be simply
proportional to the sum and difference of the phased single-particle
operators,
\begin{eqnarray}
\label{elk}
{\hat d}_{{\bf k},{\ell}}^{\dagger} =
\frac{1}{\sqrt{2}}
\left({\hat{\tilde c}}_{{\bf k}+}^{\dagger}
               + {\hat{\tilde c}}_{{\bf k}-}^{\dagger}
               \right)  ,  \\
\label{euk}
{\hat d}_{{\bf k},{u}}^{\dagger} =
\frac{-i}{\sqrt{2}}
\left({\hat{\tilde c}}_{{\bf k}+}^{\dagger}
               - {\hat{\tilde c}}_{{\bf k}-}^{\dagger}
               \right)  ,
\label{elkeuk}
\end{eqnarray}
and the corresponding energies of the lower (${\ell}$) and upper ($u$)
subbands are
\begin{equation}
\varepsilon_{{\ell},u} ({\bf k})
   = - t \left( A_{\bf k} \pm \gamma B_{\bf k} \right) \; ,
\label{epslu}
\end{equation}
where the $+$ ($-$) sign corresponds to the lower (upper) subband.
We calculate the DOSs for the two subbands by evaluating the
convolution of the DOSs corresponding to $A_{\bf k}$ and to the two
branches of $B_{\bf k}$. Strictly this requires $A_{\bf k}$ and
$B_{\bf k}$ to be uncorrelated \cite{Pet21}, which is obviously not
the case because both depend on ${\bf k}$, but we assume that this
condition becomes irrelevant in the limit $d\to\infty$, and check the
validity of this assumption afterwards (see below). We then obtain for
the subband DOSs
\begin{eqnarray}
\label{rholu}
\rho_{\ell,u}(E)&=&\frac{1}{\sqrt{2 \pi}}\frac{2}{\sigma^2 t}\;
  \mathrm{e}^{-\frac{1}{2}(\frac{2 E}{t})^2} \nonumber \\
&\times& \left\{ 1 + \sqrt{\pi}\frac{\gamma E}{\sigma t}\,
  \mathrm{e}^{\left(\frac{\gamma E}{\sigma t}\right)^2}
  \left[ \mathrm{erf}{\left(\frac{\gamma E}{\sigma t} \right)} \mp 1
  \right]\right\}  \, , \nonumber \\
\end{eqnarray}
where the upper (lower) sign corresponds to the $\ell$ ($u$) subband,
and for their sum, which we will need later,
\begin{eqnarray}
\label{rhol+u}
\rho_{\ell+u}(E)&=&\frac{2}{\sqrt{2 \pi}}\frac{2}{\sigma^2 t}\;
  \mathrm{e}^{-\frac{1}{2}(\frac{2 E}{t})^2} \nonumber \\
&\times& \left\{ 1 + \sqrt{\pi}\, \frac{\gamma E}{\sigma t}\,
  \mathrm{e}^{(\frac{\gamma E}{\sigma t})^2}
  \mathrm {erf}{\left(\frac{\gamma E}{\sigma t} \right)}
\right\}  \, .
\end{eqnarray}
These DOSs are also shown, for $\gamma=1$, in Fig.~\ref{fig:dos}(b) by
the dash-dotted red and orange lines and by the dotted red line,
respectively.

\begin{figure}[t!]
\vskip -.5cm
\includegraphics[width=1.15\columnwidth]{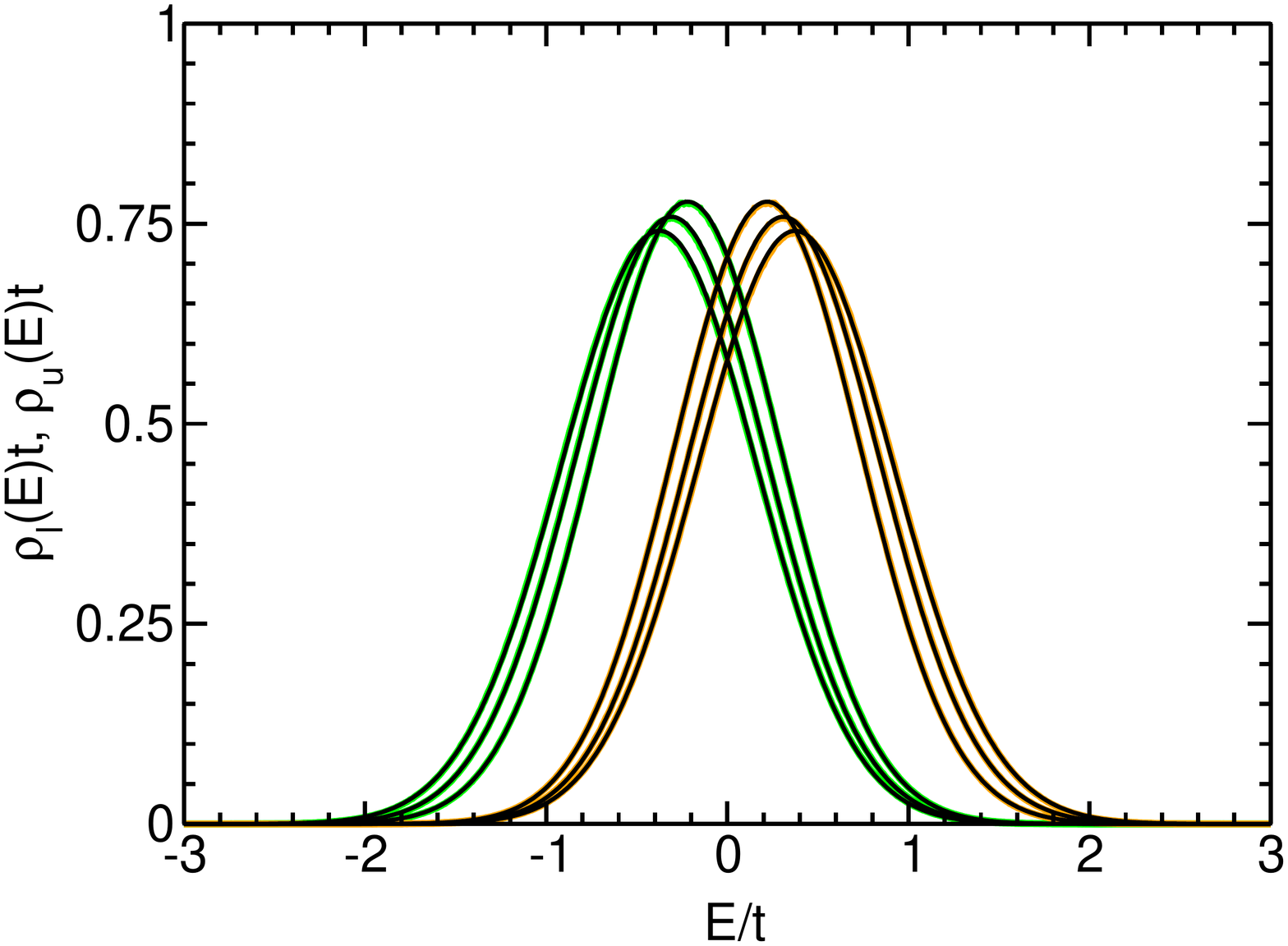}
\caption{
Densities of states $\rho_{\ell,u}(E)$ for the single-particle
eigenstates of the $e_g$ orbital model at $U=0$, for three values of
the parameter $\gamma$ ($\gamma=0.25$, $0.50$, and $0.75$, with
increasing distance between the maxima), as given by the analytical
expression (\ref{rholu}) for $d\to\infty$ (thin black lines), and as
calculated numerically at dimension $d=21$,
where green lines are used for $|{\bf k},\ell\rangle$,
orange lines are used for $|{\bf k},u\rangle$).
}
\label{fig:olfitgaa}
\end{figure}

We emphasize that the subbands
$\{\varepsilon_{\ell}({\bf k}),\varepsilon_u({\bf k})\}$ are
non-degenerate except for the ${\bf k}$-points for which $B_{\bf k}=0$.
This feature found already at dimension $d=3$ is generic and persists
at increasing dimension up to $d=\infty$, contrary to the suggestion
made in Ref. \cite{Bun98}. Figures~\ref{fig:dosdim} and
\ref{fig:olfitgaa} demonstrate this explicitly.
In Fig.~\ref{fig:dosdim} the DOSs of  the two subbands are seen to
become smoother with increasing dimension but to remain distinct
although they partly overlap. In Fig.~\ref{fig:olfitgaa} the evolution
of the subband DOSs with $\gamma$ is illustrated:
in the spin Hubbard model, {\it i.e.\/}, at $\gamma=0$ (not shown) they
necessarily coincide, compare Eq.~(\ref{epslu}), while they are seen to
get separated with their maxima moving away from each other with
increasing $\gamma$ and their maxima finally reaching a maximum
distance of $\simeq 0.84$ at $\gamma=1$ (not shown), {\it i.e.\/}, in
the full orbital model.
This figure also demonstrates the validity of our assumption made in
deriving Eq.~\eqref{rholu}: the colored lines, calculated by
${\bf k}$-space sampling of the dispersion Eq.~(\ref{epslu}) at $10^6$
${\bf k}$-values, are indistinguishable from the black lines, calculated
from the analytical $d=\infty$ expression Eq.~(\ref{rholu}).

We point out that the single-particle eigenstates given by Eqs.
(\ref{elk})-(\ref{euk}) are in fact {\it real} and could be considered
as the ``phased'' real-orbital single-particle operators which are the
counterpart to the phased complex-orbital single-particle operators
defined by Eq.~(\ref{tck}) above,
\begin{eqnarray}
\label{elkr}
{\hat d}_{{\bf k},{\ell}}^{\dagger}&=&
              \: \: \: \, \cos (\phi_{\bf k}/2) \, {\hat c}_{{\bf k} x_0}^{\dagger}
               + \sin(\phi_{\bf k}/2)\, {\hat c}_{{\bf k} \bar{x}_0}^{\dagger} ,
                \\
\label{eukr}
{\hat d}_{{\bf k},{u}}^{\dagger}&=&
               - \sin (\phi_{\bf k}/2) \, {\hat c}_{{\bf k} x_0}^{\dagger}
               + \cos(\phi_{\bf k}/2)\, {\hat c}_{{\bf k} \bar{x}_0}^{\dagger} ,
\label{elkeukr}
\end{eqnarray}
as one readily verifies from Eqs.~(\ref{tck}) and (\ref{real}), or
alternatively by recognizing that Eqs.~(\ref{elk}) and (\ref{euk}) have
the form of coherent orbital states like Eq.~(\ref{cohorb}) with
$\psi=0$ and thus are given by Eq.~(\ref{realcohorb}). So they are
represented in the ``equatorial'' plane, compare Fig.~\ref{fig:orbi}(c),
by an angle $\theta$ equal to the phase angle $\phi_{\bf k}$, which is
entirely determined by the wavevector and therefore is in general not
equal to any of the $\chi_n$. An important consequence is that each of
these single-particle states carries a finite orbital polarization,
{\it i.e.\/}, contributes at each lattice site to the value of the
pseudospin in the $xy$-plane, see Fig.~\ref{fig:orbi}(b),
\begin{eqnarray}
\label{polarkx}
\langle{\bf k},{\ell}| \hat{T}_i ^x|{\bf k},{\ell}\rangle &=&
- \langle{\bf k},u| \hat{T}_i^x |{\bf k},u\rangle =
\frac{1}{2N} \cos \phi_{\bf k} , \\
\label{polarky}
\langle{\bf k},{\ell}| \hat{T}_i ^y|{\bf k},{\ell}\rangle &=&
- \langle{\bf k},u| \hat{T}_i^y |{\bf k},u\rangle =
\frac{1}{2N} \sin \phi_{\bf k} , \\
\label{polarkz}
\langle{\bf k},{\ell}| \hat{T}_i ^z|{\bf k},{\ell}\rangle &=&
\langle{\bf k},u| \hat{T}_i^z |{\bf k},u\rangle = 0 .
\end{eqnarray}
It is noteworthy that while ``$\ell$'' and ``$u$'' may properly be
called (sub)bands, the polarization direction of the states in each band
varies with ${\bf k}$, in contrast to the familiar feature of bands of
particles carrying spin where the spin direction is the same for all
states in a band.

\begin{figure}[t!]
\vskip -.7cm
\includegraphics[width=1.17\columnwidth]{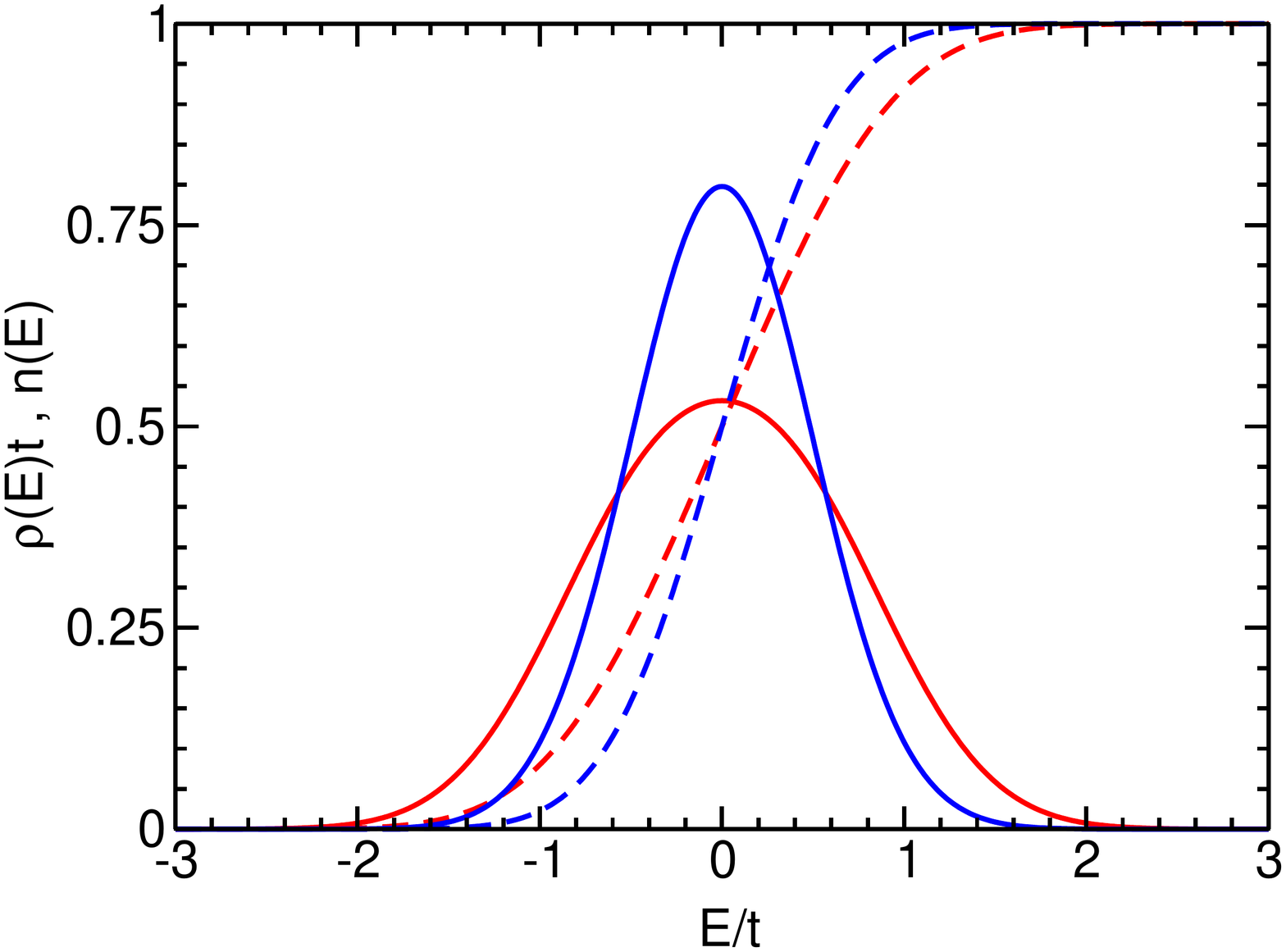}
\caption{
Total DOS $\rho_{\ell+u}(E)$ (\ref{rhol+u}) (solid lines) and the
electron density per orbital/spin flavor $n(E)$ (dashed lines) in the
unpolarized state, as obtained for the orbital $e_g$ Hubbard model
($\gamma=1$, red lines) and the spin Hubbard model
($\gamma=0$, blue lines) at $d=\infty$.
}
\label{fig:hubb}
\end{figure}

\subsection{Comparison with the spin Hubbard model}
\label{sec:compar}

The expression (\ref{rhol+u}) derived for the combined DOS of the two
subbands $\rho_{\ell+u}(E)$ gives us the opportunity to compare the
distribution of the single-particle eigenstates in the $e_g$ orbital
Hubbard with the one in the spin Hubbard model, see Fig.~\ref{fig:hubb}.
In the orbital model the total DOS is broader and the total electron
density $n(E)$ grows slower with increasing Fermi energy $E$. The plot
already indicates that the kinetic energy in the orbital model at a
particular density $n$, obtained by filling the two subbands up to the
same Fermi energy, is lower than the kinetic energy obtained at the
same density in the spin model. This difference is a manifestation of
the flavor non-conserving hopping which opens an extra channel
for the kinetic energy and thus enhances it in the orbital model.

\subsection{Kinetic energies}
\label{sec:ekin}

Before taking electron correlations into account, we examine the
kinetic energy of the various trial states $|\Phi_0 \rangle$ built from
the single-particle states discussed above, and identify the most
favorable states in the absence of Coulomb interactions.
We consider on the one hand
uniformly polarized ordered states, obtained by filling a single
band (or, in the cases of alternating order, subsequently the lower and
upper band corresponding to the same sublattice occupation), and on the
other hand unpolarized ``para-orbital'' states, obtained by filling two
degenerate bands up to the same Fermi energy, compare Eq.~\eqref{ckk}.
For each such trial state the particle density $n$ and the kinetic
energy $E_{\rm kin}$ as a function of the Fermi energy $E_{\rm F}$
are given by
\begin{eqnarray}
\label{nEF}
n(E_{\rm F}) &=&
 \int_{-\infty}^{E_{\rm F}}  \rho(E)  \, \mathrm{d}E \, , \\
\label{EkinEF}
E_\mathrm{kin}(E_{\rm F}) &=&
 \int_{-\infty}^{E_{\rm F}}  E \, \rho(E)  \, \mathrm{d}E \, ,
\end{eqnarray}
where $\rho(E)$ is the DOS of the corresponding band(s) of
single-particle states. Only the density range $0 \leq n \leq 1$ will
be considered since $n$ and $2-n$ are equivalent because of
particle-hole symmetry.

\begin{figure}[b!]
\vskip -.3cm
\begin{center}
\includegraphics[width=1.12\columnwidth]{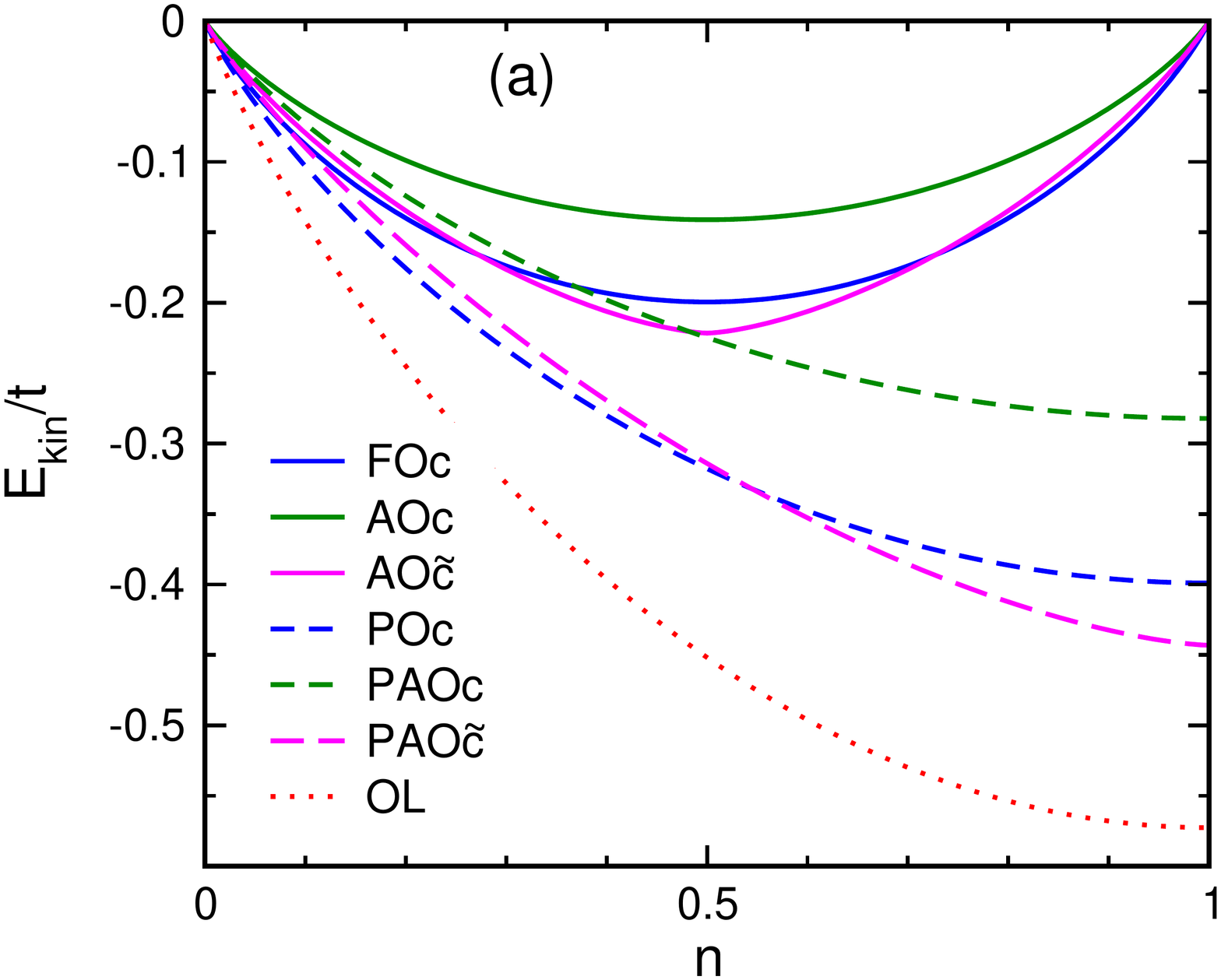}
\includegraphics[width=1.12\columnwidth]{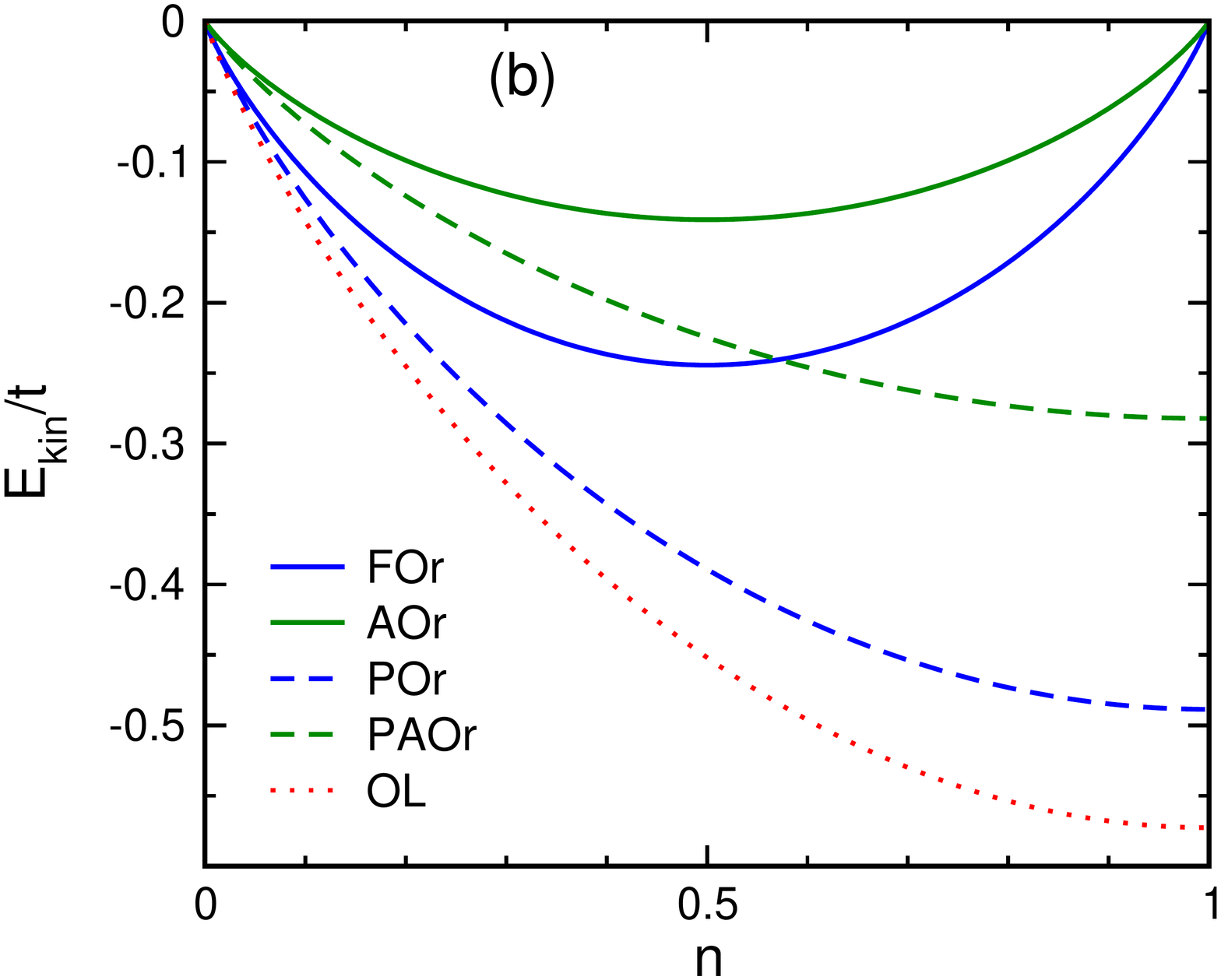}
\end{center}
\vskip -.3cm
\caption{
Kinetic energies $E_{\rm kin}$ in the limit $d\to\infty$ of the orbital
ordered and the para-orbital states of the $e_g$ Hubbard model
\eqref{Hd} ({\it i.e.\/} $\gamma=1$) at $U=0$ for increasing
electron density $n$, compared with that of the orbital liquid state:
(a) phases with {\it complex\/} orbitals:
ferro-orbital FOc (solid blue line),
alternating-orbital AOc (solid green line),
alternating phased orbital AO$\rm\tilde{c}$ (solid cyan line),
para-orbital POc (dashed blue line),
para alternating orbital PAOc (dashed green line),
para alternating phased orbital PAO$\rm\tilde{c}$ (dashed cyan line),
and orbital liquid OL (dotted red line);
(b) phases with {\it real\/} orbitals:
ferro-orbital FOr (solid blue line),
alternating-orbital AOr (solid green line),
para-orbital POr (dashed blue line),
para alternating orbital PAOr (dashed green line),
and orbital liquid OL [dotted red line, the same as in (a)].
}
\label{fig:et}
\end{figure}

We begin with the {\it ordered\/} orbital phases built from
{\it complex-orbital\/} single-particle states. The simplest one is the
complex ferro-orbital state $\big|\Phi_0^{\rm FOc}\big\rangle$ obtained
by filling the $|{\bf k}+\rangle$-band (or the $|{\bf k}-\rangle$-band).
The complex alternating-orbital state $\big|\Phi_0^{\rm AOc}\big\rangle$
is, compare Eqs.~\eqref{ckABl} and \eqref{ckABu}, obtained by filling
consecutively the $|{\bf k},+-,\ell\rangle$-band and the
$|{\bf k'},+-,u\rangle$-band, and the phased complex alternating-orbital
state $\big|\Phi_0^{\rm AO\tilde{\rm c}}\big\rangle$ is built similarly
from the $\hat{\tilde c}_{{\bf k}+-,\ell}^{\dagger}$ and
$\hat{\tilde c}_{{\bf k'}+-,u}^{\dagger}$ single-particle states.

Moving on to the {\it unpolarized\/} states we have the complex
para-orbital state POc, obtained by filling simultaneously the
$|{\bf k}+\rangle$- and the $|{\bf k}-\rangle$-band,
the complex alternating para-orbital state PAOc,
by filling simultaneously the $|{\bf k},+-,\ell\rangle$-band
and the $|{\bf k'},-+,\ell\rangle$-band, compare Eq.~\eqref{ckABl},
{\it i.e.\/}, from the lower band only, and finally the similarly
defined complex phased alternating para-orbital state
PAO$\tilde{\rm c}$. Figure~\ref{fig:et}(a) shows the kinetic energies
of these six states plotted versus the particle density.

Next we consider the trial states built from {\it real-orbital\/}
single-particle states, again beginning with the {\it ordered\/} phases.
Analogous to the complex trial states above we have the real
ferro-orbital state $\big|\Phi_0^{\rm FOr}\big\rangle$ obtained by
filling the $|{\bf k} x_0\rangle$-band (or alternatively the
$|{\bf k}{\bar x}_0\rangle$-band) and the real alternating-orbital
state $\big|\Phi_0^{\rm AOr}\big\rangle$ obtained by filling
consecutively the $|{\bf k},x_0 {\bar x}_0,\ell \rangle$-band
and the $|{\bf k'},x_0 {\bar x}_0,u \rangle$-band,
compare Eqs.~\eqref{ckABl} and \eqref{ckABu}.
Here the {\it unpolarized\/} states are the real para-orbital state
POr, with equally filled $|{\bf k} x_0\rangle$-band and
$|{\bf k}{\bar x}_0\rangle$-band, and the real alternating para-orbital
state PAOr, with the $|{\bf k},x_0 {\bar x}_0,\ell\rangle$-band
and the $|{\bf k'},{\bar x}_0 x_o,\ell \rangle$-band filled equally.
The kinetic energies of these four states plotted versus particle
density are shown in Fig.~\ref{fig:et}(b).

The $n$-dependence of the various kinetic energies can be understood
as follows. For all ordered states for which the single-particle DOS is
Gaussian, which includes the FOc, AOc, FOr, and AOr states,
the particle density and kinetic energy take the form
\begin{eqnarray}
\label{nEFw}
n(E_{\rm F}) &=& \frac{1}{2}
 \left[ 1 + \erf \! \left( \frac{E_{\rm F}}{wt \sqrt{2}} \right) \right] \, , \\
E_\mathrm{kin}(E_{\rm F}) &=& - \frac{wt}{\sqrt{2 \pi}} \;
   \mathrm{exp}\! \left(\! -\frac{1}{2} \!
   \left( \frac{E_{\rm F}}{wt} \right)^{\!\! 2} \right) \, ,
\label{EkinEFw}
\end{eqnarray}
where $\erf$ is the error function. Formally, inversion of
Eq.~\eqref{nEFw} gives
\begin{equation}
E_{\rm F}(n) = wt \sqrt{2} \: \mathrm{inverf}(2n - 1) \, ,
\label{EFwn}
\end{equation}
where $\mathrm{inverf}$ is the inverse error function,
and inserting this into Eq.~\eqref{EkinEFw} one obtains
\begin{equation}
E_\mathrm{kin}(n) = - \frac{wt}{\sqrt{2 \pi}} \;
   \mathrm{exp}\! \left( - \left( \mathrm{inverf}(2n - 1) \right)^{2} \right)  \, .
\label{EkinEFn}
\end{equation}
This demonstrates that the shape of $E_\mathrm{kin}(n)$ is the same for
the states mentioned above, as is evident in Fig.~\ref{fig:et}, and
that the only difference is in the width prefactor $w$,  which also
determines the value of the minimum, $ - wt/\sqrt{2 \pi}$, at $n=1/2$.
The only exception is the AO$\tilde{\rm c}$ state for which the Rayleigh
distribution type DOS Eq.~\eqref{rhocaltphase} leads to
\begin{eqnarray}
\label{nEFwAOw}
n^{{\rm AO}\tilde{\rm c}}(E_{\rm F}) &=& \frac{1}{2} \!
 \left[ 1 + \mathrm{sgn}(E_{\rm F}) \!  \left( \! 1 -
\mathrm{e}^{-\left(\!\frac{2E_{\rm F}}{\gamma t}\!\right)^2}
 \right) \! \right] \! \! , \\
\label{EkinEFwAOw}
E^{{\rm AO}\tilde{\rm c}}_\mathrm{kin}(E_{\rm F}) &=&
   - \frac{t}{\sqrt{2 \pi}} \frac{\gamma \pi}{4 \sqrt{2}}\;
    \left[ 1 - \erf \! \left( \! \frac{2 |E_{\rm F}|}{\gamma t} \! \right)
    \right. \nonumber \\
   & &  \hskip 1.3cm
    + \left. \frac{2}{\sqrt{\pi}}    \frac{2 |E_{\rm F}|}{\gamma t}
\mathrm{e}^{-\left(\!\frac{2E_{\rm F}}{\gamma t}\!\right)^2}  \right]  \! .
 \end{eqnarray}
As Fig.~\ref{fig:et}(a) shows, this makes the AO$\tilde{\rm c}$ state
have lower kinetic energy than the FOc state close to quarter-filling,
{\it i.e.,} in the interval $0.25<n<0.75$. This can be understood from
the fact that the DOS of the phased complex-orbital single-particle
states has more weight at more negative energies, see
Fig.~\ref{fig:dos}(a). However, as seen in Fig.~\ref{fig:et}(b), it is
the FOr state that clearly has the lowest kinetic energy of all ordered
states due to its large width, with minimum energy $-0.2443~t$, whereas
the minimum for the AO$\tilde{\rm c}$ state is only $- 0.2216~t$, see
Eq.~\eqref{EkinEFwAOw}. The basic reason for the difference is that the
FOr state benefits fully from the orbital-flavor conserving hopping
channel and also partially from the non-conserving channel, compare
Eq.~\eqref{epsrz}, while the AO$\tilde{\rm c}$ state benefits fully
from the orbital-flavor non-conserving channel but not at all from the
conserving channel, {\it cf.\/} Eq.~\eqref{epscaltphase}.

As regards the unpolarized states one may observe that for every
para-orbital state PX its kinetic energy is related to that of the
associated ordered state X by
\begin{equation}
E_\mathrm{kin}^{\rm PX}(n) = 2 \, E_\mathrm{kin}^{\rm X}(n/2) \, ,
\label{EkinP-O}
\end{equation}
which follows directly from the defining equations \eqref{nEF} and
\eqref{EkinEF}. Equation \eqref{EkinP-O} implies that the relative
magnitude of the kinetic energies of the unpolarized states is the same
as that of the ordered states, as indeed seen in Fig.~\ref{fig:et}.
In particular, the POr state has the lowest kinetic energy of all
para-orbital disordered phases, for the
same reason as the FOr state is the lowest-energy ordered state.
Also, the para-orbital states have considerably lower kinetic energy
than their ordered counterparts, but they will get renormalized at
finite $U$ whereas the ordered states will not.

Finally we consider the trial state built from the single-particle
{\it eigenstates\/} by filling the lower and the upper subband up to
the same Fermi energy, written down here explicitly, compare
Eqs.~\eqref{elk}~--~\eqref{epslu},
\begin{equation}
|\Phi_0^{\rm OL}\rangle
=\prod_{{\bf k};\varepsilon_{\ell}({\bf k})<E_{\rm F}} \,
\prod_{{\bf k'};\varepsilon_{u}({\bf k'})<E_{\rm F}} \,
\hat{d}_{{\bf k},{\ell}}^{\dagger} \,
\hat{d}_{{\bf k'},u}^{\dagger} |0\rangle \,.
\label{OL}
\end{equation}
This state we call the OL state because it is unpolarized
in the following remarkable way.

Whereas all single-particle states in $|\Phi_0^{\rm OL}\rangle$ are
polarized individually, compare Eqs.~\eqref{polarkx} and
\eqref{polarky}, the variation with ${\bf k}$ makes their contributions
to the pseudospin add up to zero at each lattice site, at any filling
$n(E_{\rm F})$ and for each subband $\ell$ and $u$ separately. This
follows from the following considerations (for a more detailed proof,
see the Appendix):
\hfill\break
(i) for every single-particle state $|{\bf k},\ell\rangle$ from the
lowest subband which is occupied in $|\Phi_0^{\rm OL}\rangle$
because it satisfies $\varepsilon_{\ell}({\bf k})<E_{\rm F}$, all
single-particle states $|{\bf k}'_n,\ell\rangle$ with wave vectors
${\bf k}'_n$ generated from ${\bf k}$ by successive cyclic permutations
of its components are also occupied; this holds because both
$A_{\bf k}$ and $B_{\bf k}$ are invariant under such permutations,
compare Eqs.~\eqref{Ak} and \eqref{Bk}, implying that
$\varepsilon_{\ell}({\bf k}'_n)=\varepsilon_{\ell}({\bf k})$, and it
follows that $\varepsilon_{\ell}({\bf k}'_n)<E_{\rm F}$;
the same reasoning applies for the upper subband $u$; \hfill\break
(ii) under each such transformation ${\bf k} \mapsto {{\bf k}'_n}$
the corresponding phase angle is transformed as
$\phi_{\bf k}\mapsto\phi_{{\bf k}}+4n\pi/d$,
so summing $\exp({i\phi_{{\bf k}'_n})}$ over all ${\bf k}'_n$
yields $0$, and the compensation of the contributions to the
pseudospin follows,
compare again Eqs.~\eqref{polarkx} and \eqref{polarky}.
\hfill\break
Therefore, even though the densities in the two partially filled
subbands are different, $n_{\ell}(E_{\rm F})>n_{u}(E_{\rm F})$,
as recognized from the DOSs shown in Fig.~\ref{fig:dos}(b),
the OL state is unpolarized, because each subband is unpolarized by
itself at any filling. This mechanism is entirely
different from the compensation of two oppositely polarized bands at
equal filling as occurs for para-orbital (or paramagnetic) states.

We emphasize that the reasoning above is not specific for $d=\infty$ but
holds in any dimension, and so the resulting absence of polarization in
both bands is a characteristic feature of the OL {\it per se\/}.
It is associated with the fact that the OL transforms as $A_1$ under the
symmetry group $C_d$ (is invariant under U(1) in the limit $d\to\infty$)
and implies that {\it the OL phase protects the (hyper)cubic symmetry\/}.
--- We further point out that there is necessarily double occupancy in
$|\Phi_0^{\rm OL}\rangle$. This is obvious from the fact that the
single-particle states in $|\Phi_0^{\rm OL}\rangle$ do not correspond
to a single common direction in the `equatorial' plane of
Fig.~\ref{fig:orbi}(c) because of the variation of $\phi_{\bf k}$ with
the wave vector ${\bf k}$.

The relevant DOS from which to obtain the particle density and the
kinetic energy in the OL state is $\rho_{\ell+u}(E)$, Eq.~\eqref{rhol+u}.
One finds
\begin{widetext}
\begin{eqnarray}
\label{nOL}
n_{\rm OL}(E_{\rm F}) &=&
  1 + \mathrm{erf}\left( \frac{\sqrt{2}E_{\rm F}}{t} \right)
  - \frac{\gamma}{\sigma \sqrt{2}}
  \mathrm{e}^{-2\left(\frac{E_{\rm F}}{\sigma t}\right)^2}
  \mathrm{erf}\left( \frac{\gamma E_{\rm F}}{\sigma t} \right) \, , \\
\label{EkinOL}
E_{{\rm kin};\rm OL}(E_{\rm F}) &=& -\frac{t}{\sqrt{2 \pi}}
  \left\{
  \, \mathrm{e}^{-2\left( \frac{E_{\rm F}}{t} \right)^2}
  + \gamma \sqrt{\pi} \left[ \frac{E_{\rm F}}{\sigma t}
  \, \mathrm{e}^{-2\left( \frac{E_{\rm F}}{\sigma t} \right)^2}
  \textrm{erf}\,\left(\frac{\gamma E_{\rm F}}{\sigma t} \right)
  + \mathrm{F} \left( \gamma; \frac{|E_{\rm F}|}{\sigma t} \right)
  \right] \right\} ,
\label{Ekin}
\end{eqnarray}
\end{widetext}
where
\begin{equation}
 \mathrm{F}(\gamma; x) = \int_x^\infty \mathrm{e}^{-2y^2}
               \mathrm{erf}(\gamma y) \, \mathrm{d} y .
\label{F}
\end{equation}
The function  $\mathrm{F}(\gamma; x)$ has to be calculated numerically.
Since
$\mathrm{F}(1; 0)=\frac{1}{\sqrt{2 \pi}}
\left[\frac{\pi}{2} - \tan^{-1}(\sqrt{2})\right]=0.245541$ \cite{Ng69},
we have rewritten Eq.~\eqref{F} at $\gamma=1$ as
\begin{equation}
 \mathrm{F}(1; x) = 0.245541 -  \int_0^x \mathrm{e}^{-2y^2}
               \mathrm{erf}(y) \, \mathrm{d} y .
\label{Fbar}
\end{equation}
to perform the calculation for the full orbital case.

The kinetic energy of the OL state is included in both
Fig.~\ref{fig:et}(a) and Fig.~\ref{fig:et}(b), and is seen to be lower
than that of any other trial state, remarkably including the
lowest-energy complex-orbital and real-orbital para states.
The explanation is similar to that above for the FOr and the POr states.
The OL state fully captures all kinetic energy available in the orbital
Hubbard model Eq.~\eqref{Hd}, both from the orbital-flavor conserving
hopping channel and from the non-conserving hopping channel, {\it cf.\/}
Eq.~\eqref{epslu}.
This is a strong hint that the OL phase could be favored also
in the presence of electron interactions. To establish the phase with
the lowest overall energy we need to consider the energy renormalization
of the para-orbital states and the OL state in the presence of the
Coulomb interaction, {\it i.e.\/}, at finite $U$.

\section{Correlations by the Gutzwiller method}
\label{sec:corr}

The original analysis of the Gutzwiller approach in the limit
$d\to\infty$, including the proof that the Gutzwiller approximation
becomes exact in that limit, was carried out specifically for the spin
Hubbard model, implying the implicit assumption of SU(2) symmetry
\cite{Vol89,Met89,Met91}. We therefore need to investigate which
modifications are required in the orbital case where we have only $C_d$
symmetry, and check if the proof of exactness still holds.

\subsection{General derivation at dimension $d$}
\label{sec:gen}

Electron correlations due to the on-site Hubbard repulsion $U$ will be
implemented using the wave function introduced by Gutzwiller,
\begin{equation}
|\Psi(g)\rangle=g^{\hat D}\left|\Phi_0\right\rangle
=\prod_{i} \left(1-(1-g)\hat D_i\right)\left|\Phi_0\right\rangle\,,
\label{psiG}
\end{equation}
with
\begin{equation}
\hat D = \sum_{i} \hat D_{i} , \quad\quad
\hat D_{i} = \hat n_{i+} \hat n_{i-} ,
\label{hatDi}
\end{equation}
where the variational parameter $g$ is determined by minimizing the
energy \cite{Gut65,Vol84,Met89,Met91},
\begin{equation}
E(g) = \frac{ \langle \Psi(g) | {\cal H}_d | \Psi(g)\rangle }
   { \langle \Psi(g) | \Psi(g)\rangle } \, .
\label{Eg}
\end{equation}
First, we consider here the general case and derive the propagator and
self-energy. In the present case the fermion (orbital) flavor is not
conserved and we introduce a propagator,
\begin{equation}
{\cal P}_{ij;\alpha\beta}= \frac{ \big\langle\Psi\big|
\hat c_{i\alpha}^{\dagger} \hat c_{j\beta}^{}\big|\Psi\big\rangle }
   { \langle \Psi(g) | \Psi(g)\rangle } \, .
\label{calP}
\end{equation}
which we allow to have, at least in principle, {\it nonzero offdiagonal
matrix elements with respect to flavor\/} (i.e., $\alpha\neq\beta$).
Following Metzner \cite{Met91}, one finds the following expression for
the matrix elements,
\begin{eqnarray}
{\cal P}_{ij;\alpha\beta}&=&\big\{c_{i\alpha}^{\dagger}c_{j\beta}^{}
[1-(1-g)(n_{i\bar{\alpha}}+n_{j\bar{\beta}})     \nonumber\\
&+& (1-g)^2(n_{i\bar{\alpha}}n_{j\bar{\beta}}
+\delta_{ij}\delta_{\alpha\beta}n_{i\bar{\alpha}})X\big\}_0^c\,,
\label{P}
\end{eqnarray}
where
\begin{equation}
X\equiv 1+\sum_{m=1}^{\infty}\frac{(g^2-1)^m}{m!}
\sum_{g_1\dots g_m}{D}_{g_1},\dots,{D}_{g_m}\,.
\label{X}
\end{equation}
Similarly, the bare propagator is defined by,
\begin{equation}
{\cal P}_{ij;\alpha\beta}^0=\big\langle\Phi_0\big|
\hat c_{i\alpha}^{\dagger}\hat c_{j\beta}^{}\big|\Phi_0\big\rangle
=\left\{c_{i\alpha}^{\dagger}c_{j\beta}^{}\right\}^c_0\,.
\label{P0}
\end{equation}
Here ${\cal P}^0$ is, like ${\cal P}$, an $N\times N\times 2\times 2$
matrix {\it with off-diagonal elements with respect to orbital flavor
generated by the off-diagonal hopping in orbital space.}
We emphasize that the curly brackets $\{\dots\}_0^c$ in Eqs.~(\ref{P})
and (\ref{P0}) denote the sum over all connected products of
anticommuting contractions, evaluated in the trial state
$|\Phi_0\rangle$, and thus the ($c_{i\alpha}$, \textit{etc.})
symbols are Grassman variables and not the fermion operators
($\hat c_{i\alpha}$, \textit{etc.}).

Next we define the self-energy ${\cal S}$.
Here ${\cal S}_{ij}$ is a
$2\times 2$ matrix {\it with off-diagonal elements in orbital space.\/}
The elements of the self-energy ${\cal S}$ are therefore different from
those for the RVB wave function \cite{Met91}, and
{\it include the specific last term\/} below,
\begin{eqnarray}
{\cal S}_{ij;\alpha\beta}\!&=&\left\{\left[(1-g)^2c_{i\alpha}^{\dagger}
n_{i\bar{\alpha}}^{}c_{j\beta}^{}n_{j\bar{\beta}}^{}\right.\right.\nonumber\\
&-&\!\left.\left.\delta_{ij}(g^2-1)\!\left(\delta_{\alpha\beta}n_{i\bar{\alpha}}\!
-\delta_{\alpha\bar{\beta}}c_{i\alpha}^{\dagger}c_{j\beta}^{}\right)\right]
\!X\right\}_0^c\!.
\label{calS}
\end{eqnarray}
The self-energy gives the double occupancy,
\begin{equation}
\overline{d}_i=\frac12\,\frac{g^2}{1-g^2}{\rm Tr}_{\alpha}
\left({\cal S}{\cal P}^0\right)_{ii}\,.
\label{di}
\end{equation}
The further derivation, following closely the case of the RVB state
considered by Metzner \cite{Met91}, leads after some algebraic
manipulations to the propagator in real space between sites $i$ and $j$,
\begin{eqnarray}
{\cal P}_{ij}^{}&=&{\cal P}_{ij}^0+\delta_{ij}\left[\frac{1-g}{1+g}
\left({\cal S}{\cal P}^0\right)_{ii}\delta_{\alpha\beta}
-\frac{{\cal S}_{ii}}{(1+g)^2}\right]    \nonumber\\
&+&\left[\left({\cal P}^0-\frac{1}{1+g}\right){\cal S}
\left({\cal P}^0-\frac{1}{1+g}\right)\right]_{ij}\,.
\label{Pij}
\end{eqnarray}
Here in the first local term at site $i$ only the diagonal elements
of the matrix ${\cal S}{\cal P}^0$ contribute, as indicated by
$\delta_{\alpha\beta}$.

The first term in Eq.~(\ref{P}) is related to the self-energy ${\cal S}$ by
\begin{equation}
\big\{c_{i\alpha}^{\dagger}c_{j\beta}^{}X\big\}_0^c\equiv\left(
{\cal P}^0+{\cal P}^0{\cal S}{\cal P}^0\right)_{ij;\alpha\beta}\,.
\label{self}
\end{equation}
Finally, we define a ``proper self-energy" ${\cal S}^*$ as the sum over
all one-particle irreducible diagrams contributing to ${\cal S}$---the
two self-energies are then related by a Dyson equation,
\begin{equation}
{\cal S}={\cal S}^*
+{\cal S}^*\circ{\cal P}^0\circ{\cal S}\,.
\label{Dyson1}
\end{equation}

After introducing a renormalized propagator
\begin{equation}
\label{Dyson2}
\overline{\cal P}={\cal P}^0
+{\cal P}^0\circ{\cal S}\circ{\cal P}^0 \,,
\end{equation}
one finds that, owing to Eq.~(\ref{Dyson1}), it satisfies
\begin{equation}
\label{Dyson3}
\overline{\cal P}={\cal P}^0
+{\cal P}^0 \circ {\cal S}^* \circ \overline{\cal P} \, .
\end{equation}
The site-diagonal matrix elements of ${\cal S}$ are related to those of
$\overline{\cal P}$ as follows,
\begin{eqnarray}
\label{SP1}
{\cal S}_{ii;\alpha\alpha}&=&\;\;
(1-g^2){\overline{\cal P}}_{ii;\bar{\alpha}\bar{\alpha}}\,, \\
\label{SP2}
{\cal S}_{ii;\alpha\bar{\alpha}}&=&
-(1-g^2)\overline{\cal P}_{ii;\alpha\bar{\alpha}}\,.
\label{ii}
\end{eqnarray}

\subsection{Single-site collapse at $d=\infty$}
\label{sec:ssa}

{\it Similar to the spin Hubbard model at $d\to\infty$\/}, the diagrams
describing the perturbation expansion of ${\cal S}^*$ with respect
to the local interaction $U$ collapse. Then the lattice sums reduce
to summations of terms in which all indices coincide on the same
lattice site, and the proper self-energy becomes site-diagonal,
{\it i.e.}, ${\cal S}^*_{ij}=\delta_{ij} {\cal S}^*_{ii}$,
The collapse permits the matrix elements of ${\cal S}^*_{ii}$
(with respect to the orbital label) to be expressed in terms of
those of the on-site matrix elements of $\overline{\cal P}_{ii}$
defined above. One finds in matrix form,
\begin{equation}
{\cal S}^*_{ii}
= f\left(\begin{array}{cc}
-\overline{\cal P}_{ii;--} & +\overline{\cal P}_{ii;+-} \\
+\overline{\cal P}_{ii;-+} & -\overline{\cal P}_{ii;++}
\end{array}\right) \, .
\label{S*}
\end{equation}
We emphasize that ${\cal S}^*_{ii}$ is not yet necessarily independent
of $i$. Here the factor $f$ is
\begin{equation}
f=\frac{1-\left[1+4(1-g^2)\det{\overline{\cal P}_{ii}}\right]^{1/2}}
  {2\det\{\overline{\cal P}_{ii}\}}\,,
\label{f}
\end{equation}
and upon inversion,
\begin{equation}
\overline{\cal P}_{ii}=\frac{1}{(1-g^2)-\det{{\cal S}^*_{ii}}}
\left( \begin{array}{cc}
+ {\cal S}^*_{ii;--} & - {\cal S}^*_{ii;+-} \\
- {\cal S}^*_{ii;-+} & + {\cal S}^*_{ii;++}
\end{array} \right)\,.
\label{Pii}
\end{equation}
Equating this to the site-diagonal part of $\overline{\cal P}$ as
obtained from Eq.~(\ref{Dyson3}) generates two equations which together
determine ${\cal S^*}$, for any given ${\cal P}^0$ and $g$. Next,
Eq.~(\ref{Dyson1}) gives the self-energy ${\cal S}$, which finally
determines the propagator via Eq.~(\ref{Pij}) and the double-occupancy
using Eq.~(\ref{di}), and thus the total energy at any given electron
filling~$n$.

\subsection{Momentum space representation}
\label{sec:inv}

As we shall be mainly interested in translation invariant states, it
is expedient to Fourier transform all variables of interest upon which
matrix products with respect to site indices turn into ordinary
products in ${\bf k}$-space, {\it e.g.\/} Eq.~(\ref{Pij}) becomes
\begin{eqnarray}
{\cal P}({\bf k})&=&{\cal P}^0({\bf k}) \nonumber\\
&+& \sum_{\bf k'}
\left[ \frac{1-g}{1+g}
\left( {\cal S}({\bf k'}){\cal P}^0({\bf k'})
\right)_{\alpha\alpha} \delta_{\alpha\beta}
-\frac{{\cal S}({\bf k'})}{(1+g)^2} \right]  \nonumber\\
&+&\left[ \left({\cal P}^0({\bf k})-\frac{1}{1+g}\right)
{\cal S}({\bf k})
\left({\cal P}^0({\bf k})-\frac{1}{1+g}\right) \right]. \nonumber\\
\label{P(k)}
\end{eqnarray}
Consider now a homogeneous state where
${\cal S}^*_{ii}\equiv S^*$ is site-independent, and we may use a
shorthand notation for its matrix elements,
\begin{equation}
{S}^*=\left( \begin{array}{cc}
S^*_{+} & R^* \\
(R^*)^+ & S^*_{-}
\end{array} \right) =
\left( \begin{array}{cc}
S^*_{+} & |R^*| \mathrm{e}^{i\phi} \\
|R^*| \mathrm{e}^{-i\phi} & S^*_{-}
\end{array} \right)\,.
\label{}
\end{equation}

Using the Dyson equation for $\overline{\cal P}$ in Eq.~(\ref{Dyson3}),
$\overline{\cal P}({\bf k})$ can be expressed (in ${\bf k}$-space)
explicitly in terms of ${\cal P}^0({\bf k})$ and ${\cal S}^*({\bf k})$,
\begin{equation}
\overline{\cal P}({\bf k})= \frac{1}{N_{\bf k}}
\left[ {\cal P}^0({\bf k}) - \det{ {\cal P}^0({\bf k}) }
\left( \begin{array}{cc}
S^*_{-}  & - R^* \\
-(R^*)^+ & S^*_{+}
\end{array} \right)
\right] \,,
\label{barP(k)}
\end{equation}
where the normalizing prefactor $N_{\bf k}$ is
\begin{equation}
\label{Nk}
N_{{\bf k}} = 1 - \mathrm{Tr}_{\alpha}
\left\{S^* {\cal P}^0({\bf k})\right\}
+ (\det{S^*}) \det{{\cal P}^0({\bf k})}  \,.
\end{equation}
It depends on the occupancy of the single-particle states at
wave vector ${\bf k}$ in the many-particle trial state
$|\Phi\rangle$. Upon solving for $S^*_{+}$, $S^*_{-}$,
$R^*$, and $(R^*)^+$, one obtains the self-energy from
\begin{equation}
{\cal S}({\bf k})= \frac{1}{N_{\bf k}}
\left[ S^* - \det{S^*}
\left( \begin{array}{cc}
{\cal P}^0({\bf k})_{--}  & -\,{\cal P}^0({\bf k}_{+-}\\
-\,{\cal P}^0({\bf k})_{-+} & {\cal P}^0({\bf k})_{++}
\end{array} \right)
\right] \,,\\
\label{S(k)}
\end{equation}
upon which it is straightforward to write down explicitly the
expression for the propagator ${\cal P}({\bf k})$ and calculate all
quantities of physical interest.

\section{Gutzwiller renormalization}
\label{sec:gure}

\subsection{Ordered states}
\label{sec:oo}

As an example, we consider an ordered, {\it i.e.\/} polarized, state with
complex orbitals, Oc,
\begin{equation}
\left|\Phi_0^{\rm Oc}\right\rangle =
\prod_{{\bf k}\in{\cal K}_+}{\hat c}_{{\bf k}+}^{\dagger}
\prod_{{\bf q}\in{\cal K}_-}{\hat c}_{{\bf q}-}^{\dagger}\,|0\rangle \, .
\label{Oc}
\end{equation}
It contains fixed numbers of electrons $n_{+}^0$ and $n_{-}^0$ in
$|i+\rangle$ and $|i-\rangle$ states ($n_+^0\geq n_-^0$), respectively,
\begin{eqnarray}
n_+^0&=&\sum_{{\bf k}\in{\cal K}_+}n_{{\bf k},+}^0,\nonumber\\
n_-^0&=&\sum_{{\bf q}\in{\cal K}_-}n_{{\bf q},-}^0,
\label{NM}
\end{eqnarray}
which determine the total electron number $n_0$ and orbital
polarization $m_0$,
\begin{eqnarray}
n_0&=&n_{+}^0+n_{-}^0, \nonumber \\
m_0&=&n_{+}^0-n_{-}^0.
\label{n0m0}
\end{eqnarray}
The state (\ref{Oc}) includes both full polarization, $n_{-}^0=0$ and
$n_0=m_0=n_{+}^0$, {\it i.e.\/}, the complex ferro-orbital state FOc,
and zero polarization, $n_{0\pm}=\frac12 n_0$, $m_0=0$,
{\it i.e.\/}, the complex para-orbital state POc.

The bare propagator is a diagonal $2\times 2$ matrix in the orbital
$\{|+\rangle,|-\rangle\}$ basis, {\it i.e.\/},
${\cal P}_{\alpha\beta}^0({\bf k})=
\delta_{\alpha\beta}n_{{\bf k}\alpha}^0$,
and thus $\det\{{\cal P}^0({\bf k})\}=n_{{\bf k}+}^0n_{{\bf k}-}^0$.
Consequently, the explicit form of the matrix elements of
$\overline{\cal P}({\bf k})$ and
$\overline{\cal P}_{ii}\equiv\overline{P}$ simplify, and the equations
following from Eqs.~(\ref{barP(k)}) and (\ref{Pii}) reduce to
\begin{eqnarray}
\label{barP++}
\overline{P}_{++}\!&=&\frac{1}{N}\!\sum_{\bf k}\overline{\cal P}({\bf k})_{++} \nonumber \\
&=&\frac{1}{N}\!\sum_{\bf k}\!\left[
\frac{n_{{\bf k}+}^0(1\!-\!n_{{\bf k}-}^0)}{1-S_+^*}\!+\!
\frac{n_{{\bf k}+}^0n_{{\bf k}-}^0(1-S_-^*)}{(1\!-\!S_+^*)(1\!-\!S_-^*)-|R^*|^2}
\right] \nonumber \\
&=&\frac{S_-^*}{1-g^2-(S_+^*S_-^*-|R^*|^2)},  \\
\label{barP+-}
\overline{P}_{+-}&=&\frac{1}{N}\sum_{\bf k}\overline{\cal P}({\bf k})_{+-} \nonumber \\
&=&\frac{1}{N}\sum_{\bf k}
\frac{n_{{\bf k}+}^0n_{{\bf k}-}^0R^*}{(1-S_+^*)(1-S_-^*)-|R^*|^2}
\nonumber \\
&=&-\frac{R*}{1-g^2-(S_+^*S_-^*-|R^*|^2)},  \\
\label{barP--}
\overline{P}_{--}\!&=&\frac{1}{N}\!\sum_{\bf k}\overline{\cal P}({\bf k})_{--} \nonumber \\
&=&\frac{1}{N}\!\sum_{\bf k}\!\left[
\frac{n_{{\bf k}-}^0(1\!-\!n_{{\bf k}+}^0)}{1-S_-^*}\!+\!
\frac{n_{{\bf k}+}^0n_{{\bf k}-}^0(1-S_+^*)}{(1\!-\!S_+^*)(1\!-\!S_-^*)-|R^*|^2}
\right] \nonumber \\
&=&\frac{S_+^*}{1-g^2-(S_+^*S_-^*-|R^*|^2)},
\end{eqnarray}
From Eq.~(\ref{barP+-}) {\it it follows that $R^*=0$\/},
and Eqs.~(\ref{barP++}) and (\ref{barP--}) are simplified to
\begin{eqnarray}
\label{P++s}
\frac{n_+^{0s}}{1-S_+^*}+\frac{d^0}{1-S_+^*}&=&
\frac{S_-^*}{1-g^2-S_+^*S_-^*},    \\
\frac{n_-^{0s}}{1-S_-^*}+\frac{d^0}{1-S_-^*}&=&
\frac{S_+^*}{1-g^2-S_+^*S_-^*},
\end{eqnarray}
where $n_\alpha^{0s}$ is the number of singly occupied ${\bf k}$-points
of flavor $\alpha \in \{+,-\}$ and $d^0$ is the number of doubly occupied
${\bf k}$-points, hence
\begin{equation}
n_\alpha^0=n_\alpha^{0s}+d^0.
\end{equation}
After some algebraic manipulation one finds
\begin{equation}
n^0_+\,\frac{S_+^*}{1-S_+^*}=n^0_-\,\frac{S_-^*}{1-S_-^*},
\end{equation}
and the solution for $S_+^*$ and $S_-^*$ can be found.

Using the simplified (because of $R^*=0$) expressions for
${\cal S}({\bf k})$ and $\overline{\cal P}({\bf k})$ one finds,
\begin{equation}
\label{Pkab}
{\cal P}({\bf k})_{\alpha\beta}=\delta_{\alpha\beta}n_{{\bf k}\alpha}
=\delta_{\alpha\beta}\left[q_{\alpha}(g)n_{{\bf k}\alpha}^0+b_{\alpha}\right],
\end{equation}
with
\begin{eqnarray}
\label{qalpha}
q_{\alpha}(g)&\equiv&
1-\frac{S_{\alpha}^*}{1+g^2}\left(1-\frac{g^2}{1-S_{\alpha}^*}\right), \\
\label{balpha}
b_{\alpha}&\equiv&\left[1-q_{\alpha}(g)\right]n_{\alpha}^0.
\end{eqnarray}
This is the Gutzwiller approximation for the ordered state Eq.~(\ref{Oc}).
Note that
\begin{equation}
n_{\alpha}=\frac{1}{N}\sum_{\bf k}n_{{\bf k}\alpha}
=q_{\alpha}n_{\alpha}^0+b_{\alpha}=n_{\alpha}^0,
\end{equation}
{\it i.e.}, Luttinger theorem is satisfied. The average double occupancy is
given by
\begin{equation}
\overline{d}=
\frac{g^2}{1-g^2}\frac{S_+^*}{1-S_+^*}\,n_{0+}=
\frac{g^2}{1-g^2}\frac{S_-^*}{1-S_-^*}\,n_{0-}\,.
\end{equation}
One recognizes that the above results are identical to those obtained
in the spin case, {\it i.e.\/}, for SU(2) symmetry.

For any other trial state built from a single pair of orthogonal
orbitals, cf. Eq.~(\ref{cohorb}),
\begin{equation}
\left|\Phi_0^{\psi\theta}\right\rangle=
\prod_{{\bf k}\in{\cal K}}{\hat c}_{\bf k}^{\dagger}(\psi,\theta)
\prod_{{\bf q}\in{\cal Q}}{\hat c}_{\bf k}^{\dagger}(\psi+\pi,\theta+\pi)|0\rangle,
\label{psitheta}
\end{equation}
one obtains exactly analogous results when similarly writing all
expressions in terms of the corresponding orbital basis. So,
{\it e.g.\/} for the real Or states built from $\{|{\bf k}x_0\rangle\}$
and $\{|{\bf q}\bar{x}_0\rangle\}$ single-particle states, compare
Eq.~\eqref{Oc}, one has
\begin{equation}
n_{{\bf k}x_0}=q_{x_0}(g)n_{{\bf k}x_0}^0+b_{x_0} \, ,
\end{equation}
compare Eq.~\eqref{Pkab}.
The renormalization $q_{x_0}(g)$ and the background constant density
$b_{x_0}$ are straightforwardly obtained from $q_{\alpha}(g)$ and
$b_{\alpha}$ by the corresponding orbital transformation.

\subsection{Orbital liquid state}
\label{sec:ol}

Next we consider the OL state introduced above, see Eq.~(\ref{OL}),
\begin{equation}
|\Phi_0^{\rm OL}\rangle
=\prod_{{\bf k};\varepsilon_{\ell}({\bf k})<E_{\rm F}} \,
\prod_{{\bf k'};\varepsilon_{u}({\bf k'})<E_{\rm F}} \,
\hat{d}_{{\bf k},{\ell}}^{\dagger} \,
\hat{d}_{{\bf k'},{u}}^{\dagger} |0\rangle \,,
\label{OL2}
\end{equation}
where ${\bf k}$ and ${\bf k'}$ label the occupied states from the lower
(${\ell}$) and upper ($u$) subband, and both subbands are occupied up
to the same Fermi energy $E_{\rm F}$.

The free propagator is, in the $(+,-)$ basis,
\begin{equation}
{\cal P}^0_{\bf k}
= \frac12 \left(\begin{array}{cc}
n_{{\bf k},{\ell}}^0+n_{{\bf k},u}^0 &
\left(n_{{\bf k},{\ell}}^0-n_{{\bf k},u}^0\right)\, e^{ i\phi_{\bf k}} \\
\left(n_{{\bf k},{\ell}}^0-n_{{\bf k},u}^0\right)\, e^{-i\phi_{\bf k}} &
n_{{\bf k},{\ell}}^0+n_{{\bf k},u}^0
\end{array}\right)  ,
\label{P0OL}
\end{equation}
This leads to the following expressions for the on-site matrix of
$\overline{\cal P}$ in real space,
$\overline{\cal P}_{ii}\equiv\overline{P}$,
\begin{eqnarray}
\label{P++}
\overline{P}_{++} &=& \frac{1}{N} \sum_{\bf k}  \left[
\frac{n_{{\bf k},\ell}^0(1-n_{{\bf k},u}^0)}
{2N_{{\bf k}s}}
+\frac{n_{{\bf k},\ell}^0 n_{{\bf k},u}^0 (1-S^*_{-})}
{N_{{\bf k}d}}  \right], \nonumber \\
\\
\label{P+-}
\overline{P}_{+-} &=& \frac{1}{N} \sum_{\bf k} \left[
\frac{n_{{\bf k},\ell}^0(1-n_{{\bf k},u}^0)
\mathrm{e}^{i\phi_{\bf k}}}{2N_{{\bf k}s}}
+\frac{n_{{\bf k},\ell}^0 n_{{\bf k},u}^0 |R^*| \mathrm{e}^{i\phi} }
{N_{{\bf k}d}} \right] , \nonumber \\
\\
\label{P--}
\overline{P}_{--} &=& \frac{1}{N} \sum_{\bf k}  \left[
\frac{n_{{\bf k},\ell}^0(1-n_{{\bf k},u}^0)}
{2N_{{\bf k}s}}
+\frac{n_{{\bf k},\ell}^0 n_{{\bf k},u}^0 (1-S^*_{+})}
{N_{{\bf k}d}}  \right] \, . \nonumber \\
\end{eqnarray}
where the normalizing prefactor $N_{\bf k}$ depends on the occupancy of
the subbands at wave vector ${\bf k}$: it equals $N_{{\bf k}s}$ if only
a single ({\it viz.} the lower) subband is occupied and equals
$N_{{\bf k}d}$ if both $|{\bf k},\ell\rangle$ and
$|{\bf k},{\rm u}\rangle$ are occupied, with
\begin{eqnarray}
\label{Ns}
N_{{\bf k}s}&=& 1-\frac{1}{2}(S^*_{+}+S^*_{-})
  - |R^*|\cos(\phi-\phi_{\bf k}) \\
\label{Nd}
N_{{\bf k}d}&=& 1-S^*_{+}-S^*_{-} + \det{S^*} \nonumber \\
   &=& (1-S^*_{+})(1-S^*_{-}) -|R^*|^2   \,.
\end{eqnarray}
Setting the expressions above
equal to those obtained in Eq.~(\ref{Pii}) one obtains a set
of four equations from which to determine $S^*_{+}$, $S^*_{-}$,
$|R^*|$, and $\phi$.

Because of the occurrence of the phases $\phi-\phi_{\bf k}$ in
Eqs.~\eqref{P++} - \eqref{P+-} these equations cannot be solved
in full generality and a simplifying assumption is necessary. Thus we
make here the ansatz that the off-diagonal elements of ${\cal S}^*$
vanish, {\it i.e.\/}, $R^*=0$. Then the terms that involve summation
over the phase factors $\{\phi_{\bf k}\}$ drop out and the structure
of the equations needed to determine ${\cal S}^*$ simplifies.
Because the OL state is not polarized the two flavors are
equivalent and the diagonal elements of  ${\cal S}^*_{ii}$
must be equal, $ S^*_{+} = S^*_{-} \equiv S^*  $.
Defining single and double occupancy {\em in ${\bf k}$-space\/}
per ${\bf k}$-point,
\begin{eqnarray}
\label{Sin}
\overline{S}&=&\frac{1}{N}
\sum_{\bf k}(1-n_{{\bf k},u}) \, n_{{\bf k},{\ell}}\,, \\
\label{D}
\overline{D}&=&\frac{1}{N}
\sum_{\bf k} n_{{\bf k},u} \, n_{{\bf k},{\ell}}\,,
\end{eqnarray}
one obtains a single equation for $S^*$ in terms of $\overline{S}$ and
$\overline{D}$,
\begin{equation}
\label{eqS*}
\frac12 \frac{\overline{S}}{1-S^*}+\frac{\overline{D}}{1-S^*}=
\frac{S^*}{1-(S^*)^2-g^2}\,. \\
\end{equation}

Using the number of electrons per site $n=\overline{S}+2\overline{D}$,
one finds finally for the orbital liquid state
\begin{equation}
S^*=\frac{1}{2-n} \left( 1-\sqrt{1-n(2-n)(1-g^2)} \right)  \,.
\label{S*fin}
\end{equation}
Inserting this result into the expressions for the renormalization
factor and double occupancy in terms of $S^*$,
\begin{eqnarray}
\label{q}
q&=&1-\frac{S^*}{(1+g)^2} \,
\left(1-\frac{g^2}{1-S^*}\right) \,, \\
\label{d}
\overline{d}&=&\frac12 \frac{g^2 n}{1-g^2}\,\frac{S^*}{1-S^*}\,,
\end{eqnarray}
one finally obtains
\begin{eqnarray}
\label{q(g)}
q(g)&=&1-\frac{1}{(1+g)^2}\,
\frac{\left(1-\sqrt{1-n(2-n)(1-g^2)}\right)^2}{n(2-n)}\,, \nonumber \\
\\
\label{d(g)}
\overline{d}(g)&=&\frac{\sqrt{1-n(2-n)(1\!-\!g^2)}+n(1\!-\!g^2)-1}{2(1-g^2)} .
\end{eqnarray}
This expression for the kinetic energy renormalization (\ref{q(g)}) is
identical to that which follows from Eq.~(\ref{qalpha}) when the
ordered state is unpolarized, {\it i.e.\/},
for $q_+(g)=q_-(g)\equiv q(g)$. At $U=\infty$ one finds $g=0$ and
\begin{equation}
\label{q(0)}
q(0)=\frac{1-n}{1-n/2},
\end{equation}
which reproduces the kinetic energy renormalization in the spin model.

\section{Brinkman-Rice transition}
\label{sec:pol}

In the following we focus on the OL state, and in particular compare
its role in the orbital Hubbard model with that of the paramagnetic
phase in the spin Hubbard model. For the latter purpose it is
convenient to introduce $t'\equiv t/2$ -- then the results obtained
here at $d=\infty$ for orbital phases are directly comparable with
those for magnetic phases \cite{Faz90}.

\begin{figure}[b!]
\includegraphics[width=\columnwidth]{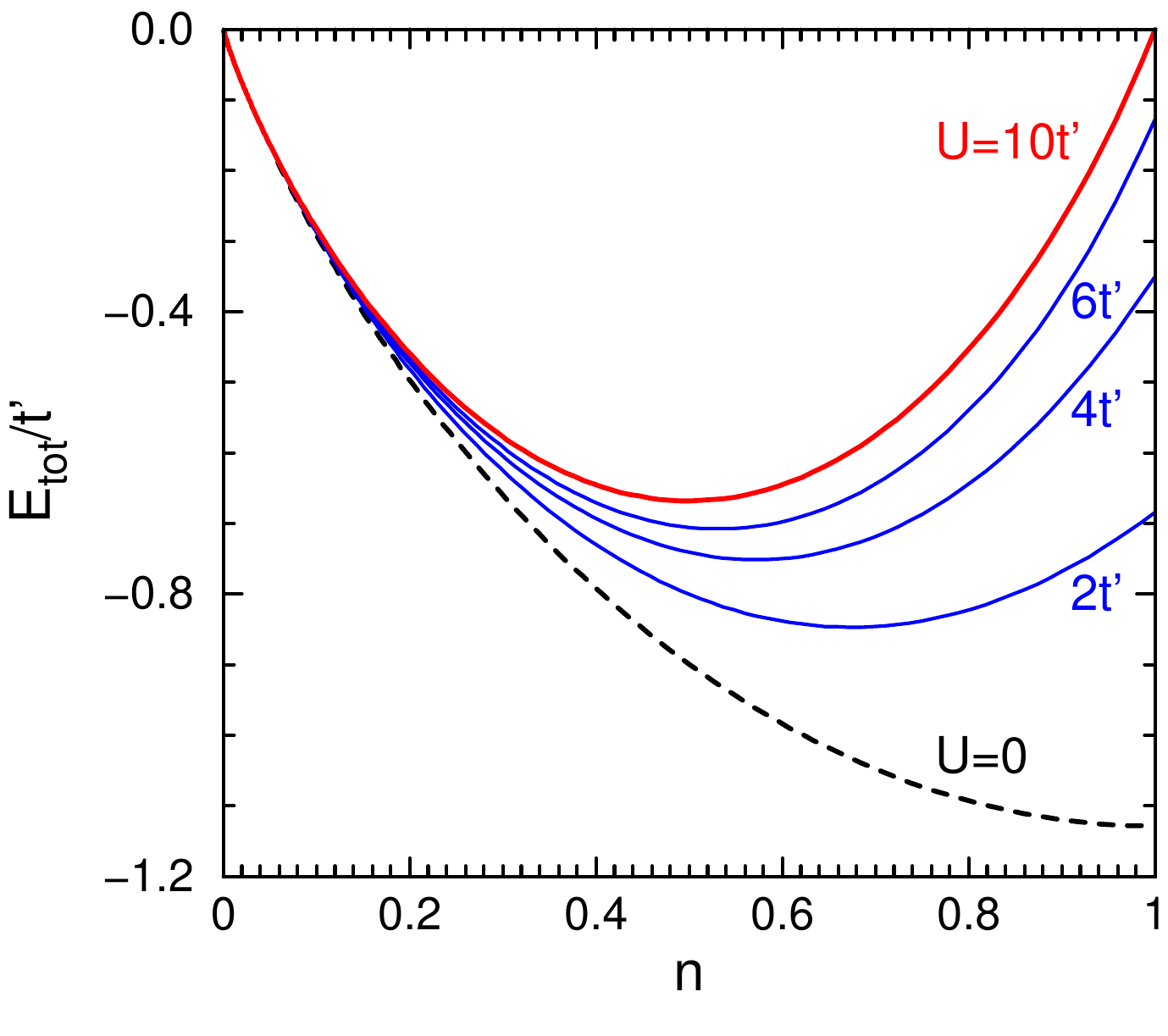}
\caption{
Total energy $E_{\rm tot}$ (\ref{Etot}) versus electron density $n$
for the OL phase in the orbital $e_g$ model with $t'\equiv t/2$.
The different curves stand for $U=0$ (black dashed line),
metallic regime with $U=2t'$, $4t'$, $6t'$ (blue solid lines) and
insulating regime $U=10t'$ (red heavy line).
}
\label{fig:etot}
\end{figure}

Altogether the energy of the OL is determined by
\begin{equation}
E_{\rm tot}^{\rm OL}(n;g)=q(n;g) E_{\rm kin}^{\rm OL}(n)
  +U \overline{d}(n;g),
\label{Etot}
\end{equation}
and one has to use Eqs.~(\ref{q(g)}) and (\ref{d(g)}) with the optimal
value of the variational parameter $g$. The result for the density
dependence of $E_{\rm tot}$ for the OL phase is shown in
Fig.~\ref{fig:etot} for a number of values of $U$. One observes that
the renormalization of the kinetic energy due to electron correlations
induced by $U$ is largest at half-filling, {\it i.e.\/}, at $n=1$.
Indeed, here one might expect an insulating state to appear eventually
at sufficiently large $U$, with completely suppressed electron dynamics,
see below.
Further, with increasing $U$ the total energy (\ref{Etot}) gradually
increases at any density $n$ and its minimum moves towards $n=0.5$.
The pattern seen in Fig.~\ref{fig:etot} is entirely similar to that
shown by the paramagnetic phase in the spin Hubbard model when treated
likewise by the Gutzwiller method, because it is dominated by the
behavior of $q(n;g)$ and $\overline{d}(n;g)$ and hardly affected by the
slight difference in the shape of $E_{\rm kin}(n)$.

\begin{figure}[t!]
\includegraphics[width=\columnwidth]{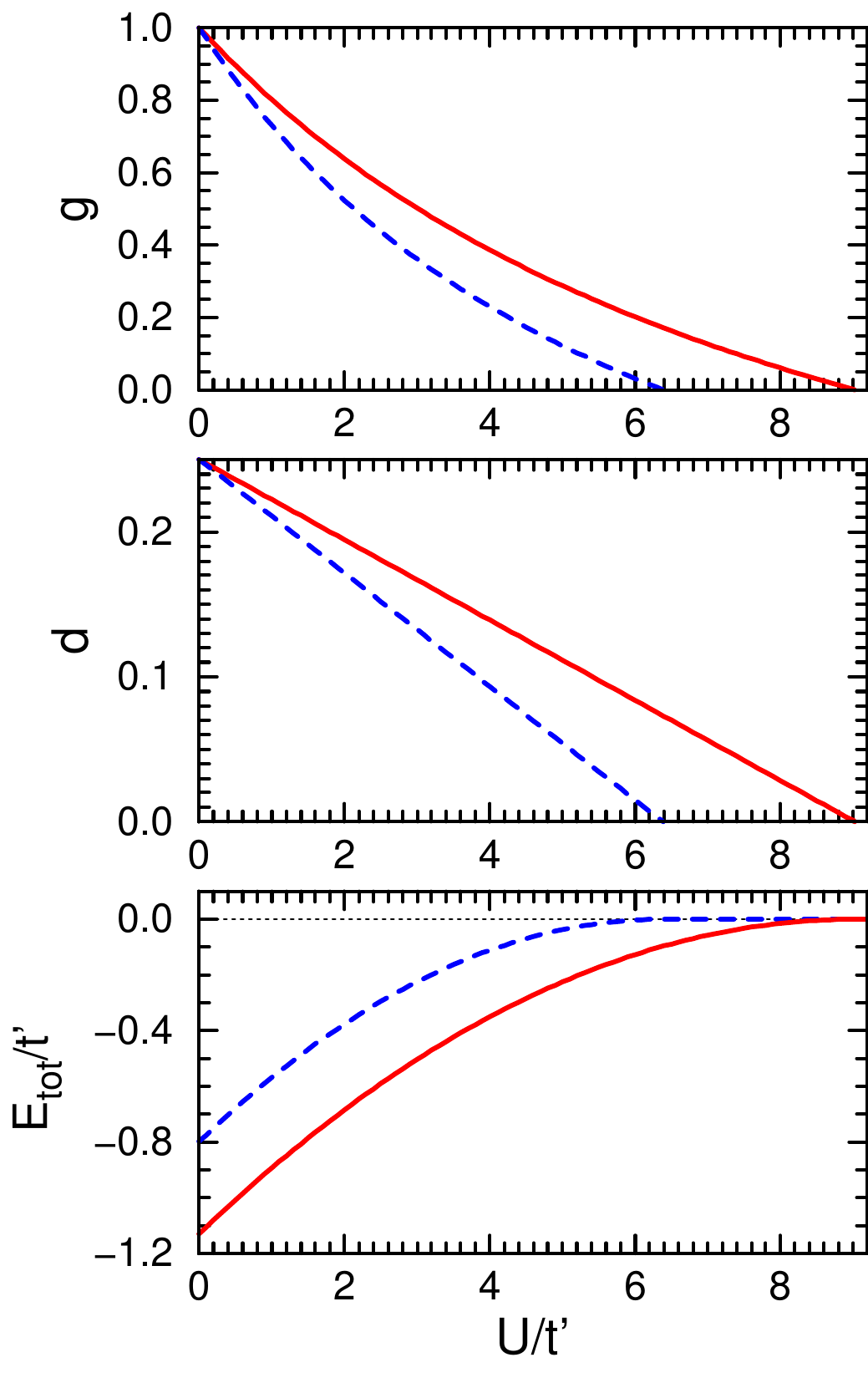}
\caption{
Dependence on $U/t'$, at electron density $n=1$, of
(a) the variational parameter $g$,
(b) the double occupancy $\overline{d}$, and
(c) the total energy $E_{\rm tot}/t'$,
for the spin Hubbard model (blue dashed lines) and for the orbital
$e_g$ Hubbard model (red full lines), at $d=\infty$.
For easier comparison with the spin case, we use here the hopping parameter
$t'\equiv t/2$ for both spin and orbital Hubbard model.
}
\label{fig:mit}
\end{figure}

Figure \ref{fig:mit} shows in some more detail what happens at $n=1$,
demonstrating that for the OL in the orbital model the dependence of
the variational parameter $g$, the double occupancy $\overline{d}$,
and the total energy $E_{\rm tot}$ on $U$ is qualitatively the same as
for the paramagnetic state in the spin model. In particular double
occupancy gets fully suppressed ($\overline{d}=0$) at and above a
critical value $U_{\rm BR}$, while simultaneously $g=0$ implying $q=0$,
{\it i.e.\/}, the electrons are entirely localized and itinerancy
disappears. This has been recognized in the spin case by Brinkman and
Rice \cite{Bri70} as a qualitative criterion for the onset of the
transition from a metallic to an insulating state \cite{exci},
which in fact becomes exact at $d=\infty$ \cite{vDo89}.

\begin{figure}[t!]
\includegraphics[width=\columnwidth]{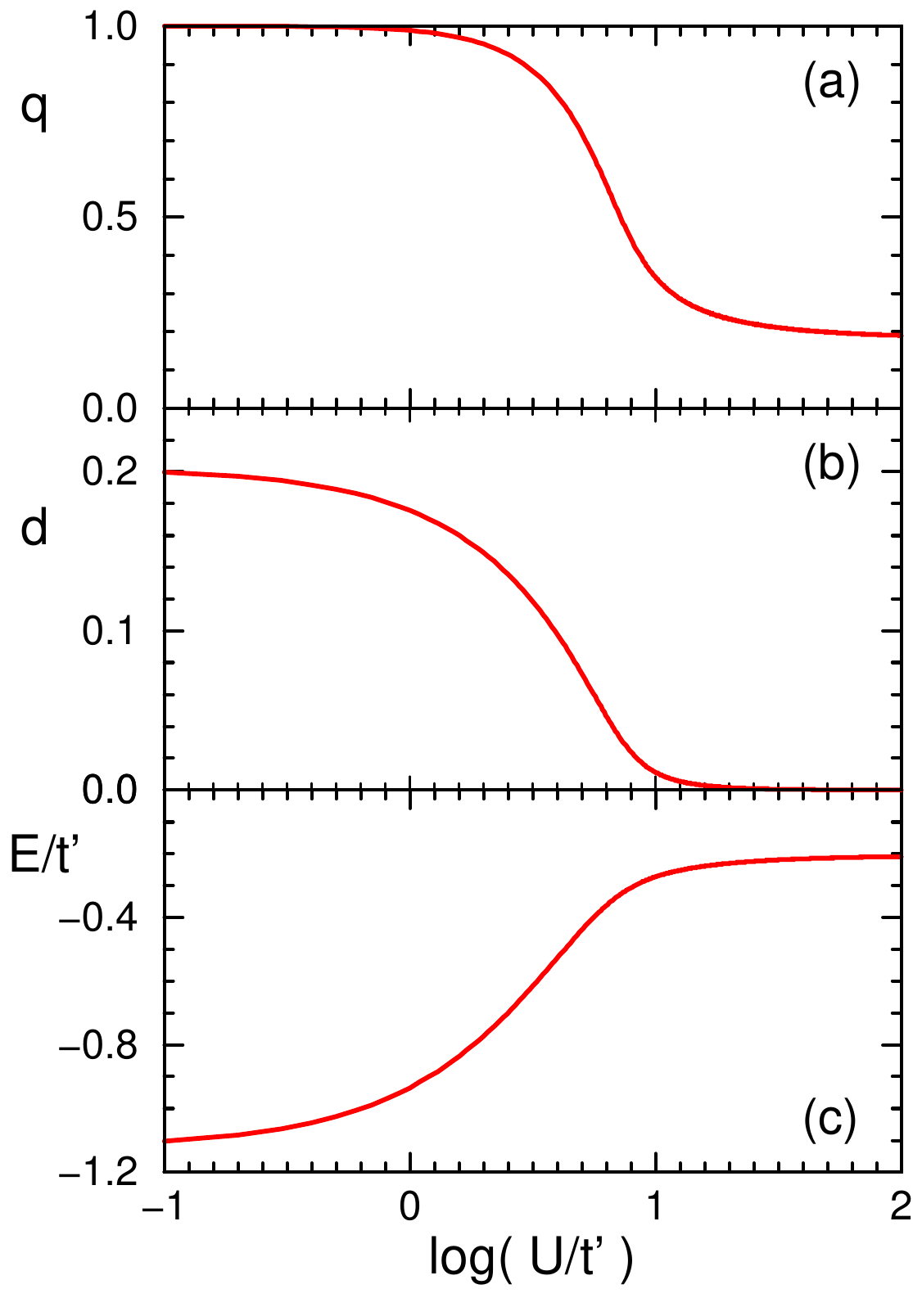}
\caption{
Evolution of the OL phase from free electrons to a strongly correlated
metal at $n=0.9$ for increasing $U/t'$ as obtained for the orbital $e_g$
Hubbard model (solid lines) at $d=\infty$:
(a) the Gutzwiller renormalization factor $q$,
(b) the double occupancy $\overline{d}$, and
(c) the total energy $E_{\rm tot}$.
For easier comparison with the spin case, we use here the hopping parameter
$t'\equiv t/2$ for the orbital Hubbard model \eqref{Hd}.
}
\label{fig:0.9}
\end{figure}

However, quantitatively there is a considerable difference:
for the OL in the orbital $e_g$ Hubbard model this transition occurs at
$U_{\rm BR}= 9.161 \, t'$, enhanced from $U_{\rm BR}=6.383 \, t'$ for
the paramagnetic state in the spin Hubbard model (equivalent to the
($\gamma=0$) POc state in the orbital model), because $U_{\rm BR}$ is
generally given by
\begin{equation}
U_{\rm BR}= 8 \, |E_{\rm kin}(E_{\rm F}=0)| \, ,
\label{BR}
\end{equation}
which is larger by the factor $1+\sqrt{\pi} F(1;0) = 1.435$ for the OL,
compare Eqs.~\eqref{EkinOL} and \eqref{EkinEFw}. Clearly this is
entirely due to the extra contribution to the kinetic energy coming
from the orbital-flavor non-conserving hopping at $\gamma=1$.
We note in passing that the real para-orbital phase POr shows a
Brinkman-Rice transition at the intermediate value
$U_{\rm BR}=7.818\, t'$, enhanced only by a factor
$w_{\rm r}/w_{\rm c}=\sigma(\gamma=1)=\sqrt{3/2}= 1.225$, because it
profits only partially from the additional kinetic energy available
in the orbital model.

At $n<1$ the renormalization factor $q(g)$, Eq.~\eqref{q(g)}, remains
finite at large $U$, and thus the electrons remain itinerant, be it
with reduced kinetic energy, see Fig.~\ref{fig:0.9}.
Increasing $U/t'$ defines here three distinct regimes: \hfill\break
(i) at very small $U$ ($U\lesssim t'$) correlations are virtually
absent, $q\approx 1$ and $\overline{d}\approx (n/2)^2$, \hfill\break
(ii) at small $U$ ($U<U_{\rm BR}$) the correlations gradually develop,
$q(g)$ is reduced from 1.0 down to $\approx 0.2$, double occupancy is
strongly reduced ($\overline{d}\to 0$), and the total energy increases,
and \hfill\break
(iii) at large $U$ ($U>U_{\rm BR}$) the correlations dominate, double
occupancy becomes excluded and the electrons move in the strongly
correlated OL without creating double occupancies.

\section{Phase diagram}
\label{sec:phd}

\begin{figure}[b!]
\begin{center}
\includegraphics[width=1.17\columnwidth]{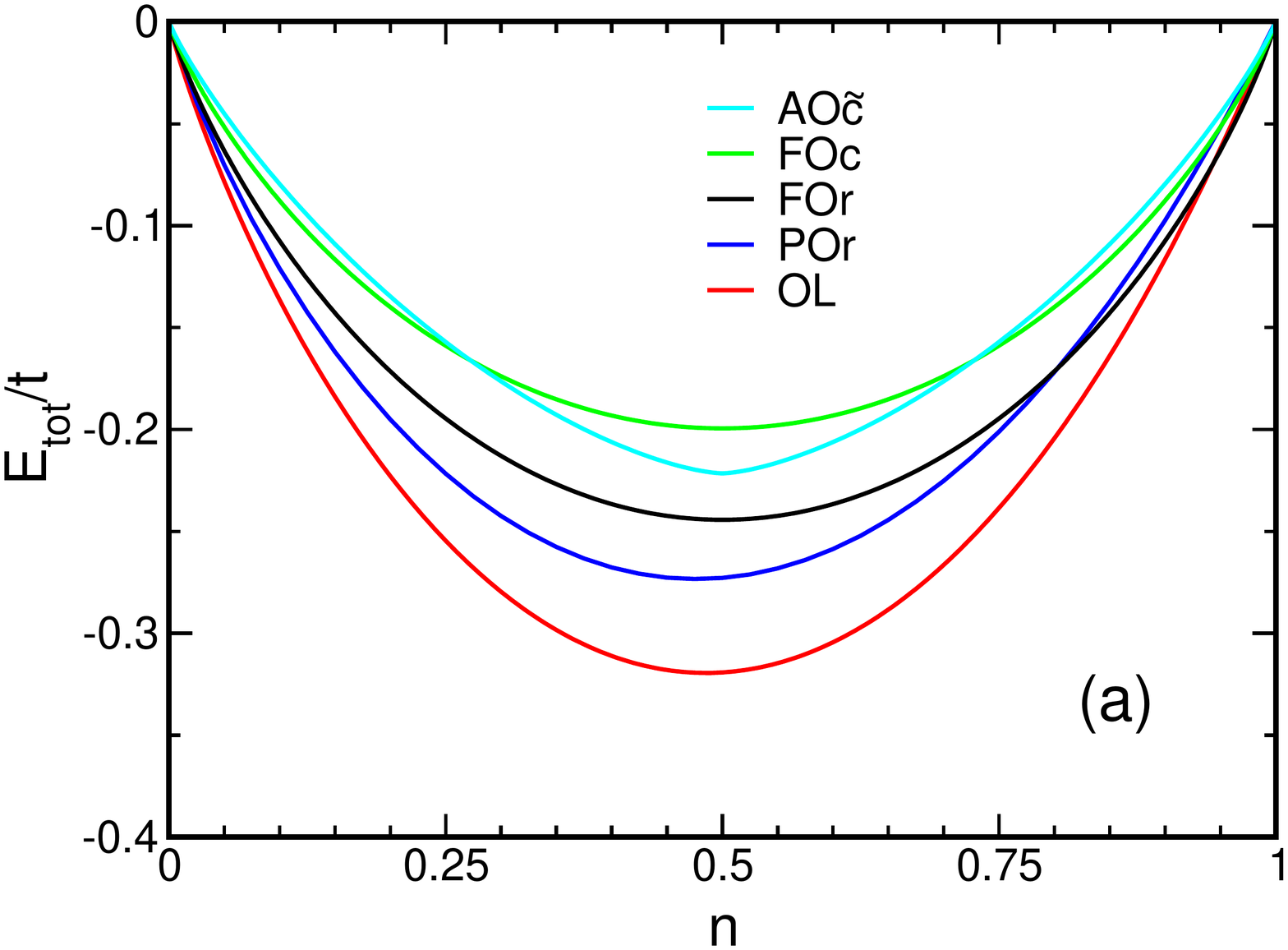}
\vskip -.2cm
\includegraphics[width=1.17\columnwidth]{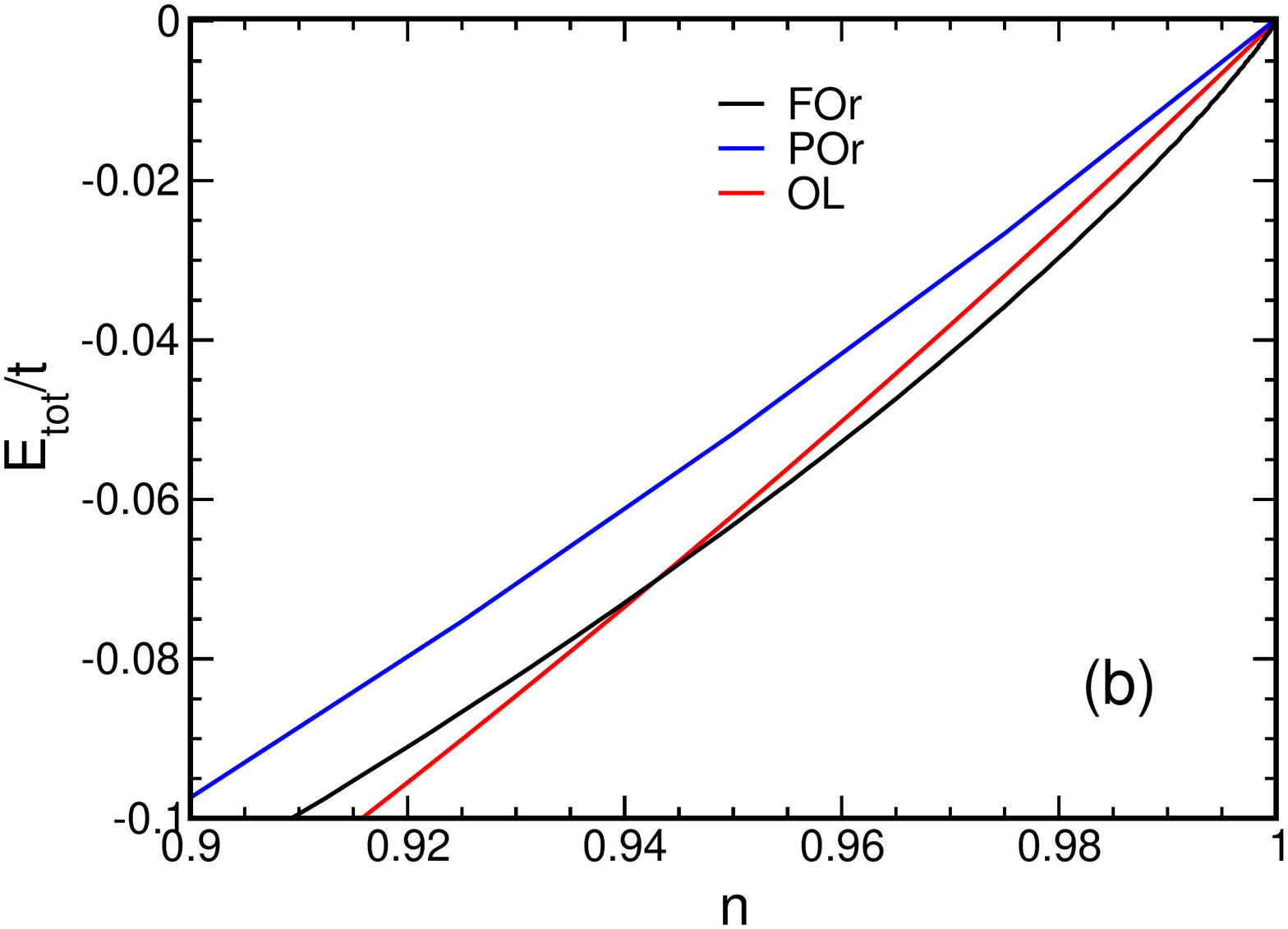}
\end{center}
\vskip -.4cm
\caption{
Total energy $E_{\rm tot}$ as a function of density $n$ at
\mbox{$U=10t$} for the OL state (red line),
the real-orbital POr state (blue line),
the real-orbital FOr state (black line),
the complex-orbital FOc state (green line),
and the phased complex-orbital AO$\tilde{{\rm c}}$ state (aubergine line):
(a) for OL, POr, FOr, FOc, and AO$\tilde{{\rm c}}$ in the range
$0 \leq n \leq 1$;
(b) zoom-in on the range close to $n=1$ for OL, POr, and FOr only.
}
\label{fig:all}
\end{figure}

From the unrenormalized kinetic energies shown in Fig.~\ref{fig:et} we
concluded that the uniformly polarized FOr phase and possibly the
disordered POr phase are the only candidates competing with the OL
phase for being the ground state after the energies of the POr and OL
phase have been renormalized by correlations, for the following reasons:
First, as double occupancies do not occur in the FOr phase or the other
uniform phases with broken symmetry, these phases are not affected by
renormalization and so the sequence of their energies is fixed.
Second, although the energies of the disordered phases are changed by
accounting for correlations, their sequential order also cannot change
if all of them are subjected to the same renormalization procedure
(at least if the procedure is only sensitive to energies as is the case
for the Gutzwiller method). This holds because for any phase its total
renormalized energy is given by an expression like Eq.~\eqref{Etot}
with the variational parameter(s) optimized for the lowest outcome; if
the kinetic energy in this equation is replaced by a lower value
associated with a different phase, the total energy is lowered even if
the variational parameter $g$ is kept fixed; subsequently allowing $g$
to optimize can only lead to a further lowering of the total energy.
This argument should not only hold for the para-orbital phases but also
be valid between the POr and OL phases, implying that the only
competition of interest should be between the FOr phase and the OL
phase.

\begin{figure}[t!]
\vskip -.1cm
\includegraphics[width=1.14\columnwidth]{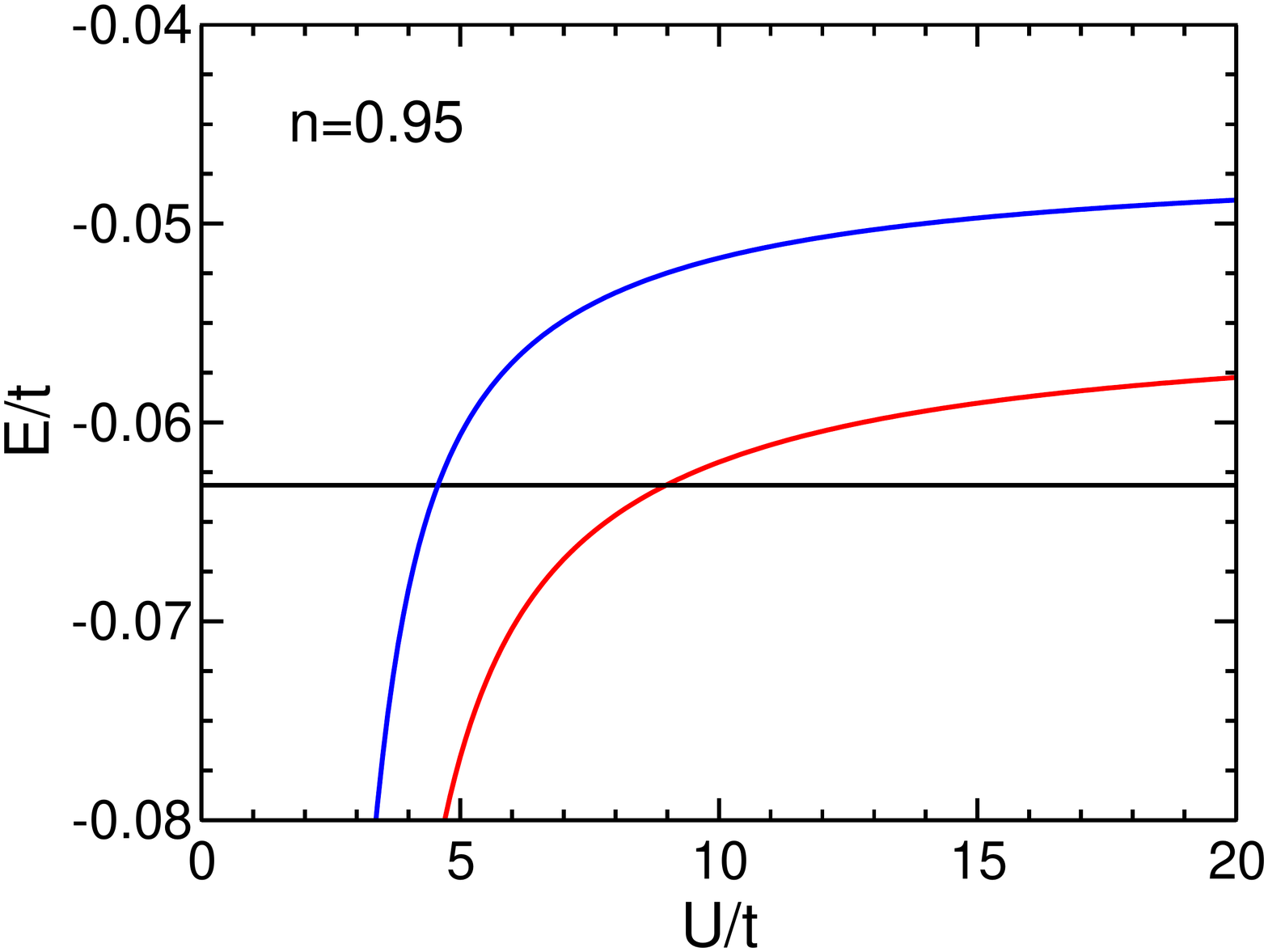}
\caption{
Total energy $E_{\rm tot}$ as a function of the Hubbard
interaction parameter
$U/t$ for the OL phase (red line), real-orbital FOr state (black
horizontal line, independent of $U$),
and real-orbital POr state (blue line) for fixed $n=0.95$.}
\label{fig:0.95}
\end{figure}

Numerical results are shown in Fig.~\ref{fig:all}(a) for the dependence
upon density of the total energies of various phases at a particular
value of $U$, {\it viz.\/} $U=10 t$. The states other than the FOr and
the OL states are included here just to demonstrate that they do not
play a role, as argued above. Yet indeed, a competition between the FOr
phase and the OL does take place and the total energies of these phases
are seen to be quite close to one another. In fact they cross and,
whereas the energy of the OL is well below that of the FOr state over
most of the density range even for this rather large value of $U$, the
OL phase is unstable against the FOr phase close to $n=1$,
as shown in more detail in Fig.~\ref{fig:all}(b). We point out that the
latter result is different from that obtained for the 3D orbital
Hubbard model (at $d=3$) where the OL is stable in the entire regime of
$n$ also at large $U$ \cite{Fei05}.

The transition between the OL phase and the FOr phase can be determined
more accurately by calculating at fixed $n$ the total energy of the OL
as a function of $U$ and comparing this with the $U$-independent total
energy of the FOr state, as shown in Fig.~\ref{fig:0.95}. One thus
obtains $U_c(n)$ where the two phases have equal energy. This defines
the OL-FOr phase boundary because the transition is first-order, since
the states $|\Psi^{\rm OL}\rangle$ and $|\Psi^{\rm FOr}\rangle$ cannot
transform continuously into one another because of the globally
different structure of $|\Phi_0^{\rm OL}\rangle$ and
$|\Phi_0^{\rm FOr}\rangle$. This stands in clear contrast to the
paramagnetic-ferromagnetic transition in the spin Hubbard model,
which is second order because the magnetization can evolve gradually
from the paramagnetic state.

\begin{figure}[t!]
\vskip -.5cm
\includegraphics[width=1.18\columnwidth]{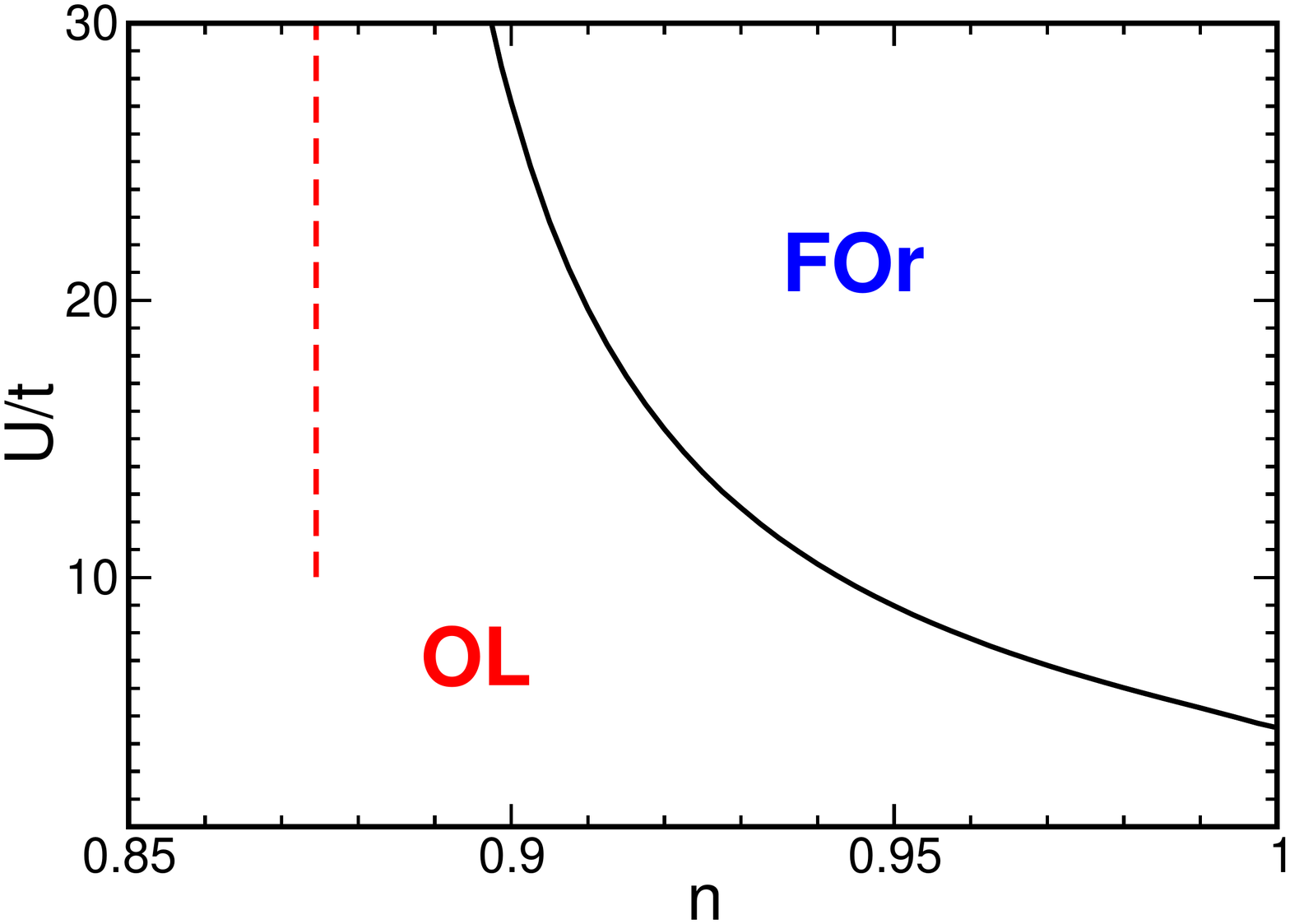}
\caption{Phase diagram of the OL state versus the FOr state at
$d=\infty$. The FOr phase is more stable than the OL phase for
$U>U_c(n)$ if $n>n_c$. The vertical red dashed line indicates the
phase boundary for $U=\infty$ at $n_c$; \mbox{$U_c(1)=4.5805\, t$}
is equal to the critical value of the interaction for the
Brinkman-Rice transition \cite{Bri70} in the OL state.
}
\label{fig:PhD}
\end{figure}

The resulting phase diagram of the orbital $e_g$ Hubbard model in the
($n$, $U$) plane as obtained in the Gutzwiller approximation is shown
in Fig.~\ref{fig:PhD}. The FOr phase is more stable than the OL phase
for $U>U_c(n)$ if $n>n_c$. Here $n_c=0.8746$ is the critical value of
the particle density at which the energies of the phases are equal at
$U=\infty$, and below which the OL is therefore always stable, see
Fig.~\ref{fig:PhD}. It is obtained from a numerical comparison using
the analytical expressions for the kinetic energies of the two phases,
Eq.~\eqref{EkinEFn} and Eq.~\eqref{EkinOL}, and for the renormalization
factor at $U=\infty$, Eq.~\eqref{q(0)}.

Next we consider the intersection of the phase boundary with the
$U/t$-axis at $n=1$. With our simple representation of real-orbital
ferro order by the state
$\big|\Psi^{\rm FOr}\big\rangle = \big|\Phi_0^{\rm FOr}\big\rangle$,
which has fully saturated orbital order such that
$E_{\rm tot}^{\rm FOr}(1)=E_{\rm kin}^{\rm FOr}(1)=0$ independent of
$U$, this intersection necessarily coincides with the Brinkman-Rice
transition point where the energy of the OL attains $0$, so
$U_c(1)=U_{\rm BR}= 4.5805\, t$.
However, as pointed out by Fazekas {\it et al.\/} for the spin case
\cite{Faz90}, precisely close to $n=1$ a small reduction of the
polarization must occur involving the exponential tails of the DOSs
of the $x_0$- and ${\bar x}_0$-bands. If this would be taken into
account it would produce some lowering of the energy of the FOr phase
thus leading to a slight decrease of $U_c(1)$.

Allowing some polarization of the FOr state at general density, at
the price of creating double occupancy, would similarly give some
energy lowering. However, this is severely limited by the restriction
that upon transferring one particle from the majority to the minority
band the gain in kinetic energy must exceed $U$. This condition is
only met as long as the Fermi energy of the minority band is in the
exponential tail of the DOS, {\it i.e.\/} only for a very small
fraction of the particles, and also $U \lesssim 4 t $, compare
Fig.~\ref{fig:dos}(b). Thus our use of the fully ordered state
$\big|\Psi^{\rm FOr}\big\rangle$ neglects at most a minor
expansion of the FOr regime in the phase diagram.

We have not attempted to construct a more sophisticated trial state for
any of the AO phases, such as has been shown to be important to obtain
a good description of the AF phase in the spin Hubbard model
\cite{Met89,Faz90}. So, strictly we cannot exclude that alternating
orbital order would show up in the orbital model at and/or close to
half-filling by outcompeting the ferro orbital order, as occurs between
antiferromagnetism and ferromagnetism in the spin case. However, this
seems unlikely in the orbital case because of the large difference in
kinetic energy between the fully ordered states, see Fig.~\ref{fig:et},
{\it viz.\/} by a factor
$w_{\rm r}/w_{\rm r,ao}=\sigma \sqrt{2}=\sqrt{3}=1.73$ between the FOr
and AOr states, and (because the kinetic energy of the AO${\tilde c}$
state is very close to that of the FOc state) by a factor
$\simeq w_{\rm r}/w_{\rm c}=\sigma=\sqrt{3/2}=1.22$ between the FOr and
AO${\tilde c}$ states. As shown above, this difference is due to the
more efficient use of both pseudospin-conserving and non-conserving
hopping channels by real-orbital ferro order than by alternating
orbital order. What we do have established is that the OL state is the
ground state over most of the ($n,U$) phase diagram,
as illustrated by Fig.~\ref{fig:PhD}.

Finally, we point out that the phase diagram obtained here at $d=\infty$
is qualitatively different from that found before at $d=3$ \cite{Fei05}.
In the latter case the lowest-energy ordered FOx phase was completely
eliminated from the phase diagram by the non-conserving hopping, and the
OL phase was the only phase present. We propose that this is caused by
the different energy ranges: $\left[-3t,3t\right]$ at $d=3$ versus
$\left[- \infty,\infty \right]$ at $d=\infty$.
The argumentation is as follows:

Let us consider, at $U=\infty$, the derivatives with respect to $n$
of the total energy of the candidate phases ({\it viz.\/} FOr and OL at
$d=\infty$, FOx and OL at $d=3$) at $n=1$, as these determine which
phase has the lowest energy in the immediate vicinity of half-filling,
compare Fig.~\ref{fig:all}. Since $\overline{d}=0$ at $U=\infty$, we
have
$E_{\rm tot}^{\rm FO}(n) = E_{\rm kin}^{\rm FO}(n)$ and
$E_{\rm tot}^{\rm OL}(n) = q(n) E_{\rm kin}^{\rm OL}(n)$,
compare Eq.~\eqref{Etot}, and therefore
\begin{eqnarray}
\label{derFO}
 \frac{dE_{\rm tot}^{\rm FO}}{dn}(n) &=&
   \frac{dE_{\rm kin}^{\rm FO}}{dn}(n) = E_{\rm F}^{\rm FO}(n) \, , \\
\label{derOL}
 \frac{dE_{\rm tot}^{\rm OL}}{dn}(n) &=&
   \frac{dq}{dn}(n) E_{\rm kin}^{\rm OL}(n)
    + q(n) \frac{dE_{\rm kin}^{\rm OL}}{dn}(n)  \nonumber \\
    &=&
   \frac{dq}{dn}(n) E_{\rm kin}^{\rm OL}(n)
    + q(n) E_{\rm F}^{\rm OL}(n) \, .
\end{eqnarray}
Both expressions can be evaluated,
\hfill\break
(i) at $d=3$: by consulting \cite{Fei05},
for the FOx phase using Eq.~(3.23) therein (from which one obtains
$E_{\rm kin}^{\rm FOx}(1-x) \eqsim 3 t (1 - x + \cdots)$ and thus
$E_{\rm F}^{\rm FOx}(1) = 3 t $),
for the OL phase making use of the data plotted in Fig.~5 therein
[from which one obtains
$E_{\rm kin}^{\rm OL}(1) \eqsim -  1.52 t $ and
$E_{\rm F}^{\rm OL}(1) = 0 $];
\hfill\break
(ii) at $d=\infty$: for the FOr phase using
Eqs.~\eqref{EFwn} in the present paper with $w_{\rm r}=\sqrt{3/2}/2$,
for the OL phase using Eq.~\eqref{EkinOL} and so obtain
$E_{\rm kin}^{\rm OL}(1) \eqsim -  0.5726 t $  ;
\hfill\break
(iii) inserting $q(1)=0$ and $dq/dn(1)=-2$.

The results are as follows:
\begin{eqnarray}
d&=&3:	 	 \frac{dE_{\rm tot}^{\rm FOx}}{dn}(1) = 3.00 \, t,  \\
d&=&3:	 	 \frac{dE_{\rm tot}^{\rm OL}}{dn}(1) =  3.04 \, t,  \\
d&=&\infty\!: \frac{dE_{\rm tot}^{\rm FOr}}{dn}(1-x)\! =\!0.8660 \: \mathrm{inverf}(1-2x) \, t , \\
d&=&\infty\!: \frac{dE_{\rm tot}^{\rm OL}}{dn}(1) = 1.1451 \, t  \, .
\end{eqnarray}
So, for $d=3$ the derivative is slightly larger for the OL, implying
that close to $n=1$ this is the lowest state and then will remain so in
the entire phase diagram, as we found before \cite{Fei05}. However, for
$d=\infty$ the slope for the FOr phase initially diverges (the function
$\mathrm{inverf}(y)\to\infty$ at $y\to 1$), so the FOr state is the
ground state close to $n=1$, as we found in the present paper. This
divergence is not seen in the plots because it is very steep. Thus
already for $x=1-n$ larger than $0.03$ the derivative is of order 1
and with further increasing $x$ the FOr phase begins to be overtaken by
the OL. So the reason for the FOr phase being stable at $d=\infty$
apparently lies in the exponential tails in its DOS, particle density
and kinetic energy. As this is a mathematical rather than a physical
phenomenon, the above result suggests that stability of ordered phases
at $d=\infty$ is mathematically correct but physically spurious, and
that the earlier theoretical result at \mbox{$d=3$} \cite{Fei05}
describes the generic physics.

We thus conclude that regarding the phase diagram the $d=\infty$ case is
not representative for all dimensions: the $d=3$ case is qualitatively
different. This actually demonstrates that the effect of the
non-conserving hopping terms themselves is qualitatively different for
different dimensions, which is not the case for the conserving terms,
since FM and AF phases appear in the phase diagram of the spin Hubbard
model in all dimensions.

\section{Discussion and conclusions}

\subsection{General aspects of the $d=\infty$ orbital model}
\label{sec:summa}

A few aspects of the $e_g$ orbital Hubbard model at $d=\infty$ deserve
some further comments. They mainly concern the limitations of the
Gutzwiller approach in comparison with more powerful recent methods,
in particular selfconsistent DMFT. As mentioned briefly in Sec.
\ref{sec:intro} the development of the DMFT approach \cite{Geo96} was
made possible by the discovery by Metzner and Vollhardt that in the
limit $d=\infty$ only on-site correlations survive \cite{Met89}.
Nowadays DMFT is recognized as a standard method to study electron
correlation effects in the electronic structure
\cite{Vol11,Vol14,Vol18,Kot18}. This method has been successfully
applied, \textit{inter alia}, to the Falicov-Kimball model \cite{Fre03}
and to nonequilibrium dynamics \cite{Wer14}. The ideas employed using
the Gutzwiller wave function helped to formulate the DMFT method in
spin systems \cite{Vol14}.

First of all, one may wonder to what extent the description of the
metal-insulator transition, as presented in Sec.~\ref{sec:pol},
{\it i.e.\/}, as a Brinkman-Rice transition, is realistic.
An additional aspect for the orbital model are
possible crystal-field splittings but it has been established that
orbital-selective Mott phases occur then for large enough interaction
$U$ \cite{deM05,Jak13}. However, when the orbitals remain equivalent as
in the OL state considered in the present paper, one may expect that
the transition occurs in a similar way as in the paramagnetic state in
the spin model. So our treatment of the metal-insulator transition in
the orbital model at $d=\infty$ by the Gutzwiller approach, leading to
the Brinkman-Rice transition, then has the same status as the treatment
of the metal-insulator transition in the spin model at $d=\infty$ by
that same approach, and gives valuable insight how correlations induce
the system to approach the localized state \cite{Vol14}.

Second, past calculations using the Gutzwiller wave function have led to
a better understanding of the strongly correlated regime of the Hubbard
model. In one dimension the metal-insulator transition is absent but
several quantities have been calculated exactly \cite{Kol02}: the double
occupation, the momentum distribution, as well as its discontinuity at
the Fermi surface. These quantities determine the expectation value of
the 1D Hubbard Hamiltonian for any symmetric and monotonically
increasing dispersion. The Gutzwiller wave function has also been found
to predict ferromagnetic behavior for sufficiently large interaction
$U$ \cite{Fre97,Obe97}. This agrees with the present result that the
orbital Hubbard model also has a range of ferro-orbital states close
to $n=1$.

However, it has been known for some time that the Brinkman-Rice picture
of the metal-insulator transition, as given by the Gutzwiller approach
both at $d=3$ and at $d=\infty$, is oversimplified, in particular
yielding a poor description of the dynamics on the insulating side
\cite{Faz99}. Recent studies of the metal-insulator transition in the
spin Hubbard model at $d=\infty$ by DMFT have demonstrated the
existence of two different transitions, a metal-to-insulator transition
at $U_{c2}$ where the quasiparticle weight becomes zero and an
insulator-to-metal transition at $U_{c1}$ where the gap closes, both
first-order, and the occurrence of hysteresis between the two critical
interaction strengths \cite{Bul99,Roz99,Bul01}. This holds from $T=0$
up to a critical temperature $T_c$ where $U_{c1}$ and $U_{c2}$ coalesce,
while for $T>T_c$ there is a smooth crossover between metallic and
insulating behavior. Notably, the thermodynamics of the hysteresis
region was recently explained at the two-particle level \cite{Kat20}.
Obviously, the above more-detailed understanding of the Mott-Hubbard
transition implies that this transition is considerably more complex
than the Brinkman-Rice picture suggests.

These differences between the Mott-Hubbard transition in the spin
Hubbard model as calculated by DMFT and the Brinkman-Rice picture are
clearly due to the fact that DMFT is coping more effectively with
correlations than the Gutzwiller approach. One would therefore expect
that a DMFT treatment of the $e_g$ orbital Hubbard model would produce
similar changes to the metal-insulator transition as in the spin
Hubbard model, because this transition is typically a
correlation-induced phenomenon. By contrast, one would expect the more
prominent role of the disordered OL in the phase diagram (as compared
to the paramagnetic phase in the spin case) to remain largely
unaffected, because it is not due to correlations but to the additional
kinetic energy provided by the non-conserving hopping channel.

However, these are just speculations. We emphasize that orbital models
such as that in Eq.~\eqref{Hd} [or Eq.~\eqref{Hoo}] have as yet not
been studied in the context of self-consistent DMFT calculations.
Therefore, even the corresponding effective impurity model for such
calculations is not known, and its construction appears not to be
entirely trivial. A prime candidate is a two-orbital Anderson impurity
model with on-site $U$ and selfconsistently generated orbital flipping
terms. In view of the overriding importance of the orbital
non-conserving hopping channel and the associated $C_d$ symmetry
demonstrated above, these features should apparently be incorporated
somehow. The only way seems to be by imposing a specific structure on
the bath modes and their coupling to the impurity. It would be
desirable if a DMFT treatment along these lines were performed, both to
see if the above speculations on the $e_g$ orbital Hubbard model are
born out and to further explore the application of the DMFT method to
a wider range of models.

The early ideas of Kugel and Khomskii \cite{Kug82} culminate in systems
where spins and orbitals become almost equivalent. This idealization
happens for face-sharing octahedra and one finds indeed a highly
symmetric SU(4) model \cite{Kug15}. So far, studies of such systems
have only been performed theoretically and these have identified an
interesting evolution of spin-orbital entanglement with increasing
spin-orbit coupling \cite{Got20}. In spite of the absence of long-range
spin-orbital order such systems exhibit features typical of those
manifesting themselves at phase transitions \cite{Val19}. The
experimental search for quantum spin-orbital liquids identified
FeSc$_2$S$_4$ as a quantum material with long-range entanglement
\cite{Lau15}. Such states are primarily realized in the Kitaev materials
where spins and orbitals lose their identity and are strongly entangled
by large spin-orbit coupling \cite{Tak19}. The challenge in the theory
is to find a quantum spin-orbital liquid at infinite dimension.

\subsection{Summary}
\label{sec:con}

Summarizing, we have presented a generalization of the $e_g$ orbital
model to infinite dimension $d=\infty$, preserving the two-fold
degeneracy of the orbitals and turning the lattice symmetry from cubic
into hypercubic. It is quite remarkable that the two-flavor orbital
model is manifestly different from the corresponding spin model at
$d=\infty$ by the presence of orbital-flipping hopping terms.
At the same time it is somewhat surprising that the two orbital
flavors are equivalent in the limit of $d=\infty$ when they are so
different in the 3D model, including the physics of the manganites
\cite{Tok06}. We have further shown that the Gutzwiller approximation
becomes exact in the limit $d\to\infty$ for the orbital Hubbard
model in perfect analogy with the spin Hubbard model.

We conclude that the peculiar features of the orbital Hubbard model
\eqref{Hd} are due to the fact that the extra hopping terms
$\propto\gamma t$ do not conserve the pseudospin as they mix the two
orbital flavors. They thus induce the following distinctive features:
\hfill\break
 (i) there are two classes of single-particle plane-wave states: those
 with complex orbitals and those with real orbitals, which behave
 differently; the same holds for the Fermi-sea-like multiparticle
 states built from them;
\hfill\break
 (ii) the single-particle {\it eigenstates\/} of the kinetic energy
 form two distinct bands with different dispersions;
\hfill\break
 (iii) each such eigenstate carries a  nonzero polarization,
 {\it i.e.\/}, contributes to the pseudospin at all sites; yet when
 summed over all eigenstates in an energy shell, these contributions
 add up to zero; by this mechanism the {\it orbital liquid state\/}
 (OL),  which is obtained by filling the two bands of single-particle
 eigenstates to the same Fermi energy, is unpolarized at any filling.
\hfill\break
The above qualitative features are generic for the $e_g$ orbital
Hubbard model, {\it i.e.\/}, they hold at any dimension, from $d=3$
up to $d=\infty$.

There are also a number of features induced by the non-conserving
hopping channel that are to some extent quantitative and are most
pronounced at $d=\infty$.
\hfill\break
 (iv) The extra hopping channel lowers the kinetic energy with respect
 to the spin case and thus makes the Brinkman-Rice transition to an
 insulator occur at a larger value of $U$;
\hfill\break
 (v.a) phases with alternating orbitals are much less favored than
 antiferromagnetism for fermions on a bipartite lattice like the
 hypercubic lattices considered here, because they benefit relatively
 less from the pseudospin-conserving channel, whereas
\hfill\break
 (v.b) unlike ferromagnetic states in the spin Hubbard model,
 ferro-orbital states are not eigenstates of the orbital Hubbard model
 \eqref{Hd} and benefit energetically more from the orbital-mixing term;
\hfill\break
 (v.c) together this makes the FOr state the main competitor of the
 orbital liquid phase, and the $d=\infty$ phase diagram confirms that
 FOr order indeed  occurs at large $U$ close to half-filling
 \cite{Bun07}, while the orbital liquid phase is the ground state
 elsewhere;
\hfill\break
 (v.d) the stability of the FOr is apparently due to the exponential
 tails in the DOS, particle density and kinetic energy, and so is
 specific for $d=\infty$, whereas at $d=3$ the OL phase is the
 lowest-energy state in the entire phase diagram.
\hfill\break
Perhaps feature (v.a) is the most remarkable quantitative consequence
of orbital physics. Having richer hopping processes both with and
without the restriction that the orbital flavor is conserved, one has
to accept that alternating orbital order is more difficult to realize
than in the spin model \cite{Faz90}.
Feature (v.b) makes it more difficult to realize fully polarized
ferro-orbital states so it may sound surprising that nevertheless the
FOr phase is more stable than the orbital liquid in a range of electron
densities $n\ge n_c$. Note that the theorem formulated by Nagaoka
\cite{Nag66} at $U=\infty$ for the spin case does not apply to the
orbital Hubbard model.

One of the attractive ideas in this field has been the possible
existence of polarized ferro-orbital states with partly filled
\textit{complex} orbitals \cite{Kho01}. We have established that
unfortunately such states cannot be realized. We have shown that
they are unstable and one has to consider instead FOr states.

\acknowledgments
We thank Walter Metzner for his interest and for very insightful
discussions.
L.F.F. thanks the Max Planck Institute for Solid State Research
in Stuttgart and the Jagiellonian University in Krak\'ow for their kind
hospitality.
\mbox{A.M.O. kindly} acknowledges Narodowe Centrum Nauki (NCN, Poland)
Project No. 2016/23/B/ST3/00839 and is grateful for support via the
Alexander von Humboldt \mbox{Foundation Fellowship} \cite{AvH}
\mbox{(Humboldt-Forschungspreis).}

\vfill
\appendix*

\section{ Proof that the orbital liquid phase is unpolarized }
\label{OLunpol}

Consider the cyclic permutation $\mathrm{C}$ defined by
\begin{eqnarray}
   \mathrm{C} {\bf k} &=& \mathrm{C} (k_{-m}, k_{-m+1}, ...., k_0,...., k_m)  \nonumber \\
                & & \equiv (k_m, k_{-m}, ...., k_{-1},...., k_{m-1}).
\label{Cdef}
\end{eqnarray}
Then we have, compare Eqs.~\eqref{Gk}--\eqref{Bk},
\begin{eqnarray}
   G_{{\mathrm{C} \bf k}} &=&  B_{\mathrm{C} {\bf k}} \,
   \mathrm{exp}( i \phi_{\mathrm{C} {\bf k}} )
  = \sum_n  \cos (\mathrm{C} {\bf k})_n \,  \mathrm{e}^{i\chi_n}   \nonumber \\
  &=& \sum_n  \cos k_{n-1} \,  \mathrm{e}^{i\chi_n}
  = \sum_n  \cos k_n \,  \mathrm{e}^{i\chi_{n+1}}       \nonumber \\
  &=& \mathrm{e}^{i 4\pi/d} \,  G_{ \bf k}
  = B_{\bf k} \, \mathrm{e}^{i 4\pi/d} \, \mathrm{exp}( i \phi_{\bf k} ),
\label{GCk}
\end{eqnarray}
and it follows that $B_{\mathrm{C} {\bf k}} =  B_{\bf k}$ and
$\phi_{\mathrm{C} {\bf k}}=\phi_{\bf k} + 4\pi/d$. The set of
${\bf k}$-vectors generated from ${\bf k}$ by successive permutations
is $\Pi({\bf k})=\{\mathrm{C}^n{\bf k}\, |-m\leq n\leq m\}$, so
\begin{eqnarray}
  \sum_{{\bf k}' \in  \Pi({\bf k})} \! \! \! \mathrm{e}^{i\phi_{{\bf k}'}}
  &=& \sum_n \mathrm{exp}(i\phi_{\mathrm{C}^n{\bf k}})
  = \sum_n \mathrm{exp}(i (\phi_{\bf k} + 4n\pi/d))      \nonumber \\
  &=& \mathrm{e}^{i \phi_{\bf k}} \, \sum_n  \mathrm{e}^{i  4n\pi/d}
  = \mathrm{e}^{i \phi_{\bf k}} \, \sum_n  \mathrm{e}^{i  \chi_n} = 0 \, .
\end{eqnarray}



\begin{thebibliography}{00}

\bibitem{Hub63} J. Hubbard,
                   Electron Correlations in Narrow Energy Bands,
                   Proceedings of the Royal Society of London
                   \textbf{A 276}, 238 (1963).

\bibitem{And59} P. W. Anderson,
                   New Approach to the Theory of Superexchange Interactions,
                   Phys. Rev. \textbf{115}, 2 (1959).

\bibitem{Vol89} W. Metzner and D. Vollhardt,
                   Correlated Lattice Fermions in $d=\infty$ Dimension,
                   Phys. Rev. Lett. \textbf{62}, 324 (1989).

\bibitem{Met88} W. Metzner and D. Vollhardt,
                   Analytic calculation of ground-state properties of
                   correlated fermions with the Gutzwiller wave-function,
                   Phys. Rev. B \textbf{37}, 7382 (1988).

\bibitem{Gut63} M. C. Gutzwiller,
                   Effect of correlation on the ferromagnetism
                   of transition metals,
                   Phys. Rev. Lett. \textbf{10}, 159 (1963).

\bibitem{Gut65} M. C. Gutzwiller,
                   Correlation of Electrons in a Narrow $s$ Band,
                   Phys. Rev. \textbf{137}, A1726 (1965).

\bibitem{Vol84} D. Vollhardt,
                   Normal He$^3$ --- An almost localized Fermi liquid,
                   Rev. Mod. Phys. \textbf{56}, 99 (1984).

\bibitem{Met89} W. Metzner,
                   Variational theory for correlated lattice fermions
                   in high dimension,
                   Z. Phys. B \textbf{77}, 253 (1989).

\bibitem{Met91} W. Metzner,
                   Analytic evaluation of resonating valence bond states,
                   Z. Phys. B \textbf{82}, 183 (1991).

\bibitem{And73} P. W. Anderson,
                   Resonating valence bonds: A new kind of insulator?,
                   Mater. Res. Bull. \textbf{8}, 153-160 (1973).

\bibitem{Faz90} P. Fazekas, B. Menge, and E. M\"uller-Hartmann,
                   Ground state phase diagram of the infinite dimensional
                   Hubbard model: A variational study,
                   Z. Phys. B \textbf{78}, 69 (1990).

\bibitem{Ish97} S. Ishihara, M. Yamanaka, and N. Nagaosa,
                   Orbital liquid in perovskite transition-metal oxides,
                   Phys. Rev. B \textbf{56}, 686 (1997).

\bibitem{Fei05} L. F. Feiner and A. M. Ole\'s,
                   Orbital liquid in ferromagnetic manganites:
                   The orbital Hubbard model for $e_g$ electrons,
                   Phys. Rev. B \textbf{71}, 144422 (2005).

\bibitem{Tan05} T. Tanaka, M. Matsumoto, and S. Ishihara,
                   Randomly diluted $e_g$ orbital-ordered systems,
                   Phys. Rev. Lett. \textbf{95}, 267204 (2005).

\bibitem{Tan07} T. Tanaka and S. Ishihara,
                   Dilution effects in two-dimensional quantum orbital systems,
                   Phys. Rev. Lett. \textbf{98}, 256402 (2007).

\bibitem{Cza17} P. Czarnik, J. Dziarmaga, and A. M. Ole\'s,
                   Overcoming the sign problem at finite temperature:
                   Quantum tensor network for the orbital $e_g$ model
                   on an infinite square lattice,
                   Phys. Rev. B \textbf{96}, 014420 (2017).

\bibitem{Bal10} L. Balents,
                   Spin liquids in frustrated magnets,
                   Nature (London) \textbf{464}, 199-208 (2010).

\bibitem{Sav17} L. Savary and L. Balents,
                   Quantum spin liquids: \mbox{A review,}
                   Rep. Prog. Phys. \textbf{80}, 016502 (2017).

\bibitem{Kug82} K. I. Kugel and D. I. Khomskii,
                   The Jahn-Teller effect and magnetism:
                   Transition metal compounds,
                   Usp. Fiz. Nauk \textbf{136}, 621 (1982)
                   [Sov. Phys. Usp. \textbf{25}, 231 (1982)].

\bibitem{Kho14} D. I. Khomskii,
                   \textit{Transition Metal Oxides}
                   (Cambridge University Press, Cambridge, 2014).

\bibitem{Ima98} M. Imada, A. Fujimori, and Y. Tokura,
                   Metal-insulator transitions,
                   Rev. Mod. Phys. \textbf{70}, 1039 (1998).

\bibitem{Tok00} Y. Tokura and N. Nagaosa,
                   Orbital Physics in Transition-Metal Oxides,
                   Science \textbf{288}, 462-468 (2000).

\bibitem{Gor10} A. V. Gorshkov, M. Hermele, V. Gurarie, C. Xu, P.~S.~Julienne,
                   J. Ye, P. Zoller, E. Demler, \mbox{M. D. Lukin,} and A. M. Rey,
                   Two-orbital SU(N) magnetism with ultracold alkaline-earth atoms,
                   Nature Phys. \textbf{6}, 289 (2010).

\bibitem{Kog04} A. Koga, N. Kawakami, T. M. Rice, and M. Sigrist,
                   Orbital-Selective Mott Transitions in the Degenerate Hubbard Model,
                   Phys. Rev. Lett. \textbf{92}, 216402 (2004).

\bibitem{Fei97} L. F. Feiner, A. M. Ole\'s, and J. Zaanen,
                   Quantum Melting of Magnetic Order due to Orbital Fluctuations,
                   Phys. Rev. Lett. \textbf{78}, 2799 (1997).

\bibitem{Fei98} L. F. Feiner, A. M. Ole\'s, and J. Zaanen,
                   Quantum disorder versus order-out-of-disorder
                   in the Kugel-Khomskii model,
                   J.~Phys.: Condens. Matter \textbf{10}, L555 (1998).

\bibitem{Kha97} G. Khaliullin and V. Oudovenko,
                   Spin and orbital excitation spectrum in the Kugel-Khomskii model,
                   Phys. Rev. B \textbf{56}, R14243 (1997).

\bibitem{Karlo} P. Corboz, M. Lajk\'o, A. M. La\"uchli, K. Penc, and F.~Mila,
                   Spin-Orbital Quantum Liquid on the Honeycomb Lattice,
                   Phys. Rev. X \textbf{2}, 041013 (2012).

\bibitem{Kha00} G. Khaliullin and S. Maekawa,
                   Orbital liquid in three-dimensional Mott insulator: LaTiO$_3$,
                   Phys. Rev. Lett. \textbf{85}, 3950 (2000).

\bibitem{Kha05} G. Khaliullin,
                   Orbital order and fluctuations in Mott insulators,
                   Prog. Theor. Phys. Suppl. \textbf{160}, 155 (2005).

\bibitem{Kit18} K. Kitagawa, T. Takayama, Y. Matsumoto, A. Kato, R.~Takano,
                   Y. Kishimoto, S. Bette, R. Dinnebier, G.~Jackeli, and H. Takagi,
                   A spin-orbital-entangled quantum liquid on a honeycomb lattice,
                   Nature \mbox{(London)} \textbf{554}, 341 (2018).

\bibitem{Fei99} L. F. Feiner and A. M. Ole\'s,
                   Electronic origin of magnetic and orbital ordering
                   in insulating LaMnO$_3$,
                   Phys. Rev. B \textbf{59}, 3295 (1999).

\bibitem{vdB99} J. van den Brink, P. Horsch, F. Mack, and A. M. Ole\'s,
                   Orbital dynamics in ferromagnetic transition-metal oxides,
                   Phys. Rev. B \textbf{59}, 6795 (1999).

\bibitem{Ole05} A. M. Ole\'s, G. Khaliullin, P. Horsch, and L. F. Feiner,
                   Fingerprints of spin-orbital physics in cubic Mott insulators:
                   Magnetic exchange interactions and optical spectral weights,
                   Phys. Rev. B \textbf{72}, 214431 (2005).

\bibitem{Rei05} A. Reitsma, L. F. Feiner, and A. M. Ole\'s,
                   Orbital and spin physics in LiNiO$_2$ and NaNiO$_2$,
                   New J. Phys. \textbf{7}, 121 (2005).

\bibitem{Cha08} J. Chaloupka and G. Khaliullin,
                   Orbital order and possible superconductivity
                   in LaNiO$_3$/LaMO$_3$ superlattices,
                   Phys. Rev. Lett. \textbf{100}, 016404 (2008).

\bibitem{Sol08} V. I. Solovyev,
                   Spin-orbital superexchange physics emerging
                   from interacting oxygen molecules in KO$_2$,
                   New J. Phys. \textbf{10}, 013035 (2008).

\bibitem{Nor08} B. Normand and A. M. Ole\'s,
                   Frustration and entanglement in the $t_{2g}$
                   spin-orbital model on a triangular lattice:
                   valence-bond and generalized liquid states,
                   Phys. Rev. B \textbf{78}, 094427 (2008).

\bibitem{Zaa09} F. Kr\"uger, S. Kumar, J. Zaanen, and J. van den Brink,
                   Spin-orbital frustrations and anomalous metallic
                   state in iron-pnictide superconductors,
                   Phys. Rev. B \textbf{79}, 054504 (2009).

\bibitem{Lv10}  W. Lv, F. Kr\"uger, and P. Phillips,
                   Orbital ordering and unfrustrated $(\pi,0)$ magnetism
                   from degenerate double exchange in the iron pnictides,
                   Phys. Rev. B \textbf{82}, 045125 (2010).

\bibitem{Nor11} B. Normand,
                   Multicolored quantum dimer models, resonating valence-bond
                   states, color visons, and the triangular-lattice $t_{2g}$
                   spin-orbital system,
                   Phys. Rev. B \textbf{83}, 094427 (2011).

\bibitem{Cha11} J. Chaloupka and A.~M.~Ole\'s,
                   Spin-orbital resonating valence-bond liquid on a triangular
                   lattice: Evidence from finite cluster diagonalization,
                   Phys. Rev. B \textbf{83}, 094427 (2011).

\bibitem{WohEPL} K. Wohlfeld, M. Daghofer, and A. M. Ole\'s,
                   Spin-orbital physics for $p$ orbitals in alkali
                   hyperoxides---Generalization of the Goodenough-Kanamori rules,
                   EPL (Europhysics Letters) \textbf{96}, 27001 (2011).

\bibitem{Woh11} K. Wohlfeld, M. Daghofer, S. Nishimoto, G. Khaliullin,
                   and J. van den Brink,
                   Intrinsic Coupling of Orbital Excitations
                   to Spin Fluctuations in Mott Insulators,
                   Phys. Rev. Lett. \textbf{107}, 147201 (2011).

\bibitem{Sch12} J. Schlappa, K. Wohlfeld, K. J. Zhou, M. Mourigal, M. W. Haverkort,
                   V.N. Strocov, L. Hozoi, C. Monney, S. Nishimoto, S. Singh,
                   A. Revcolevschi, J.-S. Caux, L. Patthey, H. M. Ronnow,
                   J. van den Brink, and T. Schmitt,
                   Spin-orbital separation in the quasi-one-dimensional
                   Mott insulator Sr$_2$CuO$_3$,
                   Nature (London) \textbf{485}, 82-U108 (2012).

\bibitem{Woh13} K. Wohlfeld, S. Nishimoto, M. W. Haverkort, and J. van den Brink,
                   Microscopic origin of spin-orbital separation in Sr$_2$CuO$_3$,
                   Phys. Rev. B \textbf{88}, 195138 (2013).

\bibitem{Bis15} V. Bisogni, K. Wohlfeld, S. Nishimoto, C. Monney,
                   J.~Trinckauf, K. Zhou, R. Kraus, K. Koepernik,
                   C.~Sekar, V. Strocov, B. B\"uchner, T. Schmitt,
                   J. van den Brink, and J. Geck,
                   Orbital Control of Effective Dimensionality:
                   From Spin-Orbital Fractionalization to Confinement
                   in the Anisotropic Ladder System CaCu$_2$O$_3$,
                   Phys. Rev. Lett. \textbf{114}, 096402 (2015).

\bibitem{CCC15} C.-C. Chen, M. van Veenendaal, T. P. Devereaux,
                   and K. Wohlfeld,
                   Fractionalization, entanglement, and separation:
                   Understanding the collective excitations in a spin-orbital chain,
                   Phys. Rev. B \textbf{91}, 165102 (2015).

\bibitem{Brz15} W. Brzezicki, A. M. Ole\'s, and M. Cuoco,
                   Spin-Orbital Order Modified by Orbital Dilution in
                   Transition-Metal Oxides: From Spin Defects to
                   Frustrated Spins Polarizing Host Orbitals,
                   Phys. Rev. X \textbf{5}, 011037 (2015).

\bibitem{Brz16} W. Brzezicki, M. Cuoco, and A. M. Ole\'s,
                   Novel Spin-Orbital Phases induced by Orbital Dilution,
                   J. Supercond. Novel Magn. \textbf{29}, 536 (2016).

\bibitem{Brz17} W. Brzezicki, M. Cuoco, and A. M. Ole\'s,
                   Exotic Spin-Orbital Physics in Hybrid Oxides,
                   J. Supercond. Novel Magn. \textbf{30}, 129 (2017).

\bibitem{Goode} J. B. Goodenough,
                   \textit{Magnetism and the Chemical Bond}
                   (Interscience, New York, 1963).

\bibitem{Ole06} A. M. Ole\'s, P. Horsch, L. F. Feiner, and G. Khaliullin,
                   Spin-Orbital Entanglement and Violation
                   of the Goodenough-Kanamori Rules,
                   Phys. Rev. Lett. \textbf{96}, 147205 (2006).

\bibitem{Ole12} A. M. Ole\'s,
                   Fingerprints of spin-orbital entanglement
                   in transition metal oxides,
                   J. Phys.: Condensed Matter \textbf{24}, 313201 (2012).

\bibitem{Brz12} W. Brzezicki, J. Dziarmaga, and A. M. Ole\'s,
                   Noncollinear Magnetic Order Stabilized
                   by Entangled Spin-Orbital Fluctuations,
                   Phys. Rev. Lett. \textbf{109}, 237201 (2012).

\bibitem{Brz13} W. Brzezicki, J. Dziarmaga, and A. M. Ole\'s,
                   Exotic spin orders driven by orbital fluctuations
                   in the Kugel-Khomskii model,
                   Phys. Rev. B \textbf{87}, 064407 (2013).

\bibitem{Sna19} M. Snamina and A. M. Ole\'s,
                   Magnon dressing by orbital excitations
                   in ferromagnetic planes of K$_2$CuF$_4$ and LaMnO$_3$,
                   New J. Phys. \textbf{21}, 023018 (2019).

\bibitem{Brz20} W. Brzezicki,
                   Spin, orbital and topological order in models
                   of strongly correlated electrons,
                   J. Phys.: Condensed Matter \textbf{32}, 023001 (2020).

\bibitem{Lee06} P. A. Lee, N. Nagaosa, and X.-G. Wen,
                   Doping a Mott insulator:
                   Physics of high-temperature superconductivity,
                   Rev. Mod. Phys. \textbf{78}, 17 (2006).

\bibitem{Oga08} M. Ogata and H. Fukuyama,
                   The $t$-$J$ model for the oxide high-$T_c$ superconductors,
                   Rep. Prog. Phys. \textbf{71}, 036501 (2008).

\bibitem{Kei15} B. Keimer, S. A. Kivelson, M. R. Norman, S. Uchida,
                   and \mbox{J. Zaanen,}
                   High Temperature Superconductivity in the Cuprates,
                   Nature \textbf{518}, 179 (2015).

\bibitem{Geo96} A. Georges, G. Kotliar, W. Krauth, and M. J. Rozenberg,
                   Dynamical mean-field theory of strongly correlated
                   fermion systems and the limit of infinite dimensions,
                   Rev. Mod. Phys. \textbf{68}, 13 (1996).

\bibitem{Mer00} J. Merino and R. H. McKenzie,
                   Transport properties of strongly correlated metals:
                   A dynamical mean-field approach,
                   Phys. Rev. B \textbf{61}, 7996 (2000).

\bibitem{Kyu06} B. Kyung, S. S. Kancharla, D. S\'en\'echal,
                   A.-M. S. Tremblay, M. Civelli, and G. Kotliar,
                   Pseudogap induced by short-range spin correlations
                   in a doped Mott insulator,
                   Phys. Rev. B \textbf{73}, 165114 (2006).

\bibitem{Kot06} G. Kotliar, S. Y. Savrasov, K. Haule, V. S. Oudovenko,
                   O. Parcollet, and C. A. Marianetti,
                   Electronic structure calculations with dynamical
                   mean-field theory,
                   Rev. Mod. Phys. \textbf{78}, 865 (2006).

\bibitem{Got20} D. Gotfryd, E. P\"arschke, J. Chaloupka, A. M. Ole\'s,
                   and K. Wohlfeld,
                   How Spin-Orbital Entanglement depends on the Spin-Orbit
                   Coupling in a Mott Insulator,
                   Phys. Rev. Research \textbf{2}, 013353 (2020).

\bibitem{Got20a} D. Gotfryd, E. P\"arschke, K. Wohlfeld, and A. M. Ole\'s,
                   Evolution of Spin-Orbital Entanglement
                   with Increasing Ising Spin-Orbit Coupling,
                   Condensed Matter \textbf{5}, 53 (2020).

\bibitem{Ole83} A. M. Ole\'s,
                   Antiferromagnetism and correlation of electrons in transition metals,
                   Phys. Rev. B \textbf{28}, 327 (1983).

\bibitem{Har03} A. B. Harris, T. Yildirim, A. Aharony,
                   Ora Entin-Wohlman, and I. Y. Korenblit,
                   Unusual Symmetries in the Kugel-Khomskii Hamiltonian,
                   Phys. Rev. Lett. \textbf{91}, 087206 (2003).

\bibitem{Dag08} M. Daghofer, K. Wohlfeld, A. M. Ole\'s, E. Arrigoni,
                   and P. Horsch,
                   Absence of Hole Confinement in Transition-Metal
                   Oxides with Orbital Degeneracy,
                   Phys. Rev. Lett. \textbf{100}, 066403 (2008).

\bibitem{Woh08} K. Wohlfeld, M. Daghofer, A.~M.~Ole\'s, and P. Horsch,
                   Spectral properties of orbital polarons in Mott insulators,
                   Phys. Rev. B \textbf{78}, 214423 (2008).

\bibitem{Wro10} P. Wr\'obel and A. M. Ole\'s,
                   Ferro-Orbitally Ordered Stripes in Systems
                   with Alternating Orbital Order,
                   Phys. Rev. Lett. \textbf{104}, 206401 (2010).

\bibitem{Wro12} P. Wr\'obel, R. Eder, and A.~M.~Ole\'s,
                   Optical conductivity due to orbital polarons
                   in systems with orbital degeneracy,
                   Phys. Rev. B \textbf{86}, 064415 (2012).

\bibitem{Hor99} P. Horsch, J. Jakli\v{c}, and F. Mack,
                   Optical conductivity of colossal-magnetoresistance compounds:
                   Role of orbital degeneracy in the ferromagnetic phase,
                   Phys. Rev. B \textbf{59}, 6217 (1999).

\bibitem{Hor99a} P. Horsch and F. Mack,
                   Optical conductivity in doped manganites
                   with planar $x^2-y^2$ orbital order,
                   Phys. Rev. Lett. \textbf{82}, 3160 (1999).

\bibitem{Zen51} C. Zener,
                   Interaction Between the $d$ Shells in the Transition Metals,
                   Phys. Rev. \textbf{81}, 440 (1951).

\bibitem{Dag01} E. Dagotto, T. Hotta, and A. Moreo,
                   Colossal magnetoresistant materials:
                   The key role of phase separation,
                   Phys. Rep. \textbf{344}, 1 (2001).

\bibitem{Dag05} E. Dagotto,
                   Open questions in CMR manganites, relevance of clustered states
                   and analogies with other compounds including the cuprates,
                   New J. Phys. \textbf{7}, 67 (2005).

\bibitem{Ram07} T. V. Ramakrishnan,
                   Modelling colossal magnetoresistance manganites,
                   J. Phys.: Condens. Matter \textbf{19}, 125211 (2007).

\bibitem{Tok06} Y. Tokura,
                   Critical features of colossal magnetoresistive manganites,
                   Rep. Prog. Phys. \textbf{69}, 797 (2006).

\bibitem{Kil99} R. Kilian and G. Khaliullin,
                   Orbital polarons in the metal-insulator transition of manganites,
                   Phys. Rev. B \textbf{60}, 13458 (1999).

\bibitem{Dag04} M. Daghofer, A. M. Ole\'s, and W. von der Linden,
                   \mbox{Orbital} polarons versus itinerant $e_g$
                   electrons in doped manganites,
                   Phys. Rev. B \textbf{70}, 184430 (2004).

\bibitem{Khaki} G. Khaliullin and R. Kilian,
                   Theory of anomalous magnon softening in ferromagnetic manganites,
                   Phys. Rev. B \textbf{61}, 3494 (2000).

\bibitem{Ole02} A. M. Ole\'s and L. F. Feiner,
                   Why spin excitations in metallic ferromagnetic
                   manganites are isotropic,
                   Phys. Rev. B \textbf{65}, 052414 (2002).

\bibitem{notesign} The sign convention taken here is natural for the purely
               electronic problem, and is opposite to that in \cite{Fei05} which
               is commonly used in the Jahn-Teller problem for $e_g$ doublets,
               {\it cf.\/} R.~Englman,
               {\it The Jahn-Teller Effect in Molecules and Crystals\/}
               (John Wiley, London, 1972).

\bibitem{Kla79} J. R. Klauder,
                   Path integrals and stationary-phase approximations,
                   Phys. Rev. D \textbf{19}, 2349 (1979).

\bibitem{notecoh} Note that $\theta_i$ is the azimuthal, not the polar angle.

\bibitem{Pet21} N. C. Petroni,
                   \textit{Probability and Stochastic Processes for Physicists}
                   (Springer, Berlin, 2021).

\bibitem{Bun98} J. B\"{u}nemann, W. Weber, and F. Gebhard,
                   Multiband Gutzwiller wave functions for general
                   on-site interactions,
                   Phys. Rev. B \textbf{57}, 6896 (1998).

\bibitem{Ng69}  E. W.~Ng and M.~Geller,
                   A Table of Integrals of the Error Functions,
                   J. Res. NBS B Math. Sci.  \textbf{73B}, 1 (1969).

\bibitem{Bri70} W. F. Brinkman and T.~M.~Rice,
                   Application of Gutzwiller's variational method
                   to the metal-insulator transition,
                   Phys. Rev. B \textbf{2}, 4302 (1970).

\bibitem{exci}  This is an idealization and at finite dimension the
                condition $d=0$ is satisfied only at $U=\infty$,
                because the Brinkman-Rice criterion neglects
                virtual excitations, which generate antiferromagnetism
                at finite $U$ in the spin Hubbard model.

\bibitem{vDo89} P. G. J. van Dongen, F.~Gebhard, and D. Vollhardt,
                   Variational evaluation of correlation functions
                   for lattice electrons in high dimensions,
                   Z. Phys. B \textbf{76}, 199-210 (1989).

\bibitem{Vol11} D. Vollhardt,
                   Dynamical Mean-Field Approach for Strongly
                   Correlated Materials, in: Lecture Notes of the
                   Autumn School on Correlated Electrons 2011, Vol. 1,
                   Eds. E. Pavarini, E. Koch, D. Vollhardt,
                   and A. Lichtenstein
                   (Verlag des Forschungszentrum J\"ulich, 2011).

\bibitem{Vol14} D. Vollhardt,
                   From Gutzwiller Wave Function to
                   Dynamical Mean-Field Theory, in: Lecture Notes of the
                   Autumn School on Correlated Electrons 2014, Vol. 4,
                   Eds. E. Pavarini, E. Koch, D. Vollhardt,
                   and A. Lichtenstein
                   (Verlag des Forschungszentrum J\"ulich, 2014).

\bibitem{Vol18} D. Vollhardt,
                   From Infinite Dimensions to Real Materials,
                   in: Lecture Notes of the Autumn School
                   on Correlated Electrons 2018, Vol. 8,
                   Eds. E. Pavarini, E. Koch, A. Lichtenstein,
                   and D. Vollhardt
                   (Verlag des Forschungszentrum J\"ulich, 2018).

\bibitem{Kot18} G. Kotliar,
                   Electronic Structure of Correlated Materials:
                   Slave-Boson Methods and Dynamical Mean-Field
                   Theory, in: Lecture Notes of the Autumn School
                   on Correlated Electrons 2018, Vol. 8,
                   Eds. E. Pavarini, E. Koch, A. Lichtenstein,
                   and D. Vollhardt
                   (Verlag des Forschungszentrum J\"ulich, 2018).

\bibitem{Fre03} J. K. Freericks and V. Zlatic,
                   Exact dynamical mean-field theory of the
                   Falicov-Kimball model,
                   Rev. Mod. Phys. \textbf{75}, 1333 (2003).

\bibitem{Wer14} H. Aoki, N. Tsuji, M. Eckstein, M. Kollar, T. Oka,
                   and P. Werner,
                   Nonequilibrium dynamical mean-field theory
                   and its applications,
                   Rev. Mod. Phys. \textbf{86}, 779 (2014).

\bibitem{deM05} L. de’Medici, A. Georges, and S. Biermann,
                   Orbital-selective Mott transition in multiband systems:
                   Slave-spin representation and dynamical mean-field theory,
                   Phys. Rev. B \textbf{72}, 205124 (2005).

\bibitem{Jak13} E. Jakobi, N. Bl\"umer, and P. van Dongen,
                   Orbital-selective Mott transitions in a doped
                   two-band Hubbard model with crystal field splitting,
                   Phys. Rev. B \textbf{87}, 205135 (2013).

\bibitem{Kol02} M. Kollar and D. Vollhardt,
                   Exact analytic results for the Gutzwiller wave
                   function with finite magnetization,
                   Phys. Rev. B \textbf{65}, 155121 (2002).

\bibitem{Fre97} R. Fr\'esard and G. Kotliar,
                   Interplay of Mott transition and ferromagnetism
                   in the orbitally degenerate Hubbard model,
                   Phys. Rev. B \textbf{56}, 12909 (1997).

\bibitem{Obe97} T. Obermeier, T. Pruschke, and J. Keller,
                   Ferromagnetism in the large-$U$ Hubbard model,
                   Phys. Rev. B \textbf{56}, R8479 (1997).

\bibitem{Faz99} P. Fazekas,
			       \textit{Lecture Notes on Electron Correlation and Magnetism}
			       (World Scientific, Singapore, 1999).

\bibitem{Bul99} R. Bulla,
                   Zero temperature metal-insulator transition in the
                   infinite-dimensional Hubbard model,
                   Phys. Rev. Lett. \textbf{83}, 136 (1999).

\bibitem{Roz99} M. J. Rozenberg, R. Chitra, and G. Kotliar,
                    Finite Temperature Mott Transition in the Hubbard Model
                    in Infinite Dimensions,
                    Phys. Rev. Lett. \textbf{83}, 3498 (1999).

\bibitem{Bul01} R. Bulla, T. A. Costi, and D. Vollhardt,
                   Finite-temperature numerical renormalization
                   group study of the Mott transition,
                   Phys. Rev. B \textbf{64}, 045103 (2001).

\bibitem{Kat20} E. G. C. P. van Loon, F. Krien, and A. A. Katanin,
                   Bethe-Salpeter Equation at the Critical End Point
                   of the Mott Transition,
                   Phys. Rev. Lett. \textbf{125}, 136402 (2020).

\bibitem{Kug15} K. I. Kugel, D. I. Khomskii, A. O. Sboychakov,
                   and \mbox{S. V. Streltsov,}
                   Spin-orbital interaction for face-sharing octahedra:
                   Realization of a highly symmetric SU(4) model,
                   Phys. Rev. B \textbf{91}, 155125 (2015).

\bibitem{Val19} V. E. Valiulin, A. V. Mikheyenkov, K. I. Kugel,
                   and A.~V.~Barabanov,
                   Thermodynamics of Symmetric Spin-Orbital Model:
                   One- and Two-Dimensional Cases,
                   JETP Letters \textbf{109}, 546-551 (2019).

\bibitem{Lau15} N. J. Laurita, J. Deisenhofer, L.-D. Pan, C. M. Morris,
                   M. Schmidt, M. Johnsson, V. Tsurkan, A. Loidl,
                   and N. P. Armitage,
                   Singlet-Triplet Excitations and Long-Range Entanglement
                   in the Spin-Orbital Liquid Candidate FeSc$_2$S$_4$,
                   Phys. Rev. Lett. \textbf{114}, 207201 (2015).

\bibitem{Tak19} H. Takagi, T. Takayama, G. Jackeli, G. Khaliullin,
                   and S. E. Nagler,
                   Kitaev quantum spin liquid---concept and materialization,
                   Nature Rev. Phys. \textbf{1}, 264-280 (2019).

\bibitem{Bun07} J. B\"{u}nemann, K. J\'avorne-Radn\'oczi, P. Fazekas,
                   and F.~Gebhard,
                   Orbital order in degenerate Hubbard models:
                   A variational study,
                   J. Phys.: Condensed Matter \textbf{19}, 326217 (2007).

\bibitem{Nag66} Y. Nagaoka,
                   Ferromagnetism in a narrow, almost half-filled $s$ band,
                   Phys. Rev. \textbf{147}, 392 (1966).

\bibitem{Kho01} J. van den Brink and D. Khomskii,
                   Orbital ordering of complex orbitals
                   in doped Mott insulators,
                   Phys. Rev. B \textbf{63}, 140416(R) (2001).
                   
\bibitem{AvH}   https://www.humboldt-foundation.de/web/humboldt-preis.html
               


\end{thebibliography}
\end{document}